\documentclass[12pt, letterpaper]{article}

\usepackage{setspace,graphicx,epstopdf,amsmath,amsfonts,amssymb,amsthm}
\usepackage{marginnote,datetime,enumitem,rotating, fancyvrb}
\usepackage{hyperref, float}
\usepackage{verbatim}
\usepackage{booktabs}
\usepackage{dcolumn}
\usepackage{subfigure}
\usepackage{bm}
\usepackage[longnamesfirst]{natbib}
\usdate
\usepackage{cleveref}
\usepackage{arydshln}

\usepackage{indentfirst} 
\usepackage{jfe}          
\usepackage[margin=1in]{geometry}

\newtheorem{claim}{Claim}
\newtheorem{assumption}{Assumption}[section]
\newtheorem{proposition}{Proposition}

\newtheorem{definition}{Definition}

\crefname{lemma}{Lemma}{Lemmas}
\crefname{claim}{Claim}{Claims}
\crefname{proposition}{Proposition}{Propostions}
\crefname{figure}{Figure}{Figures}
\crefname{table}{Table}{Tables}
\crefname{equation}{Eq.}{Eqs.}

\newcommand{\ind}[1]{#1_{it}}

\newcommand{\pd}{\mathcal{S}}

\newcommand\subfigsize{0.5\textwidth}

\newcolumntype{d}[1]{D..{#1}} 

\makeatletter
\renewcommand*\env@matrix[1][*\c@MaxMatrixCols c]{%
	\hskip -\arraycolsep
	\let\@ifnextchar\new@ifnextchar
	\array{#1}}
\makeatother

\begin{document}

\setlist{noitemsep}  




\title{General Equilibrium Under Convex Portfolio Constraints and Heterogeneous Risk Preferences\footnotetext{\hspace{-18pt}Sciences Po, Department of Economics, 28 rue des Saints P\`eres, Paris, 75007, France\\ \textit{E-mail address}: tyler.abbot@sciencespo.fr\\}\footnotetext{I would like to thank my advisors Nicolas C\oe{}urdacier and St\'ephane Guibaud for their support during this research. I would also like to thank Georgy Chabakauri for the insight that motivated the foundation of this paper, as well as Semyon Malamud, Julien Hugonnier, Ronnie Sircar, Gordon Zitkovic, Jean-Fran\c{c}ois Chassagneux, Thomas Pumir, Thomas Bourany, Nicolo Dalvit, Riccardo Zago, and Edoardo Giscato for helpful discussions. Finally, I should thank the participants at the EPFL Brown Bag Seminar, the Scineces Po Lunch Seminar, Princeton Informal Doctoral Seminar, 2017 RES Meeting, Paris 6/7 MathFiProNum seminar, and the SIAM MMF Conference for their questions and comments. A portion of this work was funded by an Alliance Doctoral Mobility Grant and by a Princeton-Sciences Po PhD Exchange Grant.}}

\author{Tyler Abbot}



\renewcommand{\thefootnote}{\fnsymbol{footnote}}

\singlespacing

\maketitle

\vspace{-.2in}

\begin{abstract}
	This paper characterizes the equilibrium in a continuous time financial market populated by heterogeneous agents who differ in their rate of relative risk aversion and face convex portfolio constraints. The model is studied in an application to margin constraints and found to match real world observations about financial variables and leverage cycles. It is shown how margin constraints increase the market price of risk and decrease the interest rate by forcing more risk averse agents to hold more risky assets, producing a higher equity risk premium. In addition, heterogeneity and margin constraints are shown to produce both pro- and counter-cyclical leverage cycles. Beyond two types, it is shown how constraints can cascade and how leverage can exhibit highly non-linear dynamics. Finally, empirical results are given, documenting a novel stylized fact which is predicted by the model, namely that the leverage cycle is both pro- and counter-cyclical.
\end{abstract}

\medskip



\clearpage


\onehalfspacing
\setcounter{footnote}{0}
\renewcommand{\thefootnote}{\arabic{footnote}}
\setcounter{page}{1}
\section*{Introduction}
Market incompleteness and individual heterogeneity are two important characteristics of financial markets. Many markets exhibit incompleteness, in the sense that one cannot freely choose their portfolio choices either because of constraints imposed by lenders, because of regulatory constraints, or simply because of a true incompleteness in the market. At the same time, in order to generate trade among individuals they must differ in some form. If all agents were identical then market prices would make them indifferent to their portfolio holdings and there would never be any trade. This paper seeks to combine these two facts about financial markets by combining portfolio constraints and preference heterogeneity, with a particular application to margin constraints. 

Margin constraints increase the market price of risk and decrease the interest rate, contributing to a higher equity risk premium. The interest rate is low because the constraint limits the supply of risk free bonds to the market. This limit in supply pushes up the bond price and down the interest rate. On the other hand, the market price of risk is high because constrained agents are unable to leverage up to take advantage of high returns. On the opposite side of this constraint are risk averse agents who would like to sell their risky assets to reduce the volatility of their consumption. They are unable to do so, given the counter-party is the constrained agent. Thus margin constraints create an implicit liquidity constraint which allows the market price of risk to remain high in order to compensate risk averse agents for having a riskier consumption stream.

Asset prices are higher or lower than in an unconstrained equilibrium depending on whether the income effect or the substitution effect dominates. When agents are constrained, other agents are forced to hold more risky assets. These unconstrained agents hold both risky and risk-free assets, implying that an increase in the market price of risk and a decrease in the risk-free rate represent an ambiguous change in the investment opportunity set. However, the effect tends to increase the equity risk premium. This has the effect of increasing the discount rate and simultaneously makes agents wealthier today and makes consumption tomorrow less expensive. The first effect (an income effect) causes agents to desire to consume more today. The second effect (a substitution effect) causes agents to desire to consume less today and more tomorrow. Which effect dominates depends on the EIS of unconstrained agents. If EIS is less than one then the substitution effect dominates: individuals consume less today, pushing up their wealth and thus increasing asset prices. If EIS is greater than one then the income effect dominates: individuals consume more today, pushing down their wealth and thus reducing asset prices. In this way we can see either an increase or a decrease in asset prices when some portion of agents are constrained, depending on whether the EIS is greater or less than one.

Margin constraints and preference heterogeneity generate both pro- and counter-cyclical leverage cycles. Less risk averse agents dominate the economy and the price of risky assets is high when aggregate production is high. High asset prices increase individual wealth and reduce leverage. On the contrary, risk averse agents dominate when aggregate production is low, reducing asset prices. Low asset prices cause individual wealth to be low and individual leverage to be high. With the introduction of a margin constraint less risk averse agents eventually run into a borrowing limit. Not only is borrowing reduced, but, as discussed before, asset prices can be higher under constraint. In turn, total leverage falls. In this way heterogeneous preferences and margin constraints produce both pro- and counter-cyclical leverage cycles.

Financial leverage has become an important policy variable since the crisis of 2007-2008. In particular leverage allows investors to increase the volatility of balance sheet equity, producing the possibility of greater returns. At the same time leveraged investors are exposed to larger down-side risk. In the face of negative shocks, constrained investors must sell assets to reduce their leverage. This is known as the "leverage cycle". The associated credit contraction produces large volatility in asset prices and has been the target of regulation in the post-crisis era (e.g. the Basel III capital requirement rules). However, leverage cyclicality remains a topic of debate. Leverage cyclicality is both pro- and counter- cyclical in the model presented here, depending on the aggregate state of the economy and the marginal agent. In \cref{sec:data} I document in the data that cycles are both pro- and counter-cyclical depending on the level of aggregate asset pricing variables which can be interpreted as proxies for marginal preferences. This fact could reconcile some of the empirical debates about the cyclicality of leverage and re-enforces the study of preference heterogeneity as a driver of financial trade.

Convex portfolio constraints arise quite naturally in finance. A convex constraint simply states that the portfolio weights must lie in a convex set containing zero (see \cite{stiglitz1981credit} for an example of micro-foundations to credit constraints). In macroeconomics there are countless examples of particular models with market incompleteness which can be described in this setting of convex constraints, such as \cite{aiyagari1994uninsured,kiyotaki1997credit,krusell1998income,bernanke1999financial} and many others. A margin constraint essentially states that an agent cannot borrow infinitely against their equity. This type of constraint is seen in consumer finance when borrowing money to purchase a home: one must almost always put up a down payment. In financial markets margin constraints arise in repo markets and other lending vehicles (see \cite{hardouvelis1992margin,hardouvelis2002asymmetric,adrian2010changing} for empirical studies of margins). In fact real world experience motivated the theoretical study of leverage cycles initiated by \cite{geanakoplos1996promises}. In addition limits to arbitrage and financial bubbles have been studied under margin constraints in the context of liquidity (see e.g. \cite{brunnermeier2009market}). Many of these phenomena arise in the model presented in this paper, but the predictions for leverage cycles are emphasized because of their novelty.    

In theoretical models leverage cyclicality depends greatly on the underlying assumptions producing trade. In his foundational work on the topic, \cite{geanakoplos1996promises} shows how the combination of belief heterogeneity and margin constraints produce a pro-cyclical leverage cycle. However, this finding is in opposition to the contemporary paper by \cite{kiyotaki1997credit}, where participation constraints force agents to invest through intermediaries, whose credit constraints produce a counter-cyclical leverage cycle. More recently \cite{he2013intermediary} and \cite{brunnermeier2014macroeconomic} also produce counter-cyclical leverage cycles by including intermediaries through whom constrained agents can profit from risky assets. In fact, \cite{he2013intermediary} even points out the debate in the applied literature and the fact that, "[Their] model does not capture the other aspects of this process, ... that some parts of the financial sector reduce asset holdings and deleverage." These models imply that the mechanism producing trade determines leverage's cyclicality.

The empirical literature has noted this ambiguity over the cyclicality of leverage in different.  \cite{korajczyk2003capital} study the capital structure of firms and find that leverage is counter-cyclical for unconstrained firms and pro-cyclical for constrained firms. However, \cite{halling2016leverage} contradict this by showing that target leverage is counter-cyclical once you account for variation in explanatory variables, pointing out that the effect in \cite{korajczyk2003capital} is only the "direct effect". In the cross section of the economy \cite{adrian2010liquidity} find that leverage is counter-cyclical for households, ambiguous for non-financial firms, and pro-cyclical for broker dealers. However, the authors study the relationship between leverage and changes in balance sheet assets. This comparison produces a mechanical correlation which somewhat disappears when assets are replaced by GDP growth as a proxy for the business cycle (see \cref{fig:data:scatter_sector_89}). \cite{ang2011hedge} point out that when accounting for prices broker dealer leverage is counter-cyclical, but that hedge fund leverage is pro-cyclical. These contrary studies can be reconciled when controlling for financial variables such as the price/dividend ratio or the interest rate. In fact, for several sectors studied (see \cref{sec:data}) the leverage cycle is \textit{both} pro- and counter-cyclical. This ambiguity is predicted by the model of preference heterogeneity and margin constraints presented here.

Many authors have criticized the assumption of a representative, constant relative risk aversion agent since \cite{mehra1985equity} posited the equity risk premium puzzle. The definition of new utility functions was the first major response to this puzzle, in particular Epstein-Zin preferences (\cite{epstein1989substitution,weil1989equity}) and habit formation (\cite{campbell1999force}) have been used to explain the equity premium puzzle. However, several papers have studied preferences across individuals and found them to be heterogeneous and constant in time (\cite{brunnermeier2008wealth,chiappori2011relative,chiappori2012aggregate}), contradicting both of these new branches of the theoretical literature. In addition, \cite{epstein2014much} pointed out that the assumptions necessary to match the risk premium using Epstein-Zin preferences produce unrealistic preference for early resolution of uncertainty. Beyond these criticisms, one needs heterogeneity in order to generate trade at all in any market model. In a representative agent setting one looks for the prices which make the agent indifferent to \textit{not} trading. Risk preference heterogeneity has succeeded in partially responding to these issues.

Heterogeneity in risk preferences has been used to generate trade in financial models since the foundational paper of \cite{dumas1989two}. Since then many authors have studied the problem from different angles, assuming different levels of market completeness, utility functions, participation constraints, information structures, etc., but almost always under the assumption of only two preference types (\cite{basak1998equilibrium,coen2004effects,guvenen2006reconciling,kogan2007equity,guvenen2009parsimonious,cozzi2011risk,garleanu2011margin,hugonnier2012rational,rytchkov2014asset,longstaff2012asset,prieto2010dynamic,christensen2012equilibrium,bhamra2014asset,chabakauri2013dynamic,chabakauri2015asset,garleanu2015young,santos2010habit}). \cite{cvitanic2011financial} studies the problem of $N$ agents with several dimensions of heterogeneity and focuses on the dominant agents, characterizing portfolios via the Malliavan calculus. \cite{abbot2016heterogeneous} studies a setting with $N$ heterogeneous CRRA agents in a complete financial market using a value function approach and shows how changes in the number of types can produce substantially different quantitative results and how the variance in preferences provides an additional degree of freedom for explaining the equity risk premium puzzle. However, that work produces large amounts of aggregate leverage and high individual margins. This observation points towards the need to introduce some degree of constraint or incompleteness to better match the real world. To that end, this paper studies the same type of economy with $N$ heterogeneous CRRA agents under convex portfolio constraints with an application to margin constraints.

A fundamental paper by \cite{cvitanic1992convex} studied the general case of convex portfolio constraints in partial equilibrium. The authors developed an ingenious way to embed the agent in a series of fictitious economies, parameterized via a sort of Kuhn-Tucker condition, and then to select the appropriate market to make the agent just indifferent. However, their approach was to use convex duality to characterize the solution, which relies on a strict assumption that the relative risk aversion be bounded above by one. This limitation led others to look to solve the primal problem directly, such as \cite{he1993labor,cuoco1994dynamic,cuoco1997optimal,karatzas2003optimal}. These works use dense and complex mathematical techniques which may or may not provide tractable solutions for calculation. The present paper takes a more direct approach to solve the primal problem by noticing that homogeneous preferences are associated to a value function which factors into a function of wealth and a function of the aggregate state, under the appropriate ansatz. Using this ansatz, the Hamilton-Jacobi-Bellman equation becomes a PDE over consumption weights.

\section{A Model of Preference Heterogeneity} \label{sec:Mod}
\subsection{Financial Markets}
Consider a continuous time, infinite horizon \cite{lucas1978asset} economy with one consumption good. This consumption good, denoted $D_t$, is produced by a tree whose dividend follows a geometric Brownian motion (GBM):
\begin{align*}
\frac{dD_t}{D_t} = \mu_D dt + \sigma_D dW_t
\end{align*}
where $W_t$ is a standard Brownian motion and $(\mu_D, \sigma_D)$ are constants. Agents can trade in a (locally) risk-free and a risky security, whose prices are denoted $S^0_t$ and $S_t$ respectively. These prices are assumed to follow an exponential and an It\^o process, respectively:\\
\begin{tabular}{p{6.0cm}p{6.575cm}}
	{\begin{align} 
		\frac{dS_t}{S_t} &= \mu_t dt + \sigma_t dW_t \label{eq:S}
		\end{align}}
	&
	{\begin{align} 
		\frac{dS^0_t}{S^0_t} &= r_t dt \label{eq:S0}
		\end{align} }
\end{tabular}\\
where $(\mu_t, \sigma_t, r_t)$ are determined in equilibrium. Individuals are initially endowed with a share in the per-capita tree, $\alpha_{i0}$, and a position in the risk-free asset, $\beta_{i0}$.

\subsection{Preferences and Wealth}
The economy is populated by an arbitrary number $N$ of atomistic agents indexed by $i \in \{1, \dots, N\}$. Agents have constant relative risk aversion (CRRA) preferences and differ in their rate of relative risk aversion, $\gamma_i$, such that their instantaneous utility is given by
\begin{align*}
u_i(c) = \frac{c^{1-\gamma_i}}{1 - \gamma_i}
\end{align*}
Denote by $X_{it}$ an individual's wealth at time $t$ and note that initial wealth is given by $X_{i0} = \alpha_{i0}S_0 + \beta_{i0}S^0_t$. Denote by $\pi_{it}$ the share of an individual's wealth invested in the risky stock, which implies $1 - \pi_{it}$ is the share invested in the bond. Assuming that trading strategies are self financing, an individual's wealth evolves as
\begin{align*}
dX_{it} = \left [ X_{it} \left ( r_t + \pi_{it} \left( \mu_t + \frac{D_t}{S_t} - r_t \right) \right ) - c_{it} \right ] dt + X_{it} \pi_{it} \sigma_t dW_t
\end{align*}

\subsection{Portfolio Constraints and Individual Optimization}
Individual investors solve a utility maximization problem subject to their self-financing budget constraint and a portfolio constraint:
\begin{equation*} \label{eq:maximization}
\begin{aligned}
& \underset{\{ c_{it},\pi_{it} \}_{t=0}^\infty}{\text{max}}
& & \mathbb{E} \int_{0}^{\infty} e^{-\rho t}\frac{c_{it}^{1-\gamma_i}}{1-\gamma_i}dt \\
& \text{s.t.}
& &  dX_{it} = \left [ X_{it} \left ( r_t + \pi_{it} \left( \mu_t + \frac{D_t}{S_t} - r_t \right) \right ) - c_{it} \right ] dt + X_{it} \pi_{it} \sigma_t dW_t\\
& & & \pi_{it} \in \Pi_i
\end{aligned}
\end{equation*}
where $\Pi_i \subseteq \mathbb{R}$ represents a closed, convex region of the portfolio space which contains $\{0\}$. For example $\Pi_i = \mathbb{R}$ is the unconstrained case, $\Pi_i = \mathbb{R}^+$ is a short sale constraint, $\Pi_i = \{ \pi : \pi \leq m_i \hspace{5pt} | \hspace{5pt} m_i \geq 0 \}$ is a margin constraint. This set is allowed to differ across agents, as implied by the subscript. This paper focuses on an application to  margin constraints, but the approach is applicable to any constraint which can be written as a function of the aggregate state.

\subsection{Equilibrium}
Investors are considered to be atomistic and thus I consider a \cite{radner1972existence} type equilibrium.
\begin{definition}
	An equilibrium in this economy is defined by a set of processes \\
	$\{r_t, S_t, \{\ind{c}, \ind{X}, \ind{\pi} \}_{i=1}^N\} \hspace{5pt} \forall \hspace{5pt} t$, given preferences and initial endowments, such that $\{\ind{c}, \ind{X}, \ind{\pi}\}$ solve the agents' individual optimization problems and the following set of market clearing conditions is satisfied:
	\begin{align}\label{market_clearing}
	{\sum_i}\ind{c} = D_t \text{  ,  }
	{\sum_i} (1 - \ind{\pi})\ind{X} = 0 \text{  ,  }
	{\sum_i} \ind{\pi}\ind{X} = S_t 
	\end{align}
\end{definition}
\noindent I study Markovian equilibria such that equilibrium quantities can be written as functions of some state vector. That is for some equilibrium process $Y_t$, I look for functions $f(\cdot)$ such that $Y_t = f(z_t)$ for some process $z_t$. I will look for a particular equilibrium in the vector of consumption weights defined by
\begin{align*}
\bm{\omega} = [ \omega_{1t}, \dots, \omega_{(N-1)t} ]^T = [c_{1t}/D_t, \dots, c_{(N-1)t}/D_t]^T
\end{align*}
This is in the spirit of \cite{chabakauri2013dynamic,chabakauri2015asset}, where given two agents we can take the consumption weight of a single agent as the state variable. I've only included $N-1$ consumption weights because the last is determined by market clearing. However, for some equations the full vector of weights is useful, so I will define
\begin{align*}
\bm{\Omega} = [ \omega_{1t}, \dots, \omega_{Nt} ]^T = [c_{1t}/D_t, \dots, c_{Nt}/D_t]^T
\end{align*}    
The following section will describe how to characterize equilibrium processes in terms of these quantities.

\section{Equilibrium Characterization} \label{sec:sol}
To solve this problem I begin with the approach of \cite{cvitanic1992convex}. This method uses a fictitious, unconstrained economy and a shadow cost of constraint, or Lagrange multiplier, to find the correct pricing process. Unlike in their work I do not use a duality approach, but show how the primal problem admits a Markov representative. The reason this works is because when preferences are homothetic they can be represented by a utility function which is homogeneous of some degree. In this case the value function factors and the resulting ODE is no longer a function of individual wealth. This approach will likewise work for any homogeneous utility function, including Epstein-Zin\footnote{In particular, first order conditions from a dynamic program give consumption as $c = u'^{-1}\left(\partial_X J(X, Y)\right)$, where $Y$ is any arbitrary, aggregate state vector and $X$ an individual's wealth. We would like to find $c = X/V(Y)$. Equate these and rearrange to find $\partial_XJ(X, Y) = u'(X/V(Y))$. When the utility function is homogeneous of degree $k+1$, $u'(\cdot)$ is homogeneous of degree $k$. Thus $\partial_X J(X, Y) = u'(1)V(Y)^{-k} X^{k}$. By integrating with respect to $X$ one finds a proposal for the value function such that consumption is a linear function of wealth.}. This process will be described in the following subsections.

\subsection{Optimality in Fictitious Unconstrained Economy}
In order to find the constrained equilibrium, we define new processes for individual prices, which are "adjusted" by a process $\ind{\nu}$, considered the shadow cost of constraint:
\begin{align*}
\frac{dS^0_{it}}{S^0_{it}} &= (r_t + \delta_i(\nu_{it}))dt\\
\frac{dS_{it}}{S_{it}} &= (\mu_t + \ind{\nu} + \delta_i(\ind{\nu})) dt + \sigma_t dW_t
\end{align*}
The function $\delta_i(\cdot)$ is the support function of $\Pi_i$, which is defined as
\begin{align*}
\delta_i(\nu) = \sup_{\pi \in \Pi_i}( -\nu \pi )
\end{align*}
In addition, this gives rise to the effective domain of $\ind{\nu}$ defined by $\mathcal{N}_i = \{\nu \in \mathbb{R} : \delta_i(\nu) < \infty\}$ (for examples see \cref{table:constraint_sets_example}). Finally, we have a complimentary slackness condition which states $\ind{\nu}\ind{\pi} + \delta_i(\ind{\nu}) = 0$. Each agent solves their optimization problem in the face of their individual, fictitious financial market.

\begin{table}
	\begin{center}
		\begin{tabular}{ld{4.6}d{4.6}d{4.6}}
			\hline
			& \multicolumn{1}{c}{\begin{tabular}{@{}c@{}} $\Pi_i$ \end{tabular}}
			& \multicolumn{1}{c}{\begin{tabular}{@{}c@{}} $\delta_i(\nu)$ \end{tabular}}
			& \multicolumn{1}{c}{\begin{tabular}{@{}c@{}} $\mathcal{N}_i$ \end{tabular}}\\
			\hline
			\hline
			\rule{0pt}{4ex}Unconstrained & \multicolumn{1}{c}{\begin{tabular}{@{}c@{}} $\mathbb{R}$ \end{tabular}}  & \multicolumn{1}{c}{\begin{tabular}{@{}c@{}} $0$ \end{tabular}} & \multicolumn{1}{c}{\begin{tabular}{@{}c@{}} $0$ \end{tabular}} \\
			\rule{0pt}{4ex}Margin Constraint & \multicolumn{1}{c}{\begin{tabular}{@{}c@{}} $\{ \pi : \pi \leq m_i; m_i \geq 0\}$ \end{tabular}}  & \multicolumn{1}{c}{\begin{tabular}{@{}c@{}} $-\nu m_i$ \end{tabular}} & \multicolumn{1}{c}{\begin{tabular}{@{}c@{}} $\mathbb{R}_-$ \end{tabular}} \\
			\rule{0pt}{4ex}Short-Sale Constraint & \multicolumn{1}{c}{\begin{tabular}{@{}c@{}} $\{ \pi : \pi \geq s_i; s_i \leq 0\}$ \end{tabular}}  & \multicolumn{1}{c}{\begin{tabular}{@{}c@{}} $-\nu s_i$ \end{tabular}} & \multicolumn{1}{c}{\begin{tabular}{@{}c@{}} $\mathbb{R}_+$ \end{tabular}}\rule[-2ex]{0pt}{0ex} \\
			\hline
		\end{tabular}
	\end{center}
	\vspace{-10pt}
	\caption{Examples of constraint sets, support functions, and effective domain of the adjustment $\nu_i$.}
	\label{table:constraint_sets_example}
\end{table}

Define the stochastic discount factor (SDF) of an individual agent as an It\^o process which evolves as a function of the individual's adjustment:
\begin{align}\label{eq:ind_sdf}
\frac{d\ind{H}}{\ind{H}} = -(r_t + \delta_i(\ind{\nu}))dt - \left( \theta_t + \frac{\ind{\nu}}{\sigma_t} \right)dW_t
\end{align}
By a straight-forward application of the martingale approach (\cite{karatzas1987optimal}) in this fictitious economy one finds individual consumption as a function of individual SDF's:
\begin{align}
c_{it} = \left( \Lambda_i e^{\rho t} H_{it} \right)^{-\frac{1}{\gamma_i}} \label{eq:foc}
\end{align}
for all $i$, where $\Lambda_i$ is the Lagrange multiplier associated to the static budget constraint. In the case where $\Pi_i=\mathbb{R}$ $\forall$ $i$, the SDF's coincide and the ratios of marginal utilities are constant. However, when agents are constrained in their portfolio choice this is not the case and we have
\begin{align*}
\frac{\ind{c}^{-\gamma_i}}{c_{jt}^{-\gamma_j}} = \frac{\Lambda_i \ind{H}}{\Lambda_j H_{jt}}
\end{align*}
These ratios of SDF's, which are proportional to ratios of marginal utilities, are very familiar in the theory of incomplete market equilibria. In \cite{cuoco2001dynamic}, a representative agent with state dependent preferences is studied, where the preferences are a weighted average of individual preferences. The stochastic weights are exactly equal to the ratio of marginal utilities. This is also seen in \cite{basak1998equilibrium} and \cite{hugonnier2012rational}.

\subsection{General Equilibrium Characterization} \label{subsec:geneq}
Equilibrium is characterized by first assuming the existence of a Markovian equilibrium, deriving a system of ODE's for wealth-consumptions ratios, then recovering the adjustments $\ind{\nu}$ using the complimentary slackness conditions. Given this it is possible to prove optimality of the value functions\footnote{This remains a claim at this point. The proof is ongoing.}. First, consider the interest rate and market price of risk:

\begin{proposition} \label{prop:r_theta}
	The interest rate and market price of risk can be shown to be functions of weighted averages of individuals' consumption weights, preference parameters, and adjustments such that
	\begin{align}
	\theta_t &= \frac{1}{\sum_i \frac{\ind{\omega}}{\gamma_i}}\left(\sigma_D - \frac{1}{\sigma_t } \sum_i \frac{\ind{\omega}\ind{\nu}}{\gamma_i} \right) \label{eq:prop1_theta} \\
	r_t &= \frac{1}{\sum_i \frac{\ind{\omega}}{\gamma_i}} \left( \mu_D + \rho\sum_i \frac{\ind{\omega}}{\gamma_i} - \sum_i \frac{\ind{\omega}}{\gamma_i}\delta_i(\ind{\nu}) \right.\\
	&\hspace{70pt}\left. - \frac{1}{2}\sum_i \frac{1 + \gamma_i}{\gamma_i^2} \left(\theta_t + \frac{\ind{\nu}}{\sigma_t} \right)^2 \ind{\omega} \right) \label{eq:prop1_r}
	\end{align}
\end{proposition}

The interest rate and market price of risk take a typical form, but are augmented by the adjustment to individuals' marginal utilities. First notice that the market price of risk (\cref{eq:prop1_theta}) is determined by the fundamental volatility $\sigma_D$ divided by the weighted average of elasticity of intertemporal substitution (EIS), exactly as in complete markets (\cite{abbot2016heterogeneous}). In addition the constraint will either increase or reduce the market price of risk, depending on the domain of $\ind{\nu}$. In the case of margin constraints $\ind{\nu} \leq 0$, so the market price of risk will be weakly higher under constraint. This is driven by an implicitt liquidity constraint. Constrained agents are unable to take advantage of high returns. In addition, the effect of volatility implies that in times when stock price volatility is low, greater constraint implies greater returns. This correlation is again driven by the fact that agents cannot borrow to take advantage of the returns, producing the same type of liquidity effect described in the limits-to-arbitrage literature (e.g. \cite{brunnermeier2009market} or \cite{hugonnier2012rational}). Risk neutral agents would arbitrage away the high returns, but cannot because of their margin constraint.

The interest rate similarly exhibits a familiar shape. We see a rate of time preference term, an intertemporal smoothing term, and a prudence or risk preference term:
\begin{align*}
r_t &= \underbrace{\rho}_{\text{Rate of Time Preference}}
+ \underbrace{\frac{\mu_D - \sum_i \frac{\ind{\omega}}{\gamma_i}\delta_i(\ind{\nu}) }{\sum_i \frac{\ind{\omega}}{\gamma_i}}}_{\text{Intertemporal Smoothing}} 
- \underbrace{\frac{1}{2}\frac{ \sum_i \frac{1 + \gamma_i}{\gamma_i^2} \left(\theta_t + \frac{\ind{\nu}}{\sigma_t} \right)^2 \ind{\omega}}{ \sum_i \frac{\ind{\omega}}{\gamma_i}}}_{\text{Prudence/Risk Preferences}}
\end{align*}
Both the intertemporal smoothing and prudence terms are augmented by the constraint. Under a homogeneous margin constraint, $\delta_i(\ind{\nu}) = - m\ind{\nu}$, but recall that $\ind{\nu} \leq 0$, which together imply that the constraint reduces interest rates through the intertemporal smoothing term. Constrained agents are unable to supply bonds to the market in order to transfer consumption and wealth from the future to today. A lower supply of bonds pushes up the price and down the interest rate. At the same time constraint affects the interest rate through the prudence motive by changing the demand for precautionary savings. Individuals demand more precautionary savings when their SDF is more volatile (\cite{kimball1990precautionary}). When agents are constrained, their SDF is less volatile as they are unable to increase their exposure to fundamental risk. Ceterus paribus, this reduces the demand for precautionary savings and increases the interest rate, counteracting the intertemporal motive. Together these forces produce an equity risk premium which depends on the shape of heterogeneity, the degree of constraint, and the state variable, all driven by the individual consumption weights which determine the marginal agents.

How consumption weights evolve over time is important not only from an economic perspective, but also in order to derive the solution of the model. We can study the dynamics of consumption weights by applying It\^o's lemma and matching coefficients to find their drift and diffusion:
\begin{proposition}\label{prop:weights}
	Consumption weights follow an It\^o process whose dynamics are given by:
	\begin{align}
	\frac{d\omega_{it}}{\omega_{it}} &= \mu_{\omega it}dt + \sigma_{\omega it}dW_t \nonumber \\
	\text{where} \nonumber\\
	\mu_{\omega it} &= \frac{1}{\gamma_i} \left( r_t + \delta_i(\ind{\nu}) - \rho + \frac{1}{2}\frac{1 + \gamma_i}{\gamma_i} \left( \theta_t + \frac{\ind{\nu}}{\sigma_t} \right)^2 - \sigma_D\left( \theta_t + \frac{\ind{\nu}}{\sigma_t} \right) \right)\nonumber\\
	& + \sigma_D^2 - \mu_D \label{eq:prop:mu_w}\\
	\sigma_{\omega it} &= \frac{1}{\gamma_i}\left( \theta_t + \frac{\ind{\nu}}{\sigma_t} \right) - \sigma_D
	\label{eq:prop:sig_w}
	\end{align}
	This implies that the state variable $\bm{\omega}$ follows an It\^o process such that
	\begin{align}\label{eq:prop:bmwdynamics}
	d\bm{\omega} = \bm{\mu_\omega}dt + \bm{\sigma_\omega}dW_t
	\end{align}
	where $\bm{\mu_\omega} = [\mu_{\omega 1t}\omega_{1t}, \dots, \mu_{\omega (N-1)t}]^T$ and $\bm{\sigma_\omega} = [\sigma_{\omega 1t}\omega_{1t}, \dots, \sigma_{\omega (N-1)t}]^T$
\end{proposition}
\noindent These equations are very similar to those one finds in the complete markets case (\cite{abbot2016heterogeneous}), but augmented by the constraint. In particular, consider the volatility of consumption weights given in \cref{eq:prop:sig_w}. An agent's consumption volatility is exactly zero when their preference parameter satisfies
\begin{align*}
\gamma_i = \frac{1}{\xi_t} - \frac{\Xi_t}{\xi_t}\frac{1}{\sigma_D} + \frac{\ind{\nu}}{\sigma_D \sigma_t} \hspace{10pt} \text{where} \hspace{10pt}
\xi_t = \sum_i \frac{\ind{\omega}}{\gamma_i} \hspace{10pt} \text{ , } \hspace{10pt} 		  \Xi_t = \sum_i \frac{\ind{\omega}\ind{\nu}}{\gamma_i}
\end{align*}
We can think of this as the marginal preference level in the market for consumption. However, it is possible that this preference level is not unique. Consider the case where some agents face a margin constraint, but others do not. Amongst the unconstrained agents, the marginal preference level corresponds to the first two terms, while among the constrained agents all of the terms matter. Given $\ind{\nu} \leq 0$ under margin constraints, there could very well exist both a constrained and an unconstrained agent who have zero consumption volatility. This is driven by the constrained agents being unable to leverage up to gain more exposure to aggregate risk.

Since $\theta_t$ and $r_t$ are functions of $\{\omega_{it}\}_{i=1}^N$, it remains to show that $\{\nu_{it}\}_{i=1}^N$ are as well. First, one can derive a system of PDE's for individual wealth/consumption ratios, from which one can determine the adjustments.
\begin{proposition}\label{prop:PDE}
	Given \cref{prop:r_theta,prop:weights} and assuming adjustments and volatility are functions of $\bm{\omega}$ such that $\nu_i(\bm{\omega}) = \ind{\nu}$ and $\sigma(\bm{\omega}) = \sigma_t$, it is possible to define the interest rate and market price of risk as functions of $\bm{\omega}$ such that $r_t = r(\bm{\omega})$ and $\theta_t = \theta(\bm{\omega})$. Assuming there exists a Markovian equilibrium in $\bm{\omega}$, the individuals' wealth-consumption ratios, $V_i(\bm{\omega}) = \ind{X}/\ind{c}$, satisfy PDE's given for each $i$ by
	\begin{equation}
	\begin{aligned}\label{eq:prop:pde}
	&\hspace{10pt}\frac{1}{\gamma_i} \left[ (1 - \gamma_i)(r(\bm{\omega}) + \delta_i(\nu_i(\bm{\omega}))) - \rho + \frac{1 - \gamma_i}{2\gamma_i} \left(\theta(\bm{\omega}) + \frac{\nu_i(\bm{\omega})}{\sigma(\bm{\omega})} \right)^2 \right] V_i(\bm{\omega}) + \\
	&\left[ \frac{1 - \gamma_i}{\gamma_i} \left(\theta(\bm{\omega}) + \frac{\nu_i(\bm{\omega})}{\sigma(\bm{\omega})} \right) \bm{\sigma_\omega}^T + \bm{\mu_\omega}^T \right] \nabla V_i(\bm{\omega}) + \frac{1}{2} \bm{\sigma_\omega}^T HV_i(\bm{\omega})\bm{\sigma_\omega} + 1 = 0
	\end{aligned}
	\end{equation}
	where $\nabla$ and $H_{\bm{\omega}}$ represent the gradient and hessian operators, and where $\bm{\mu_\omega}$ and $\bm{\sigma_\omega}$ are given in \cref{prop:weights}.\\
	
	\noindent Boundary conditions when a single agent dominates are given by the autarkical case, where
	\begin{equation}\label{eq:prop:eq_bound}
	\begin{aligned}
	\lim\limits_{\omega_{i} \rightarrow 1} V_j(\bm{\omega}) = \frac{\gamma_j}{\rho - (1 - \gamma_j)\left( \frac{\left( \theta(\bm{\omega}) + \nu_i(\bm{\omega})/\sigma(\bm{\omega}) \right)^2}{2\gamma_j} + r(\bm{\omega}) + \delta_j(\nu_j(\bm{\omega})) \right)} 
	\end{aligned}	
	\\
	\end{equation}
	Boundary conditions when an agent's weight goes to zero are given by the solution to an $N-1$ agent problem.
\end{proposition}
\noindent The boundary conditions when a single agent dominates represent the vertices of the state space. On the other hand, when an agent's weight goes to zero, the economy solution is equivalent to a two agent economy, with the zero agent's wealth/consumption ratio still satisfying \cref{eq:prop:pde}, but their choices having no effect on aggregate variables. These partial differential equations represent the shape of individuals' wealth/consumption ratios over the state space. Unlike in complete markets, however, the system is highly non-linear, since the coefficients depend in a complicated way on the solution itself.

Next, consider the portfolios of individuals, given in \cref{prop:portfolios}.
\begin{proposition}\label{prop:portfolios}
	Assuming adjustments and volatility can be written as functions of $\bm{\omega}$ such that $\nu_i(\bm{\omega}) = \ind{\nu}$ and $\sigma(\bm{\omega}) = \sigma_t$, it can be shown that portfolios are functions of $\bm{\omega}$ such that $\pi_i(\bm{\omega}) = \ind{\pi}$, where
	\begin{align}
	\pi_i(\bm{\omega}) &= \frac{1}{\gamma_i \sigma(\bm{\omega})}\left( \theta(\bm{\omega}) + \frac{\nu_i(\bm{\omega})}{\sigma(\bm{\omega})} + \gamma_i \frac{\bm{\sigma}_{\bm{\omega}}(\bm{\omega})^T \nabla \bm{V}(\bm{\omega})}{V_i(\bm{\omega})} \right) \label{eq:prop2_opt_sep_pi}
	\end{align}
	where $\bm{\sigma}_{\bm{\omega}}(\bm{\omega}) = [\sigma_{\omega i}(\bm{\omega}) \omega_i]_i^T$ is the vector of diffusions of $\bm{\omega}$.
\end{proposition}
\noindent One can see right away that portfolios take the typical ICAPM form (\cite{merton1971optimum}). There is first a myopic term, represented by the market price of risk scaled down by risk aversion and volatility, which gives the instantaneous portfolio demand of an individual given the market price of risk. Next is a hedging term, determined by the co-movement of an individual's wealth with the aggregate state. Finally, there is a constraint term, which compensates the individual's portfolio such that they are within the constraint set.

On an aggregate level, we can derive asset pricing variables from an application of It\^o's lemma and from market clearing for wealth.
\begin{proposition}\label{prop:asset_prices}
	Assuming adjustments can be written as functions of $\bm{\omega$} such that $\nu_i(\bm{\omega}) = \ind{\nu}$, it can be shown that volatility and the price dividend ratio are functions of $\bm{\omega}$ such that $\sigma(\bm{\omega}) = \sigma_t$ and $\mathcal{S}(\bm{\omega}) = S_t/D_t$, where
	\begin{equation}\label{eq:prop:eq_sigmat}
	\begin{aligned}
	\sigma(\bm{\omega}) = \sigma_D +\frac{ \bm{\sigma}_{\bm{\omega}}(\bm{\omega})^T \left( \nabla \bm{V}(\bm{\omega})+ \bm{J_V}(\bm{\omega})^T\bm{\Omega} \right) }{\pd(\bm{\omega})}
	\end{aligned}
	\end{equation}
	where $\bm{J_V}(\bm{\omega})$ represents the Jacobian matrix and where
	\begin{equation}\label{eq:prop:eq_pd}
	\begin{aligned}
	\pd(\bm{\omega}) = \sum_{i} \omega_i V_i(\bm{\omega})
	\end{aligned}
	\end{equation}
	represents the price dividend ratio $S_t/D_t$.
\end{proposition}
Volatility in \cref{eq:prop:eq_sigmat} is driven by the fundamental volatility, the shape of wealth consumption ratios, and the volatility of consumption weights. When agents have high volatility in consumption weights, the volatility of asset prices will be higher. At the same time, individuals' wealth will be less volatile under constraint. This will produce a reduction in volatility. We will see these two forces in the numerical simulations in \cref{sec:num}.

We need to derive an expression for $\lbrace \ind{\nu} \rbrace_{i=1}^N$ in order to close the model. The functional form depends on the type of constraint. To that end, I will focus from here only on margin constraints. The following proposition gives the functional form for the adjustments under homogeneous margin constraints when $\ind{\pi} \leq m$ for all $i$, where $m \geq 0$, which implies an effective domain of $\mathcal{N}_i = \{ \nu: \nu \leq 0 \}$ and a support function of $\delta_i(\nu) = - m \nu$.
\begin{proposition}\label{prop:adjustment_margin}
	Under margin constraints, adjustments can be written as functions of $\bm{\omega}$ such that $\nu_i(\bm{\omega}) = \ind{\nu}$, where
	\begin{equation}\label{eq:prop:eq_nu}
	\begin{aligned}
	\nu_i(\bm{\omega}) = \min \left\lbrace 0 ; m \gamma_i \sigma(\bm{\omega})^2 \left( 1 - \frac{1}{m\sigma(\bm{\omega})} \left( \frac{\theta(\bm{\omega})}{\gamma_i} + \frac{\bm{\sigma}_{\bm{\omega}}(\bm{\omega})^T \nabla \bm{V}(\bm{\omega})}{V_i(\bm{\omega})} \right) \right) \right\rbrace 
	\end{aligned}
	\end{equation}
\end{proposition}

Finally, we need a verification argument for optimality of the value functions. In particular, we would like to be sure that solution to the PDE's in \cref{prop:PDE} are indeed the wealth/consumption ratios associated to the individuals optimal choices. If we are willing to make the assumption that the wealth/consumption ratios are twice continuously differentiable, then we can easily show using It\^o's lemma that the value functions are indeed optimal (this proceeds as in \cite{chabakauri2015asset}). However, it would be preferable to relax this assumption. To do so we can make use of a powerful new result in \cite{confortola2017backward}, namely that under certain conditions the value function implied by \cref{prop:PDE} is indeed optimal\footnote{This is left as a claim, as only an outline of a proof has been completed.}.
\begin{claim}\label{prop:verification}
	Assuming that individual wealth/consumption ratios $V_i(\bm{\omega})$ are $C^1$ with bounded first derivative, then there exists a unique solution to \cref{eq:prop:pde} (in the viscosity sense) and this solution corresponds to the value functions in \cref{eq:proof_prop2_seperable}. Furthermore this represents a Markovian equilibrium satisfied by \cref{prop:r_theta,prop:weights,prop:portfolios,prop:PDE,prop:adjustment_margin,prop:asset_prices}.
\end{claim}
\noindent This claim relies only on a single degree of differentiability\footnote{I believe this condition can be relaxed to simply Lipschitz continuity.}.

\section{Numerical Solution}\label{sec:num}
This section presents numerical results for several assumptions about the distribution of preferences. First, the case of two types is evaluated and the cyclicality of the leverage cycle is emphasized. The leverage cycle is pro- or counter-cyclical depending on the marginal agent. Second, results are presented for three agents. Two key features which are not observed in the two agent case are the possibility of cascading constraints and a highly non-monotonic leverage. When one agent is constrained, other agents tend to hold more leverage. This pushes the intermediate agent closer to their own constraint. At the same time, this increase in individual leverage can partially or even fully offset the reduction in total leverage generated by the first agent's constraint. Over all simulations I hold fixed $(\mu_D, \sigma_D, \rho) = (0.01, 0.032, 0.02)$, chosen to compare to \cite{chabakauri2015asset}.

\subsection{Two Types and Leverage Cycles}\label{sec:sec:two_types}
Consider the case of two agents\footnote{The two agent model represents the boundary of the three agent problem, so is its solution is necessary to treat the three agent case. In addition, understanding the shape of functions in this simple case will help to fix ideas in the more complex case of arbitrary number of types.} with relative risk aversion 
$(\gamma_1, \gamma_2) = (1.1, 5.0)$ who face a margin constraint such that the share, $\pi_{it}$, of their wealth invested in the risky asset is less than some constant $m_i$. This is equivalent to a leverage constraint:
\begin{align*}
\pi_{it} \leq m_i \Leftrightarrow \frac{\alpha_{it}S_t}{\alpha_{it}S_t + b_{it}S_t^0} \leq m_i \Leftrightarrow \frac{1}{1 + \frac{b_{it}S_t^0}{\alpha_{it}S_t}} \leq m_i \Leftrightarrow -\frac{b_{it}S_t^0}{\alpha_{it}S_t} \leq \frac{m_i - 1}{m_i}
\end{align*}
In particular, take $m_i = m = 1.2$.

There will exist a region of the state space over which this constraint binds for the less risk-averse agent. In this region, the more risk averse agent holds a larger share of their wealth in the risky asset. In order to achieve these portfolio weights, the constrained agent holds fewer risky shares and the unconstrained agent holds more risky shares. By reducing their risky shares, the less risk-averse agent's constraint actually tightens, causing them to sell more risky-shares, making the effect more than proportional. This corresponds to a substantial decline in leverage and a tightening of credit demand, pushing down the interest rate. In addition, the market price of risk is high in order to compensate the risk averse investor for holding a larger share. Whether these two effects combine to make asset prices higher or lower depends on whether the income or wealth effect dominates.

\begin{figure}[!h]
	\subfigure[]{
		\includegraphics[width=\subfigsize]{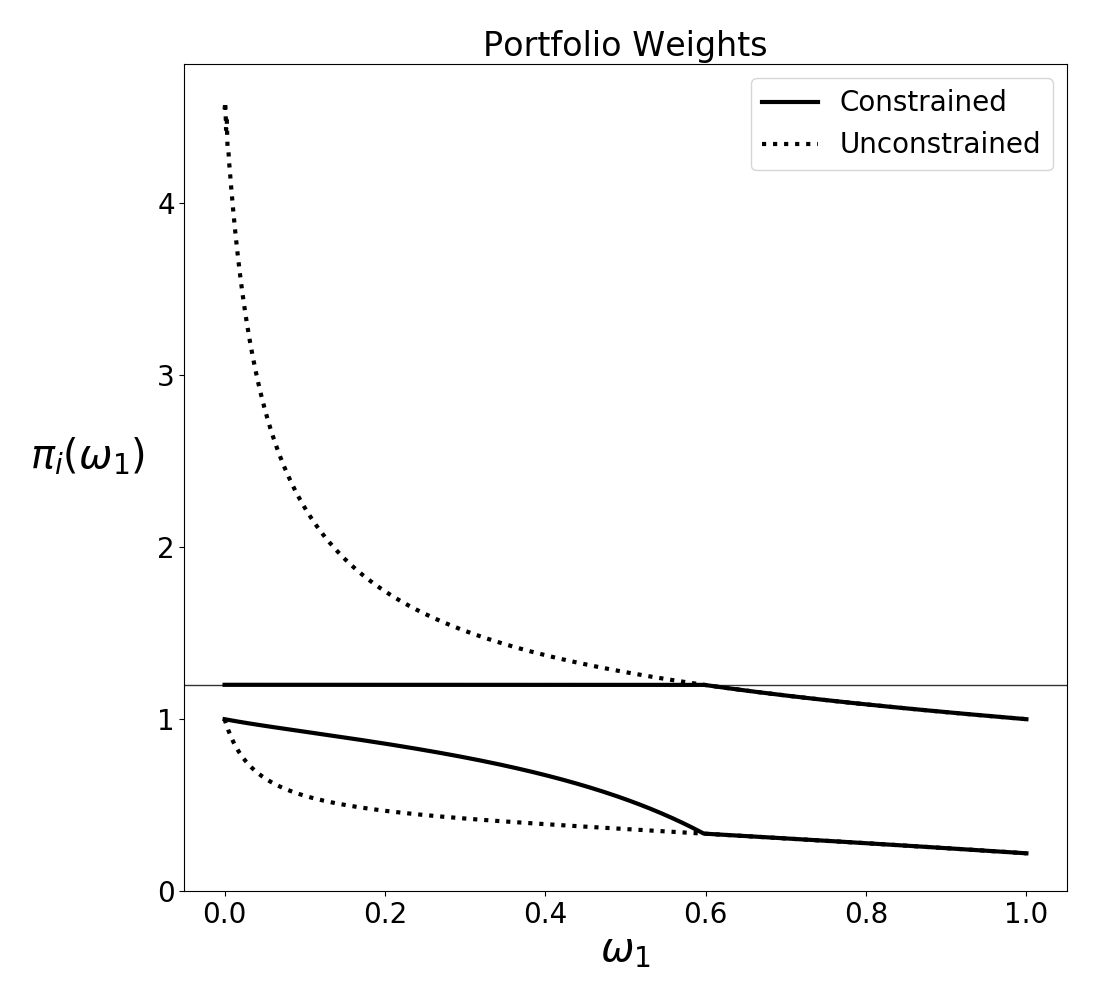}
		\label{fig:two_agents:pi}
	}\hspace{-10pt}
	\subfigure[]{
		\includegraphics[width=\subfigsize]{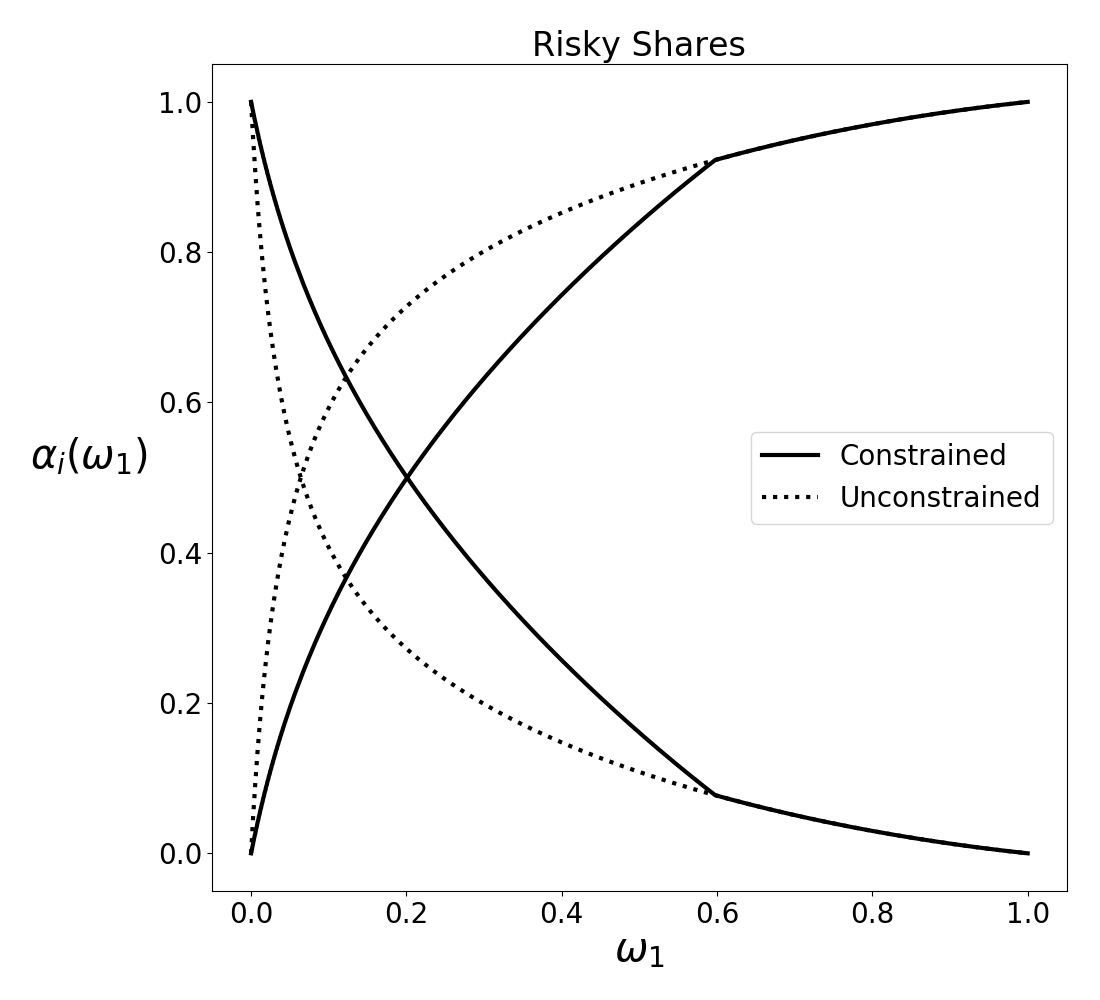}
		\centering
		\label{fig:two_agents:shares}
	}\hspace{-10pt}
	\subfigure[]{
		\includegraphics[width=\subfigsize]{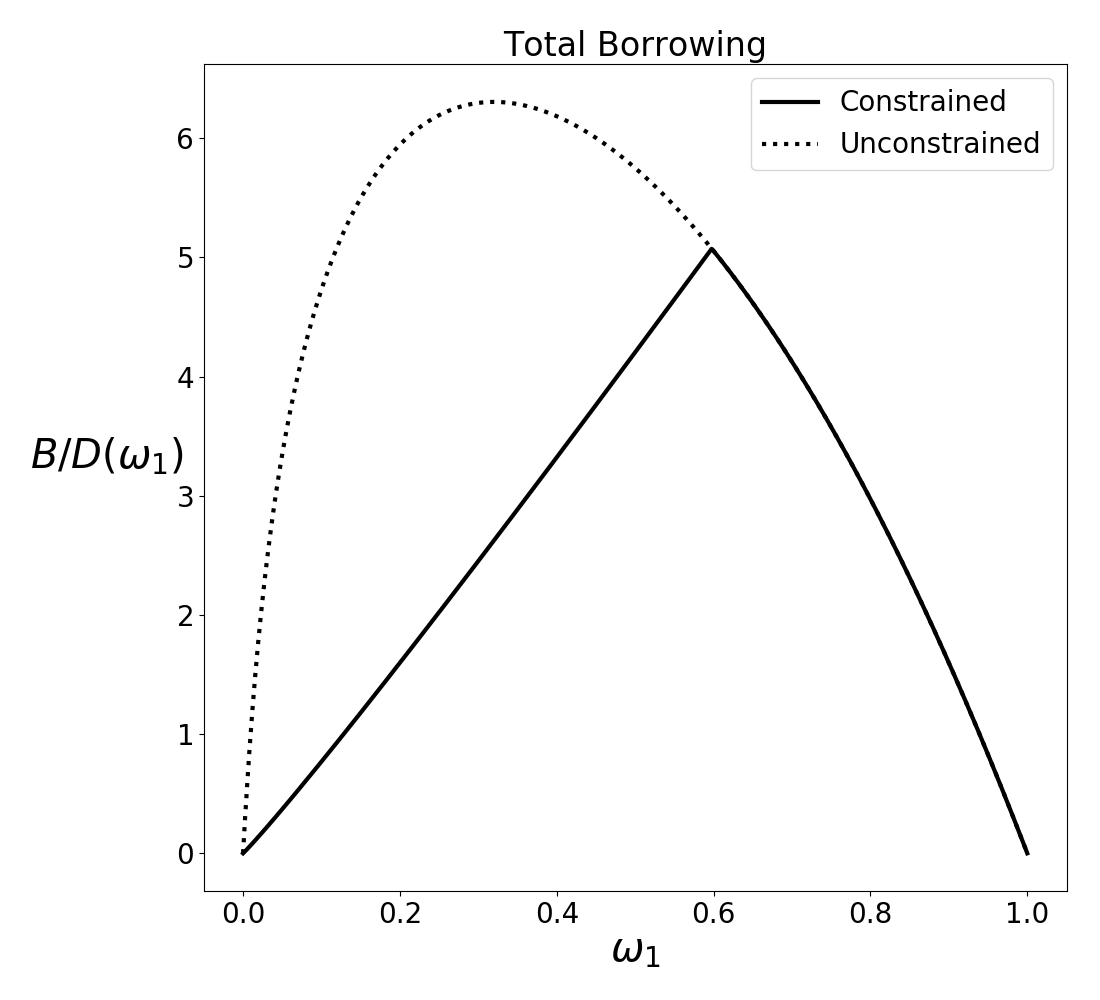}
		\centering
		\label{fig:two_agents:borrow}
	}\hspace{-10pt}
	\subfigure[]{
		\includegraphics[width=\subfigsize]{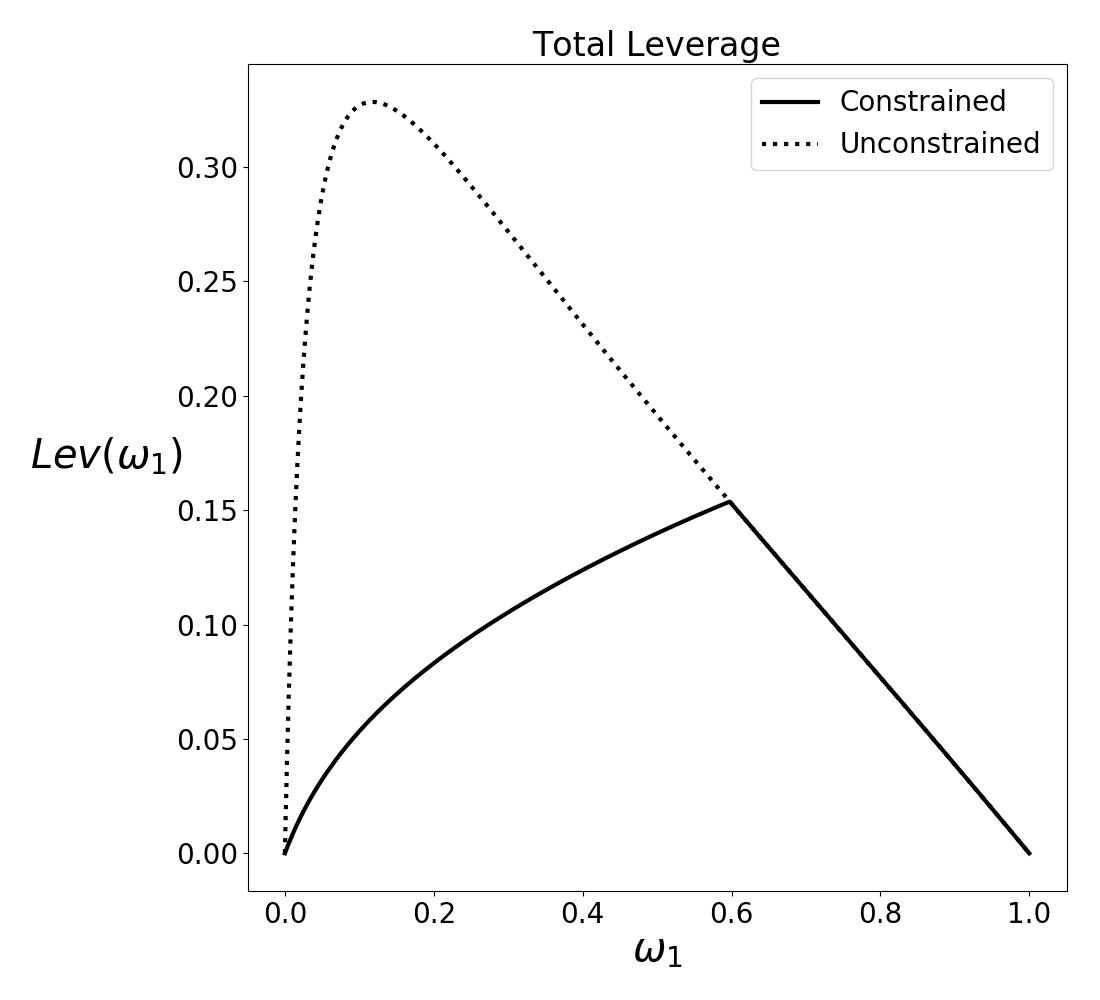}
		\centering
		\label{fig:two_agents:leverage}
	}\hspace{-10pt}
	\caption{}
	\label{fig:two_agents:1}
\end{figure}

\begin{figure}[!h]
	\subfigure[]{
		\includegraphics[width=\subfigsize]{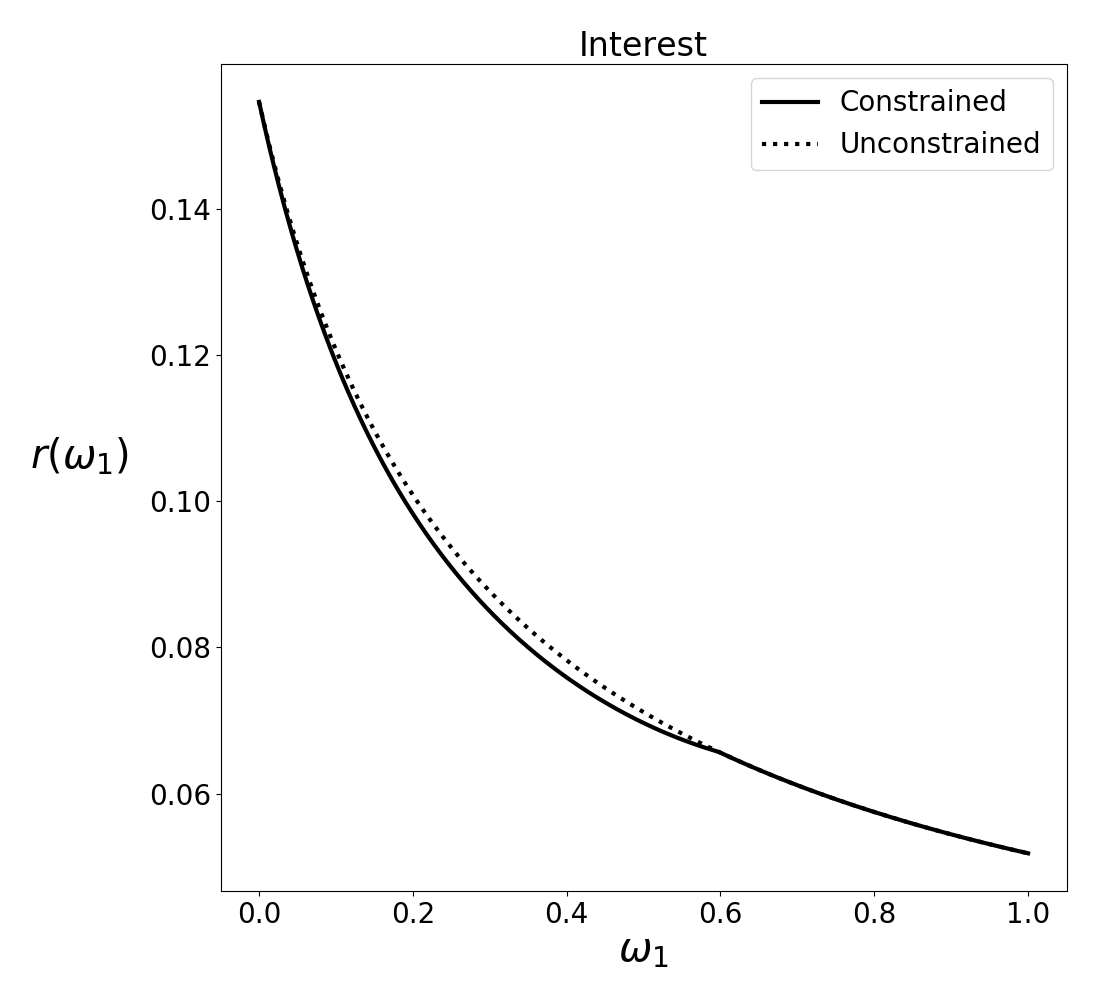}
		\centering
		\label{fig:two_agents:r}
	}\hspace{-10pt}
	\subfigure[]{
		\includegraphics[width=\subfigsize]{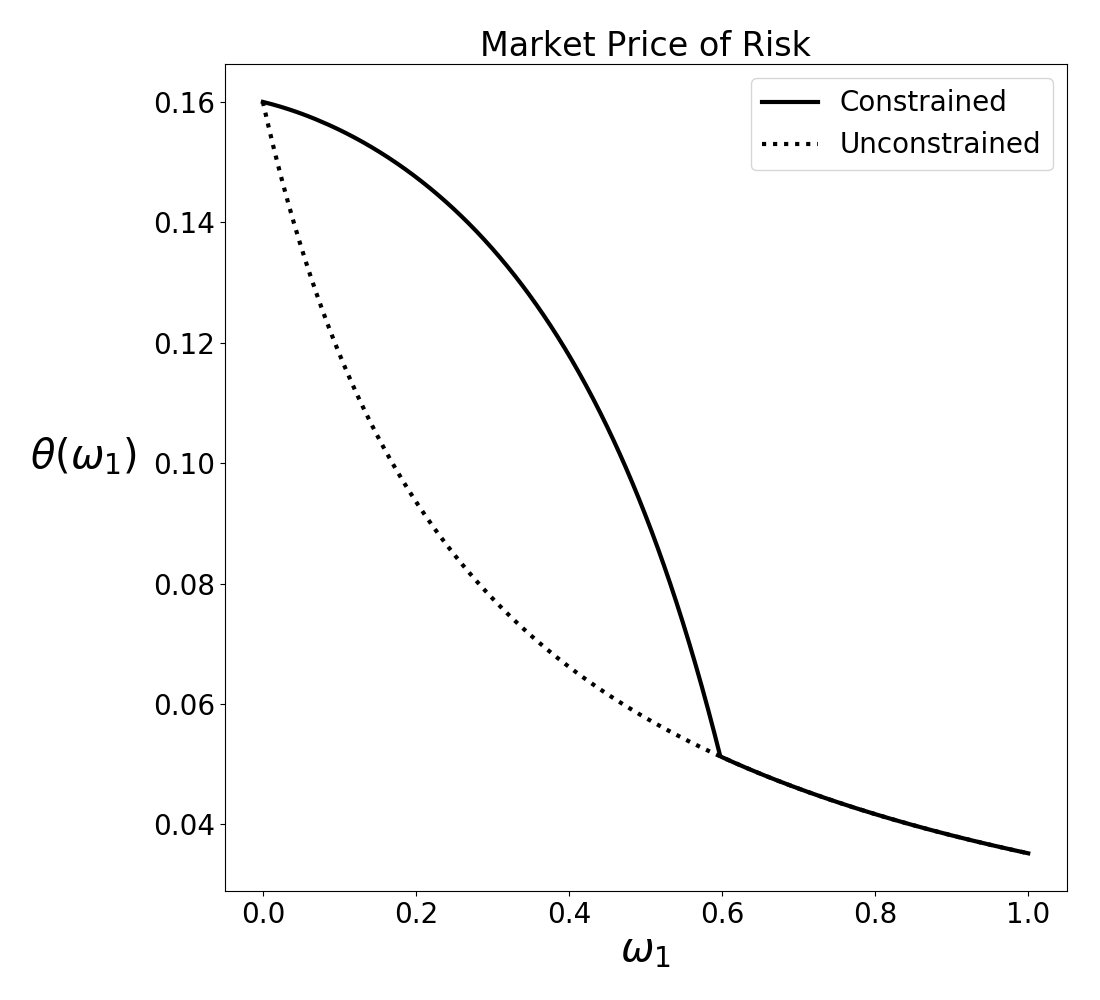}
		\centering
		\label{fig:two_agents:theta}
	}\hspace{-10pt}
	\subfigure[]{
		\includegraphics[width=\subfigsize]{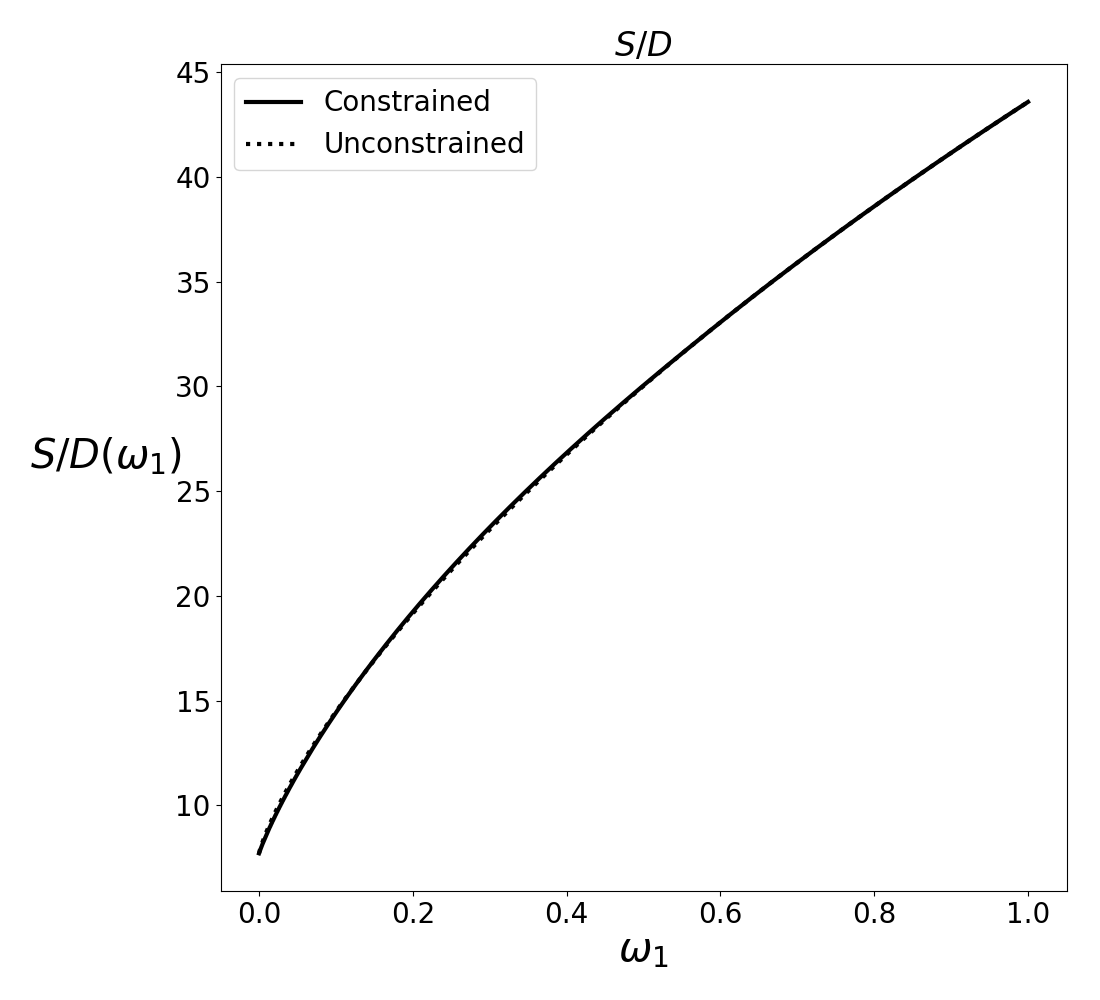}
		\centering
		\label{fig:two_agents:S}
	}\hspace{-10pt}
	\subfigure[]{
		\includegraphics[width=\subfigsize]{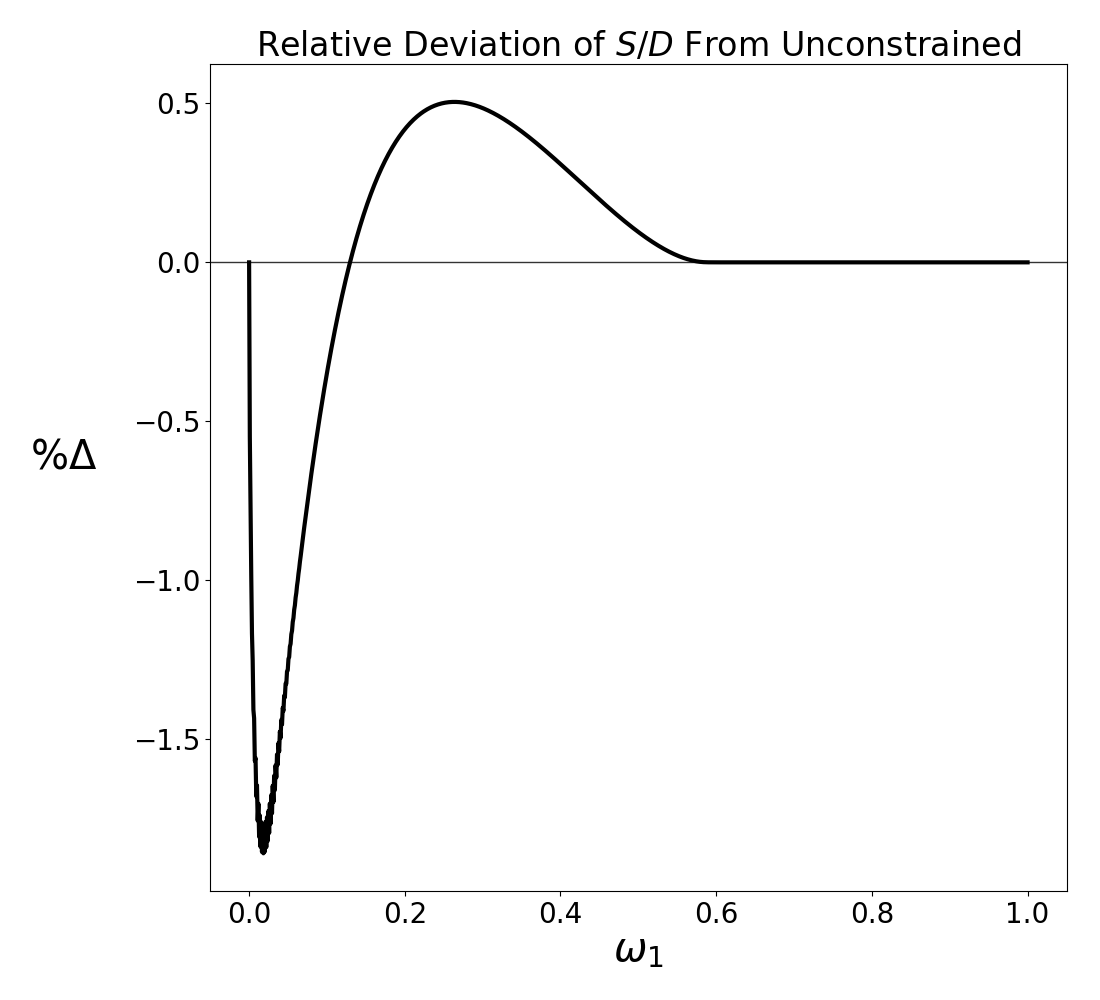}
		\centering
		\label{fig:two_agents:Spercdev}
	}\hspace{-10pt}
	
	\caption{}
	\label{fig:two_agents:2}
\end{figure}

Portfolios are represented in \cref{fig:two_agents:pi}. Moving form right to left in the state space, agent $1$, the least risk averse agent would prefer to leverage up, but runs into their constraint. In order remain below their constraint they adjust the composition of their wealth. The portfolio weight is falling in risk free borrowing, so the agent reduces their risk free borrowing. This reduction in the supply of risk free assets pushes up the price and down the interest rate, as is seen in \cref{fig:two_agents:r}. But how does the agent finance a reduction in their borrowing? They shift their wealth out of risky shares and into risk-free savings. When the agent is a borrower the portfolio weight is decreasing in risky shares, so this actually can only partially alleviate their situation. The constrained agent must further reduce their borrowing. As seen in \cref{fig:two_agents:borrow,fig:two_agents:leverage}, there is a substantial fall in borrowing and leverage as this agent shifts out of risky assets and into risk-free assets.

The market price of risk must be higher to compensate the unconstrained agent for holding more risky assets. As previously mentioned, the constrained agent is selling risky assets to the unconstrained agent, who is more risk averse. This agent requires higher returns on the risky asset and so the market price of risk is higher (\cref{fig:two_agents:theta}). The combination of a lower interest rate and higher market price of risk produces an ambiguous effect on the risky asset price.

To discuss asset prices we need to consider how individual preferences translate into choices about consumption given changes in the investment opportunity set. Given an improvement in the investment opportunity set, an agent will always have a substitution effect which reduces consumption today, as they substitute consumption from today to tomorrow. On the other hand, the agent is richer today and gets income from their wealth, implying an income effect. This income effect pushes up consumption in all periods. The interaction of these two forces determine the level of consumption today, which in turn determines the wealth/consumption ratio and asset price.

Given an improvement in the investment opportunity set, the price of the risky asset increases or decreases depending on whether the income or substitution effect dominates. When an agent has RRA of one, or EIS of one, their income and substitution effects perfectly offset. When EIS is less than one the income effect dominates and the agent chooses to increase consumption today. Thus, relative to wealth, their consumption is greater, implying a lower wealth/consumption ratio. This reduces asset prices which are a weighted average of wealth/consumption ratios. When EIS is greater than one the substitution effect dominates. Wealth/consumption ratios rise and the asset price increases.

\begin{figure}[!h]
	\subfigure[]{
		\includegraphics[width=\subfigsize]{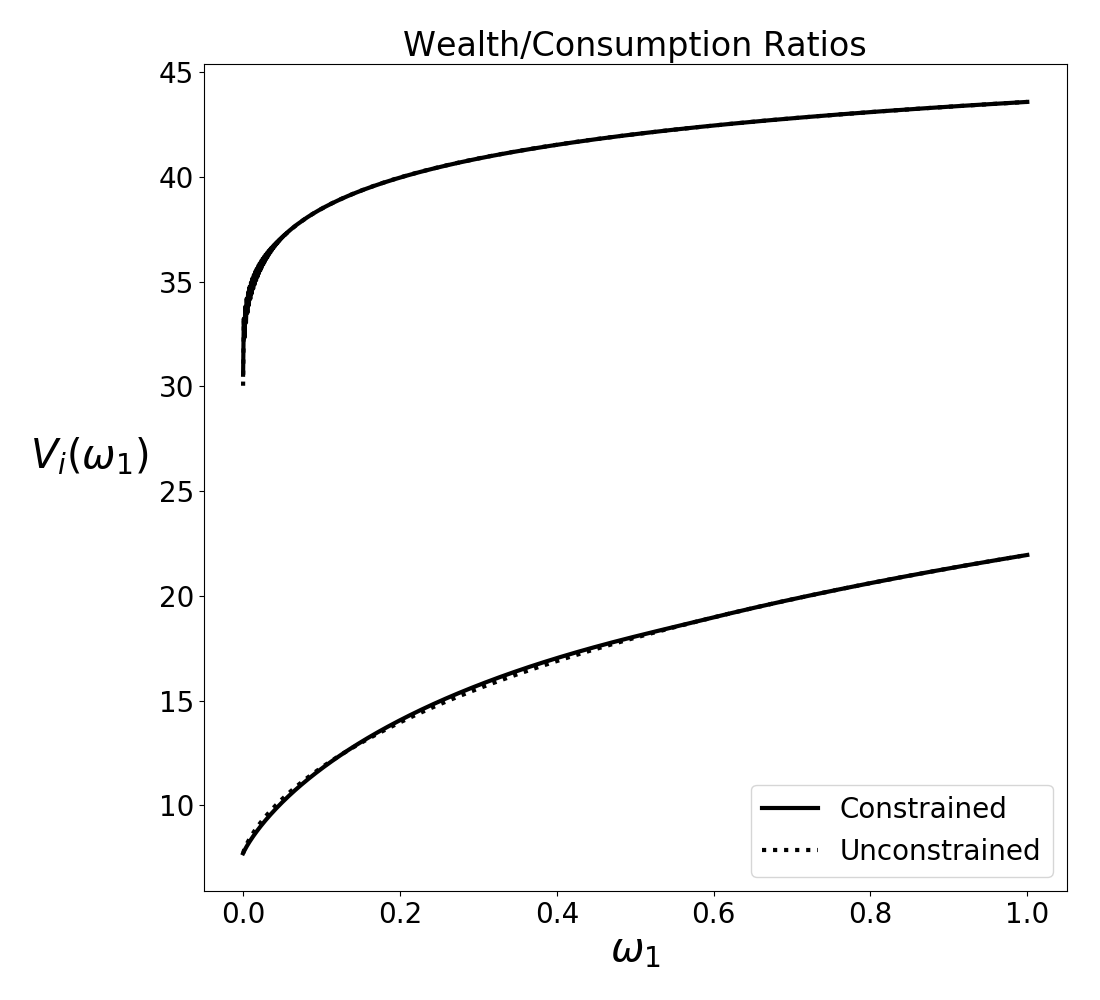}
		\centering
		\label{fig:two_agents:V}
	}\hspace{-10pt}
	\subfigure[]{
		\includegraphics[width=\subfigsize]{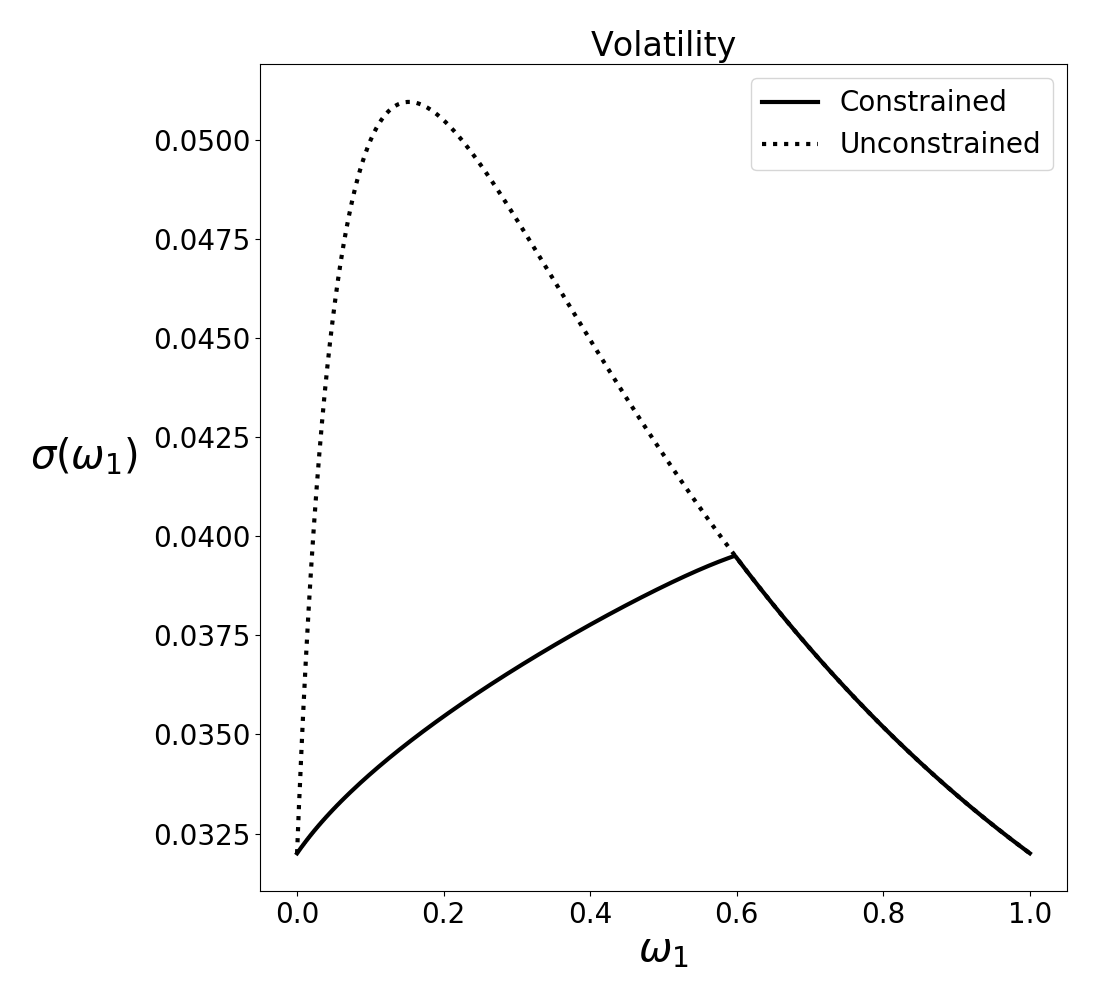}
		\centering
		\label{fig:two_agents:vol}
	}
	\subfigure[]{
		\includegraphics[width=\subfigsize]{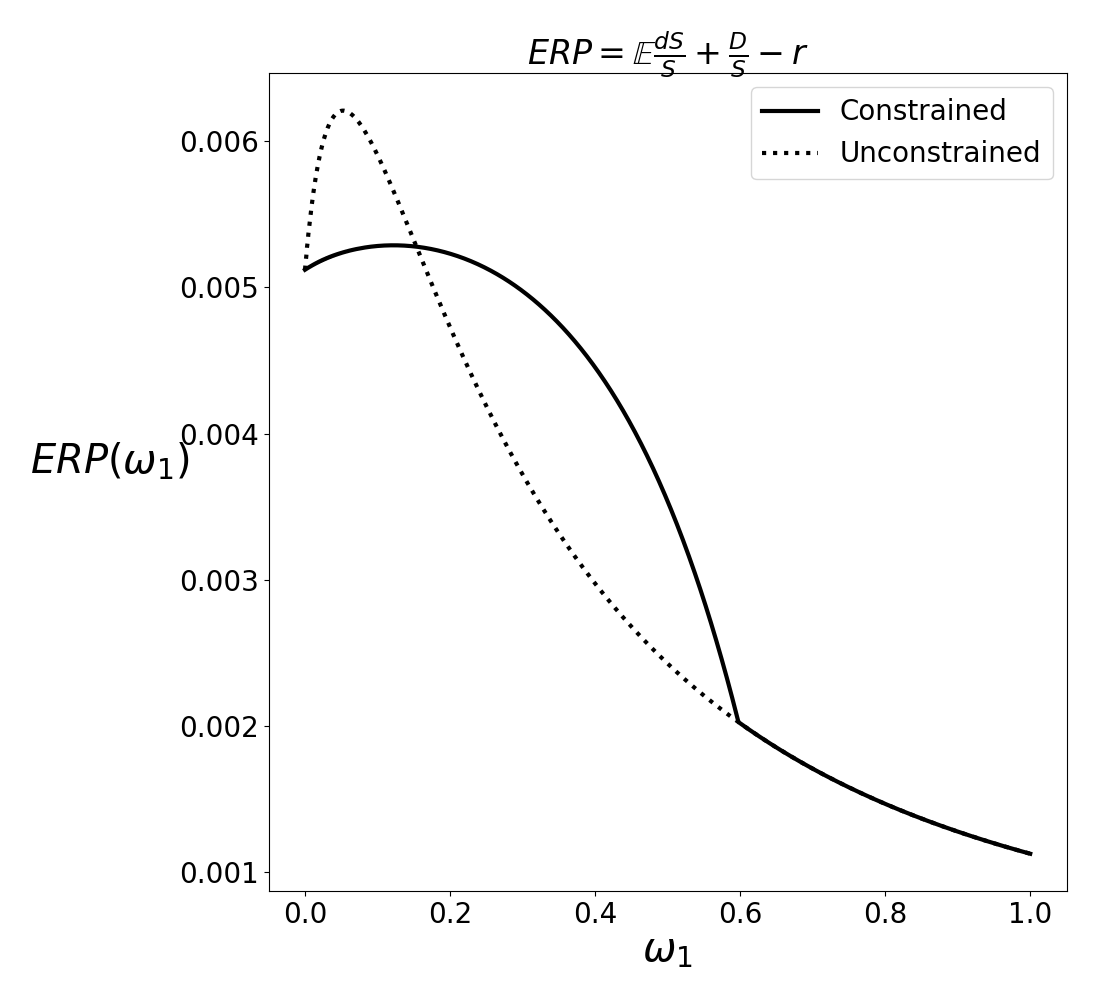}
		\centering
		\label{fig:two_agents:erp}
	}
	\caption{}
	\label{fig3}
\end{figure}

In the present setting, both agents have low EIS, so we expect that for a given improvement/deterioration in the investment opportunity set, asset prices will be lower/higher. \cref{fig:two_agents:S,fig:two_agents:Spercdev} show that over part of the state space the asset price is indeed higher under constraint. Although the market price of risk is higher, the risk free rate is lower and the constraint shifts more weight to agents who are net lenders. Thus the effect on the risk-free rate dominates and the investment opportunity set deteriorates and, since the income effect dominates, the asset price rises. However, in the lower area of the state space the asset price is lower than in the absence of constraint. This is driven by the fact that the change in the investment opportunity is not unambiguously negative. The unconstrained agent holds both risky and risk-free assets, and the return on risk-free assets has fallen. To see how the relative returns on these two assets changes we can look at the equity risk premium
\begin{align*}
ERP_t = \mu_t + \frac{D_t}{S_t} - r_t = \theta_t \sigma_t
\end{align*}
The equity risk premium is the expected capital gains plus dividend yield minus the risk free rate, which is simply the market price of risk times volatility. In \cref{fig:two_agents:erp} we see that the asset price is lower or higher under constraint exactly when the equity risk premium is lower or higher. This is because the unconstrained agent is a net lender, so is essentially short the equity risk premium. Any improvement in this premium translates to a deterioration in the investment opportunity set faced by the unconstrained agent, causing them to reduce consumption and pushing up their wealth/consumption ratios and, in turn, asset prices.

In addition to these first-order moments, the dynamics of the model are also affected by the constraint. There is a reduction in trade when one agent is constrained. In this region, shares are only exchanged in order to maintain the portfolio weight which holds the constrained agent against their constraint. This reduction in exchange dampens volatility as there is less change in the marginal agent pricing risky assets. This effect can be seen in \cref{fig:two_agents:vol}. One take-away from this observation could be an intuition for volatility frowns and smirks observed in options pricing data. Implied volatility can have a positive or negative term premium for different values of the strike, implying changes in volatility over the state space. What this model predicts is that for markets where participants face constraints in their trade of the underlying, there will be a positive term premium (or frown), while for unconstrained assets there will be a negative term premium (smirk or smile).

The effect on leverage is substantial given both a supply effect and a demand effect. The demand for credit is artificially lower under constraint when risk neutral agents cannot leverage up. The supply of credit is also reduced because risk averse agents shift wealth into risky assets. They do so because they see low volatility and high expected returns. Risk averse agents shift wealth into risky shares and the supply of credit contracts. As the economy moves between the constrained and unconstrained regions the cyclicality of leverage changes.

Leverage cycles are both pro- and counter-cyclical in both complete and incomplete markets, but the dynamics of this cyclicality is vastly different under the two regimes. In complete markets, the slope of leverage varies smoothly, moving from positive to negative as one moves through the state space. Only in very bad states does leverage exhibit pro-cyclicality, as risk averse agents begin to dominate and the interest rate becomes too high for risk neutral agents to desire to borrow. This inflection point becomes a singularity under margin constraints. In \cref{fig:two_agents:leverage} we can see a kink at the boundary between the constrained and unconstrained regions, implying a jump from pro- to counter-cyclicality. This prediction connects to the large literature on the cyclicality of leverage (\cite{geanakoplos1996promises,geanakoplos2010leverage,adrian2010liquidity}, as well as many others). These observations will be studied empirically in \cref{sec:data}.

\subsection{Three Types, Cascading Constraints, and Non-Monotonic Leverage Cycles}\label{sec:sec:three_types}
Consider next the case of three agents\footnote{For a description of the numerical solution to this problem see \cref{appendix:numerical}.}. Introducing a third agent shows how there can exist a cascade effect. As the least risk-averse agent's constraint binds, the other agents begin to leverage up. This causes the agent in the middle of the distribution of preferences to move towards their constraint. This is not evident with only two agents, as the most risk-averse agent will never hit their constraint. The increase in leverage of the intermediate agent actually leads to a full recovery of leverage. That is, in the region where the least risk-averse agent is constrained, the intermediate agent will take their place in the market for borrowing, partially or even fully offsetting the reduction in borrowing caused by constraint. This leads to a sort of double-dip in leverage: first leverage contracts as one agent becomes constrained, then rises as the intermediate agent takes up the slack, and eventually falls when the intermediate agent also becomes constrained.

All parameters are the same as in \cref{sec:sec:two_types}, except preferences which are set to $(\gamma_1, \gamma_2, \gamma_3) = (1.1, 1.5, 3.0)$ and margin constraints are set to $m_i = 1.2$ for all $i$. Graphs are plotted over the state space where $(\omega_1, \omega2) \in \{(x, y) \in \mathbb{R}_+: x + y \leq 1 \}$, the two dimensional simplex. In the extreme cases where $D_t \rightarrow 0$ or $D_t \rightarrow \infty$, $(\omega_1, \omega_2) \rightarrow (0, 0)$ and $(\omega_1, \omega_2) \rightarrow (1, 0)$, respectively. Thus we can think of negative shocks pushing in a southwest direction and positive shocks pushing in a southeast direction, with some deviation in the interior of the state space\footnote{Quiver plot of shock directions to be added.}.

\begin{figure}[h]
	\subfigure[]{
		\centering
		\includegraphics[trim=175pt 0pt 0pt 0pt, width=1.0\linewidth]{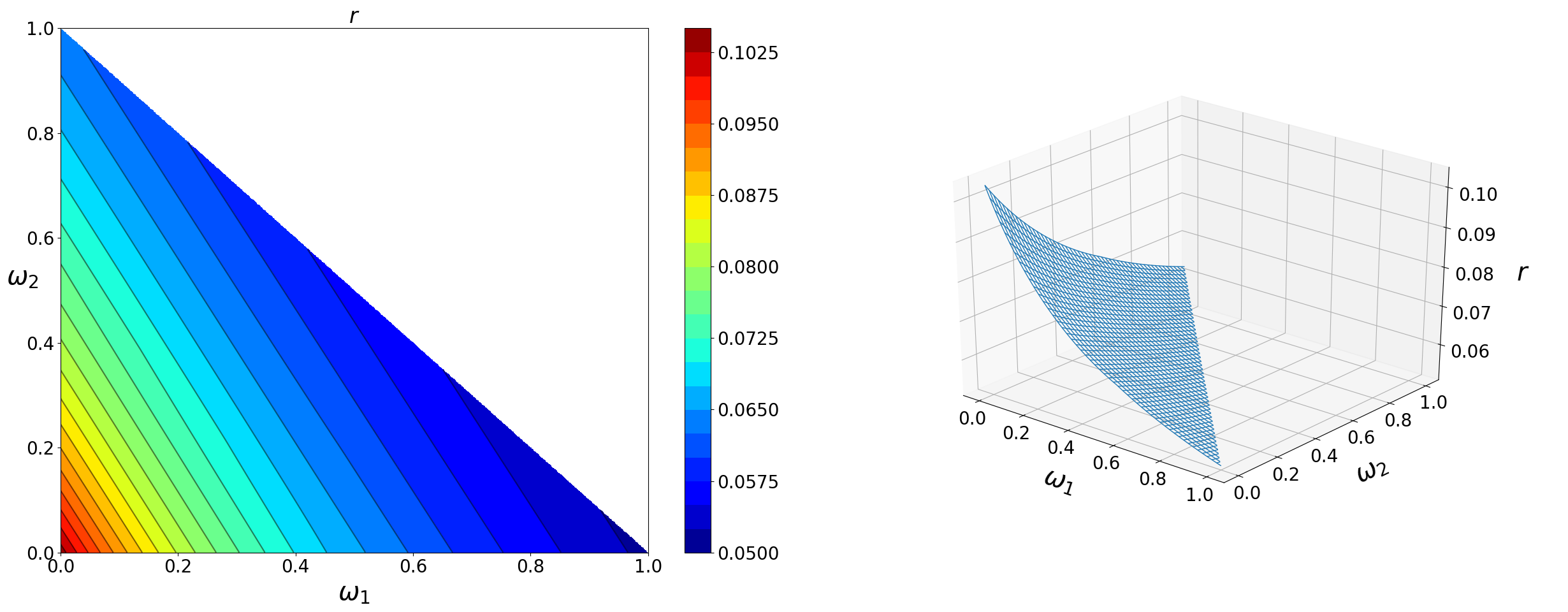}
		\label{fig:three_agents:r}
	}
	\subfigure[]{
		\centering
		\includegraphics[trim=175pt 0pt 0pt 0pt, width=1.0\linewidth]{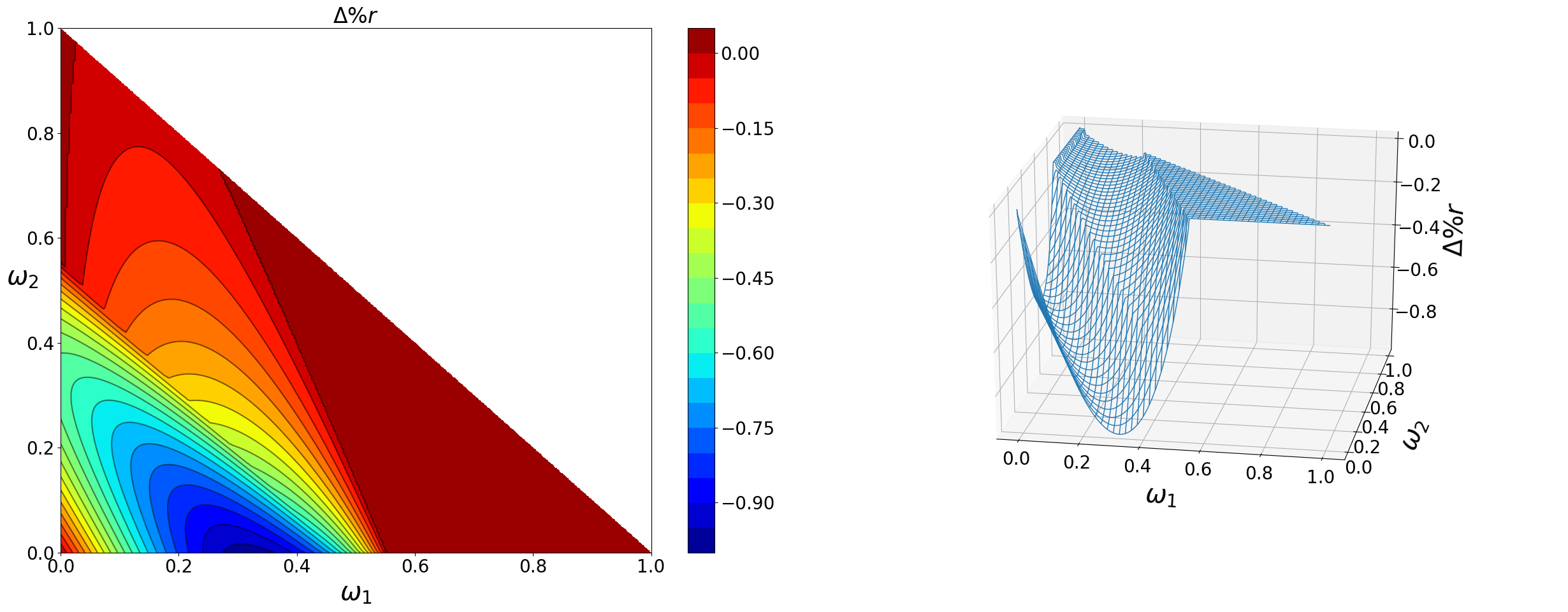}
		\label{fig:three_agents:r_dev}
	}
	\caption{Risk free rate in levels (\cref{fig:three_agents:r}) and in deviations from complete markets (\cref{fig:three_agents:lev_r}).}
	\label{fig:three_agents:r_all}
\end{figure}

\begin{figure}[h]
	\subfigure[]{
		\centering
		\includegraphics[trim=175pt 0pt 0pt 0pt, width=1.0\linewidth]{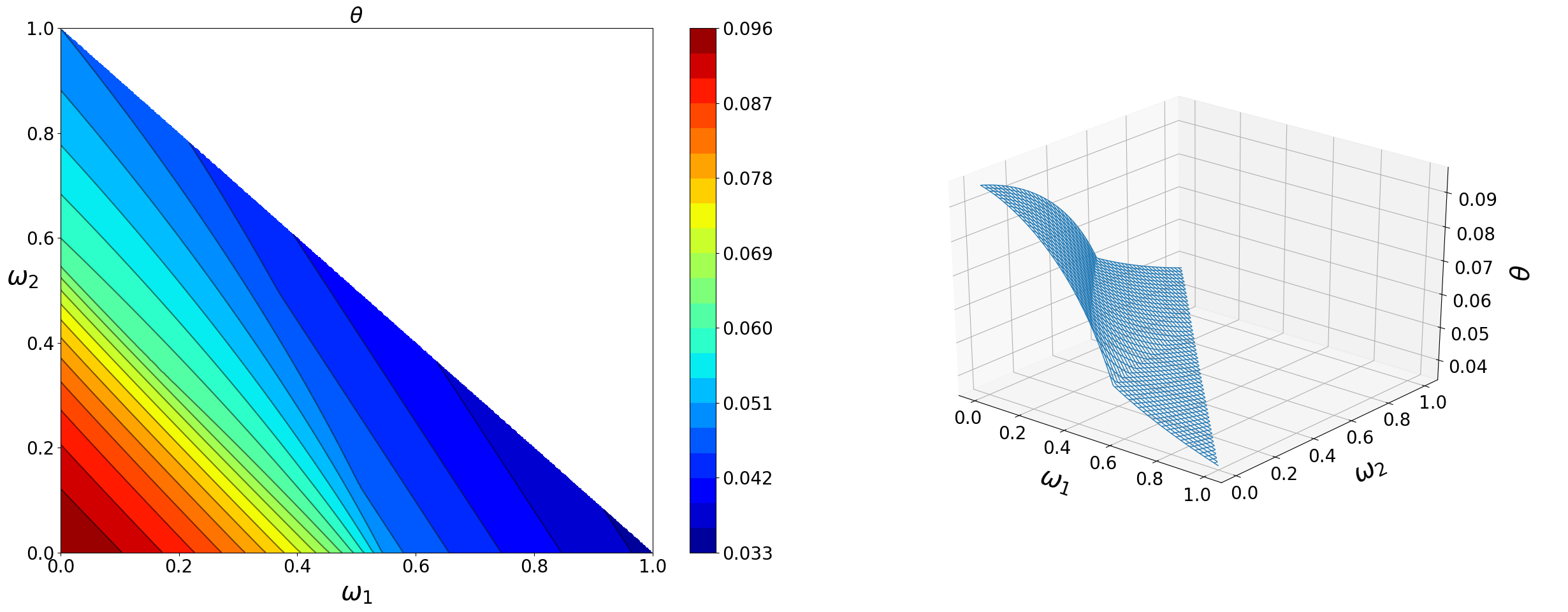}
		\label{fig:three_agents:theta}
	}
	\subfigure[]{
		\centering
		\includegraphics[trim=175pt 0pt 0pt 0pt, width=1.0\linewidth]{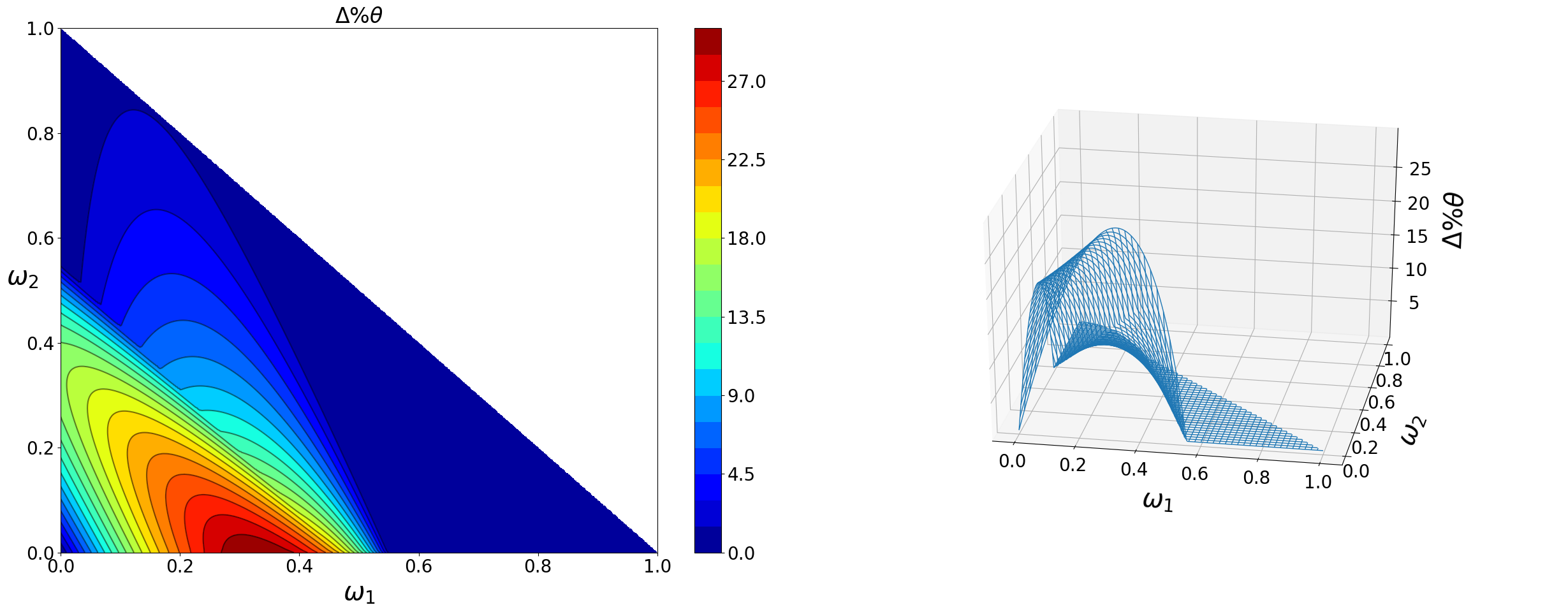}
		\label{fig:three_agents:theta_dev}
	}
	\caption{Market price of risk in levels (\cref{fig:three_agents:theta}) and in deviations from complete markets (\cref{fig:three_agents:theta_dev}).}
	\label{fig:three_agents:theta_all}
\end{figure}

Consider first the interest rate and market price of risk, depicted in \cref{fig:three_agents:r_all,fig:three_agents:theta_all}. These two variables determine the investment opportunity set, which makes them key in determining asset prices. You'll notice first that the interest rate is increasing under negative shocks. This is similar to the complete market, where negative shocks push more weight to the most risk averse agent who is very patient and in turn requires a higher interest rate. However we can see that there is a region where the interest rate is lower in the constrained equilibrium, evidenced by the negative values in \cref{fig:three_agents:r_dev}. In this region at least one agent is constrained. There is a contraction in the demand for credit, pushing up the price of bonds and down the interest rate. The market price of risk follows similar dynamics (\cref{fig:three_agents:theta_all}) for similar reasons. However, the market price of risk is higher under constraint, as seen in \cref{fig:three_agents:theta_dev}. This is driven by risk-averse agents requiring higher returns to hold a greater share of their wealth in risky assets. These effects combined have an ambiguous effect on asset prices, a priori, but tend to increase asset prices for the given parameterization.

\begin{figure}[h]
	\subfigure[]{
		\centering
		\includegraphics[trim=175pt 0pt 0pt 0pt, width=1.0\linewidth]{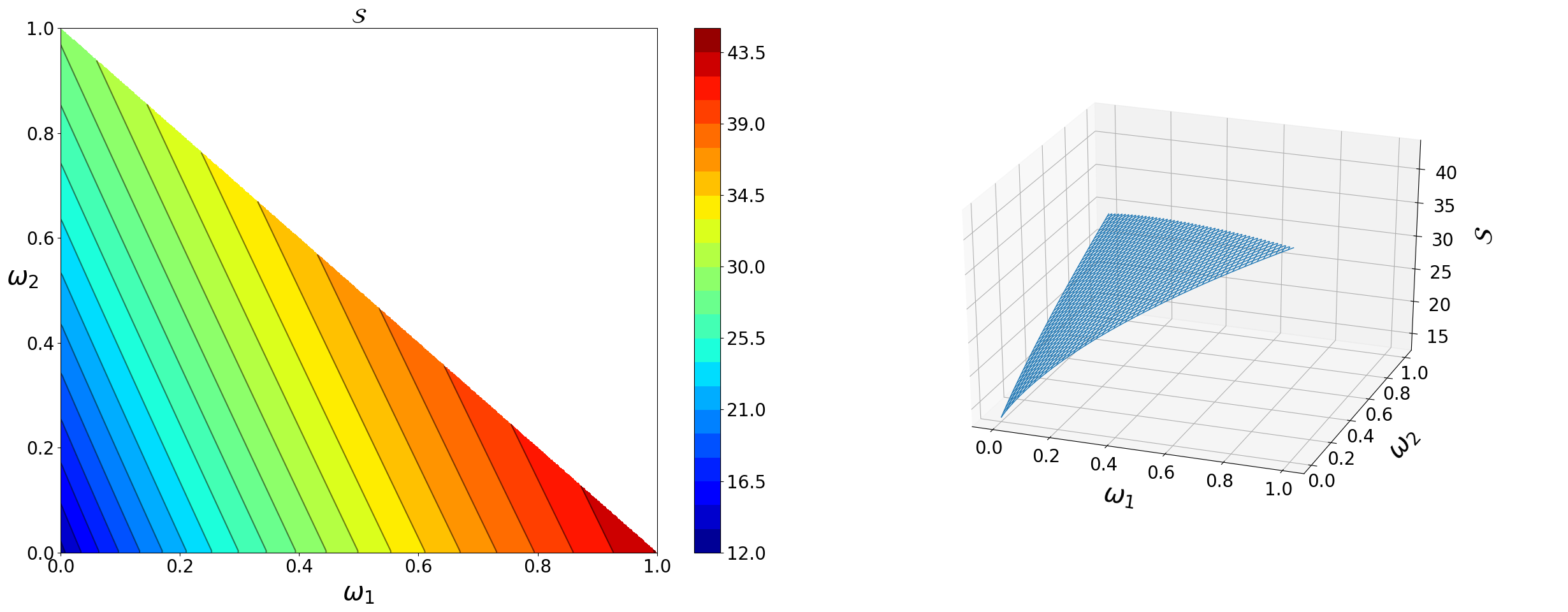}
		\label{fig:three_agents:S}
	}
	\subfigure[]{
		\centering
		\includegraphics[trim=175pt 0pt 0pt 0pt, width=1.0\linewidth]{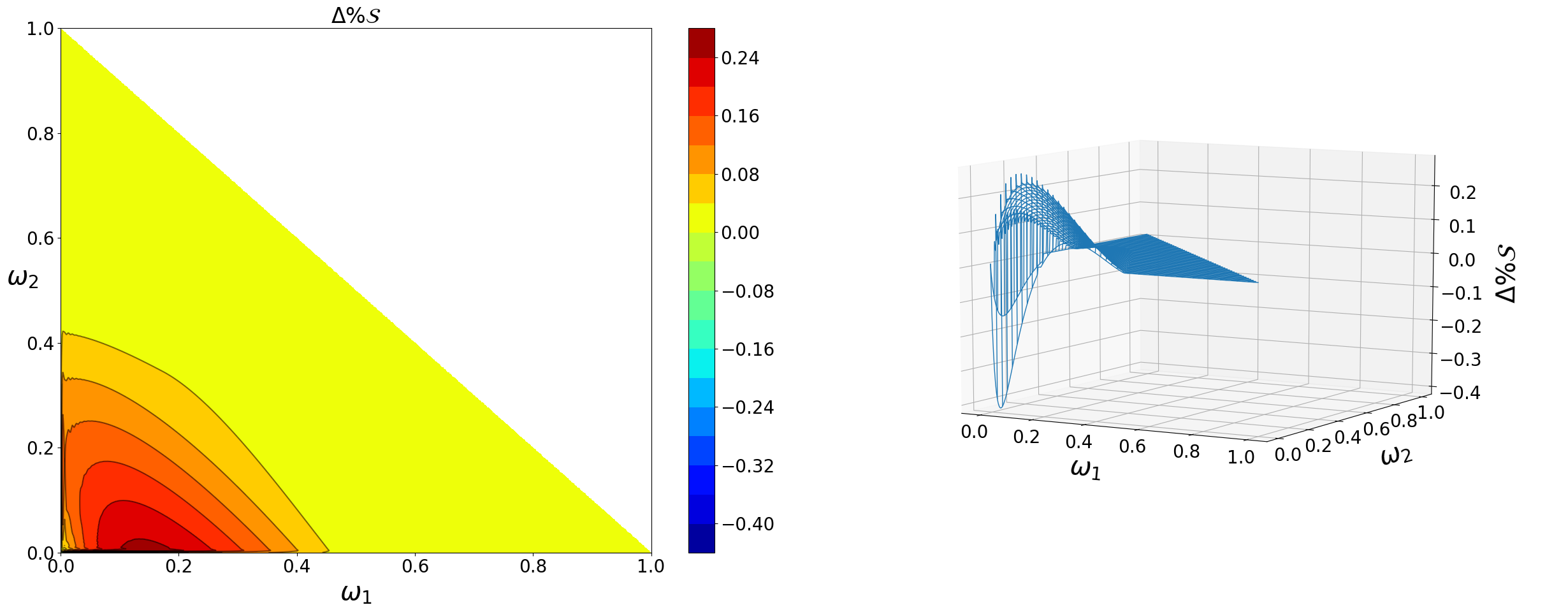}
		\label{fig:three_agents:S_dev}
	}
	\caption{Asset prices in levels (\cref{fig:three_agents:S}) and in deviations from complete markets (\cref{fig:three_agents:S_dev}).}
	\label{fig:three_agents:S_all}
\end{figure}

\begin{figure}[h]
	\subfigure[]{
		\centering
		\includegraphics[trim=175pt 0pt 0pt 0pt, width=1.0\linewidth]{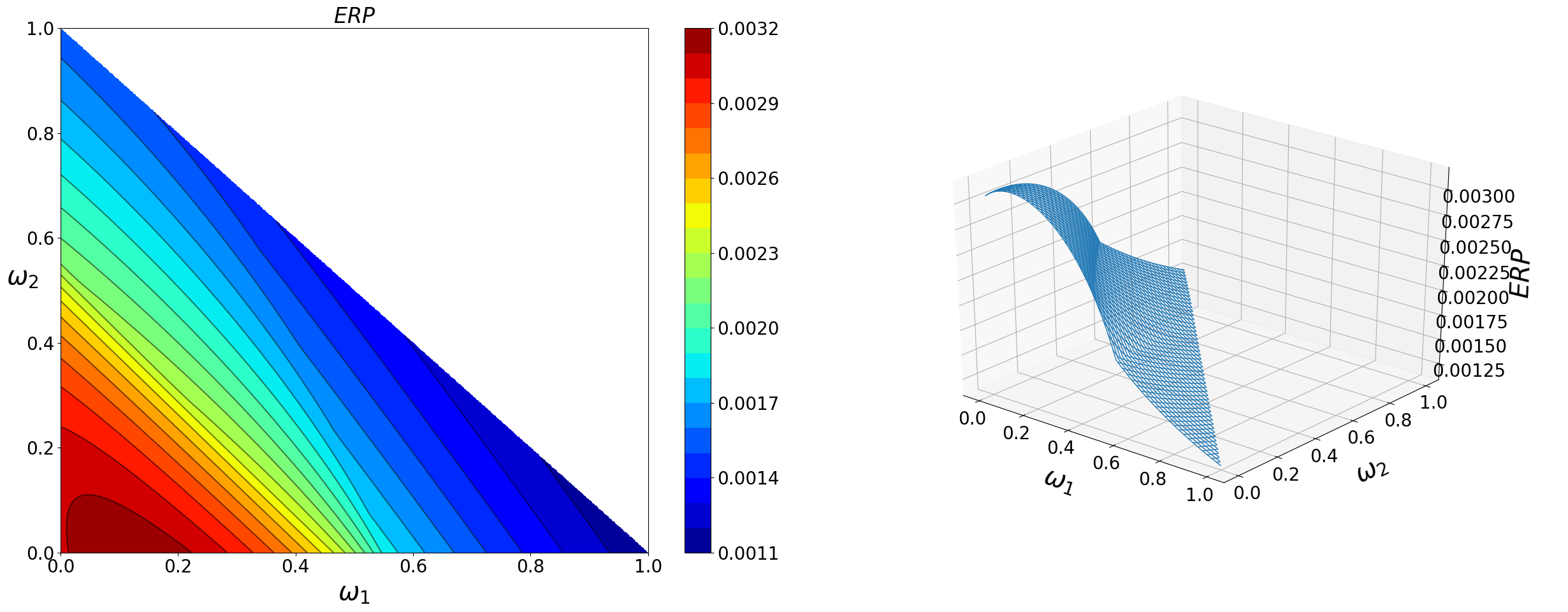}
		\label{fig:three_agents:erp}
	}
	\subfigure[]{
		\centering
		\includegraphics[trim=175pt 0pt 0pt 0pt, width=1.0\linewidth]{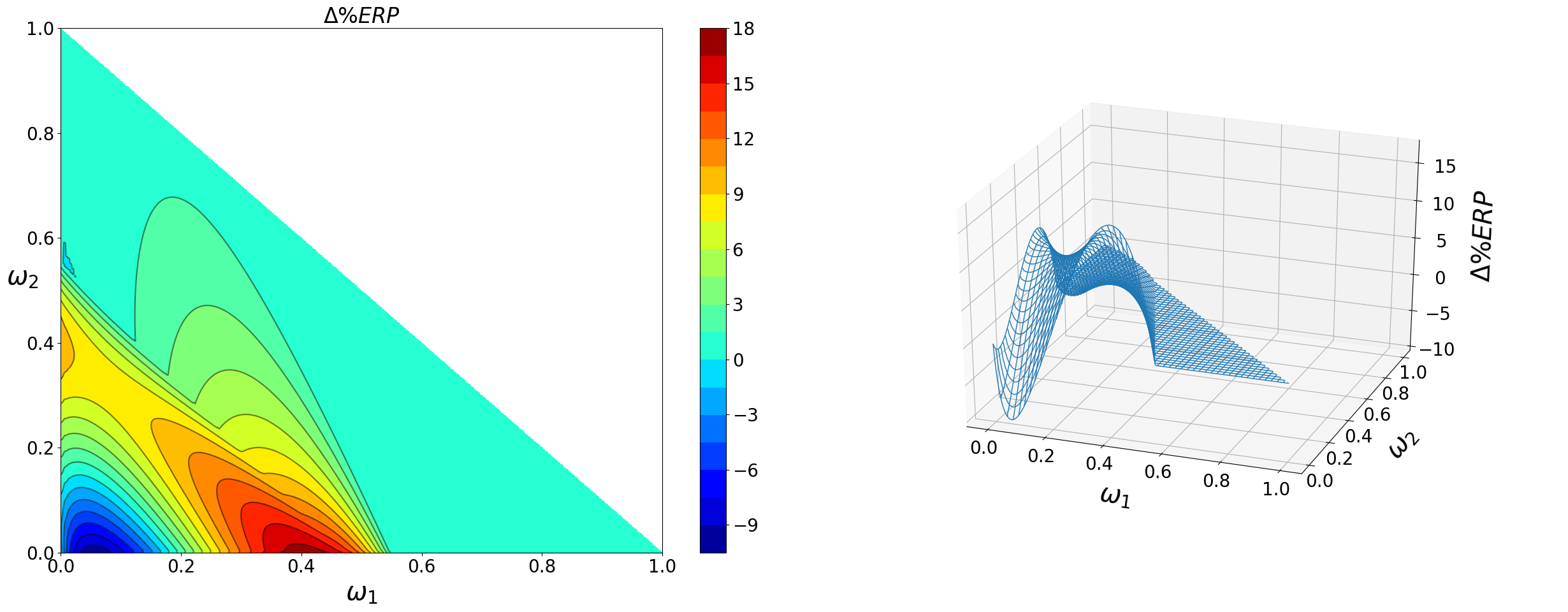}
		\label{fig:three_agents:erp_dev}
	}
	\caption{Asset prices in levels (\cref{fig:three_agents:erp}) and in deviations from complete markets (\cref{fig:three_agents:erp_dev}).}
	\label{fig:three_agents:erp_all}
\end{figure}

Asset prices are higher under constraint, as can be seen in \cref{fig:three_agents:S_all}. This effect is similar to that discussed in \cref{sec:sec:two_types}. Because agents have EIS less than one, the income effect dominates and wealth consumption ratios increase, pushing up asset prices. Interestingly, the deviation in asset prices is very steep near the boundary, driven by a rapid deterioration in the equity risk premium (\cref{fig:three_agents:erp_all}). The equity risk premium is higher in the constrained region to compensate risk-averse investors, but it falls quickly as the economy moves towards the point $(\omega_1, \omega_2) = (0,0)$.

\begin{figure}[h]
	\vspace{-50pt}
	\subfigure[]{
		\centering
		\includegraphics[trim=175pt 0pt 0pt 00pt, width=1.0\linewidth]{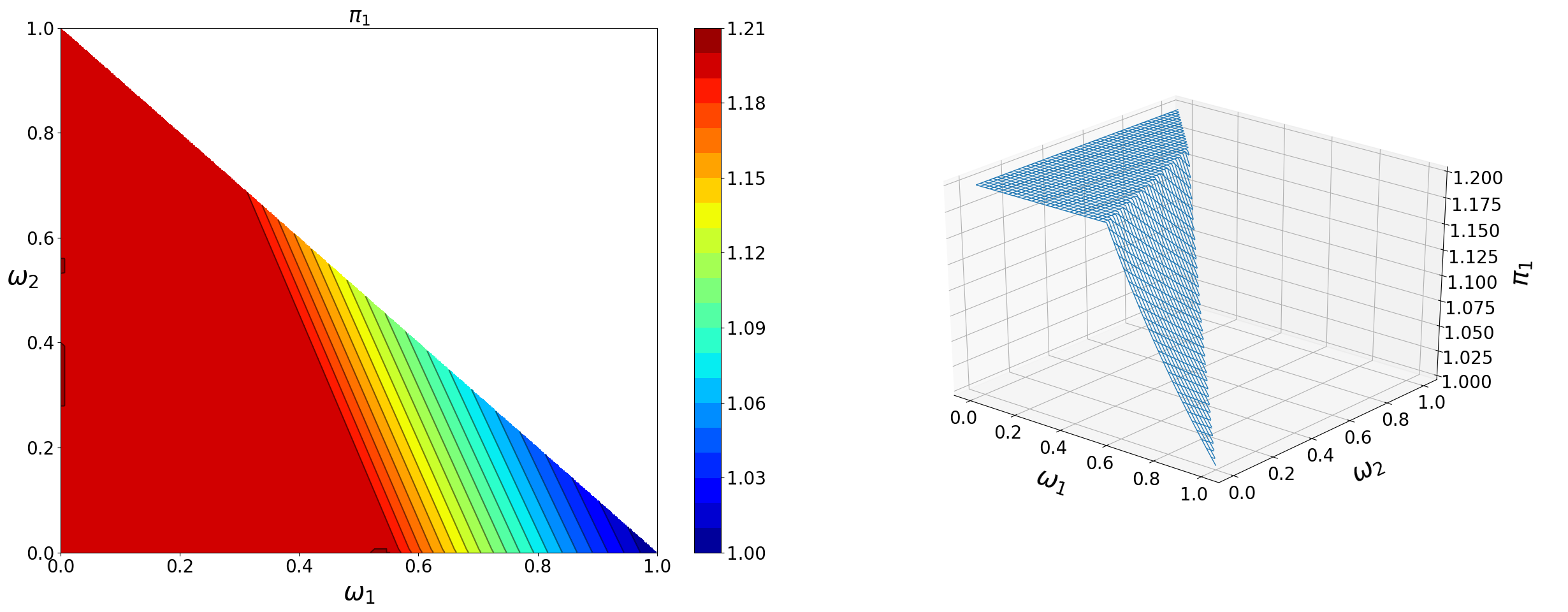}
		\label{fig:three_agents:pi1}
	}
	\subfigure[]{
		\centering
		\includegraphics[trim=175pt 0pt 0pt 0pt, width=1.0\linewidth]{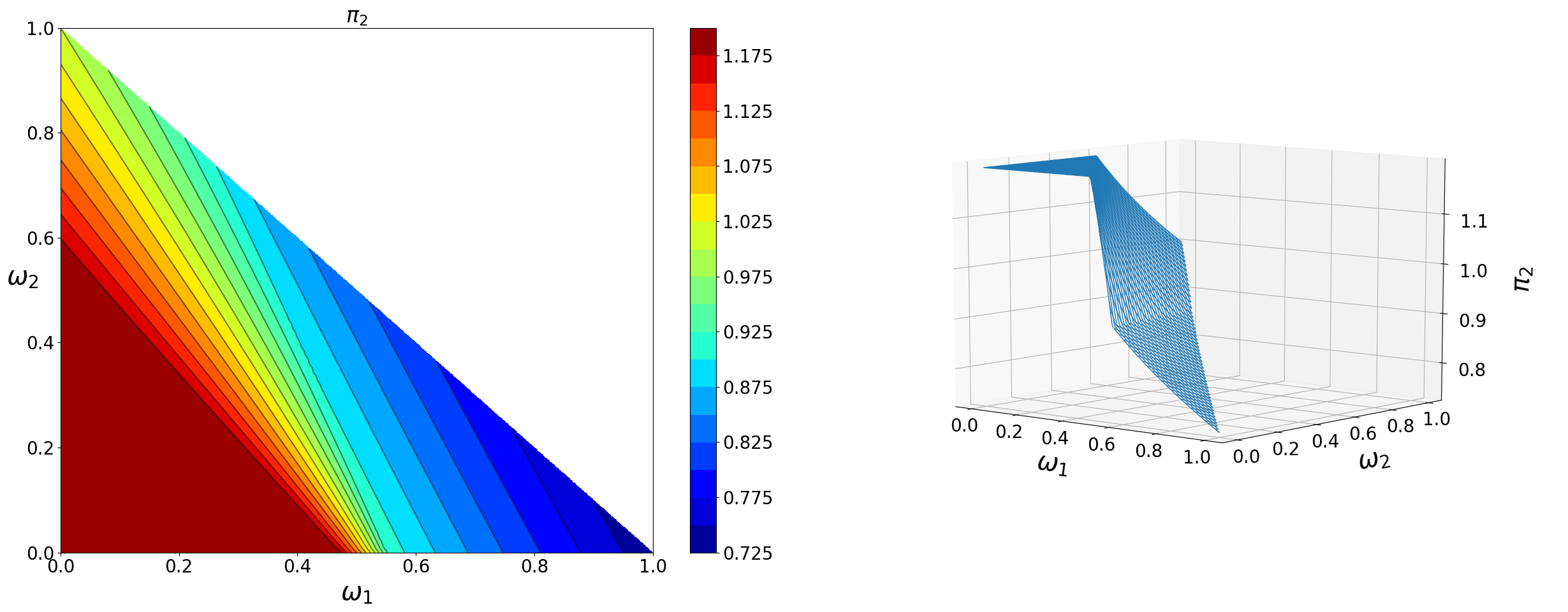}
		\label{fig:three_agents:pi2}
	}
	\subfigure[]{
		\centering
		\includegraphics[trim=175pt 0pt 0pt 0pt, width=1.0\linewidth]{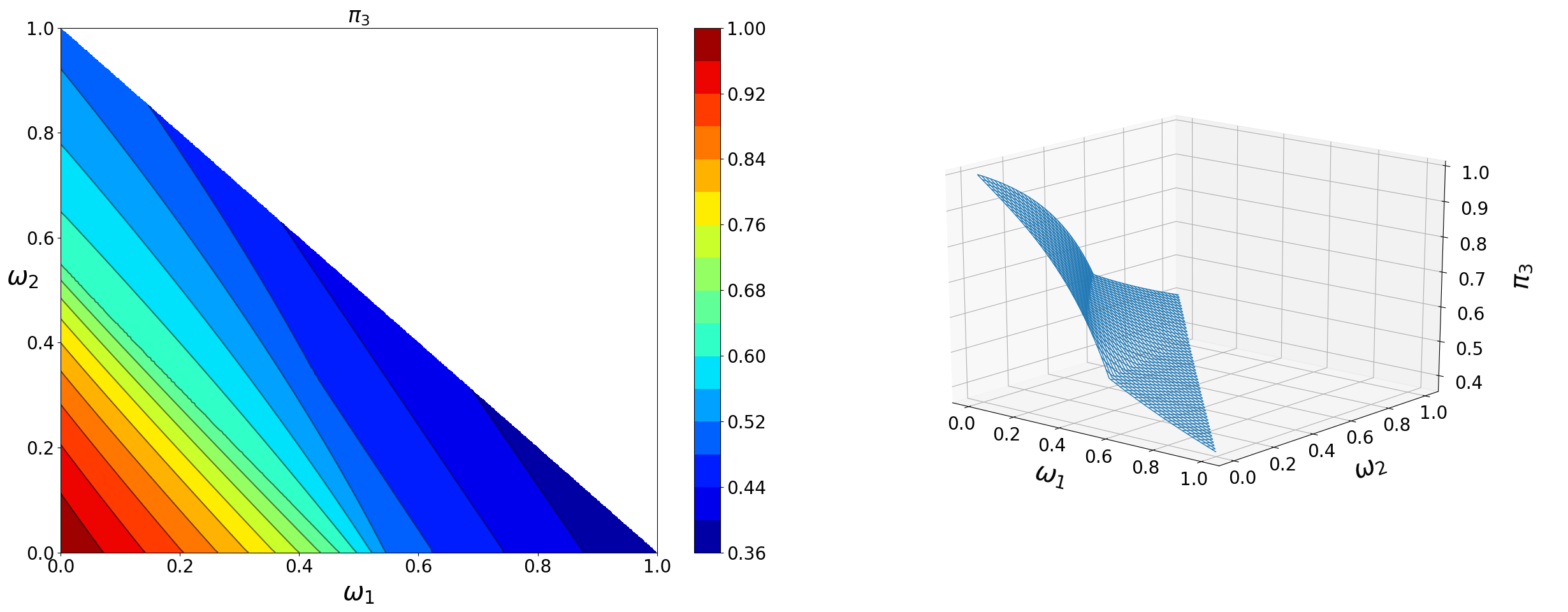}
		\label{fig:three_agents:pi3}
	}
	\caption{Portfolio weights. On the left is a countour plot and the right a surface plot of the same data.}
	\label{fig:three_agents:pi}
\end{figure}

\begin{figure}[h]
	\vspace{-50pt}
	\subfigure[]{
		\centering
		\includegraphics[trim=175pt 0pt 0pt 00pt, width=1.0\linewidth]{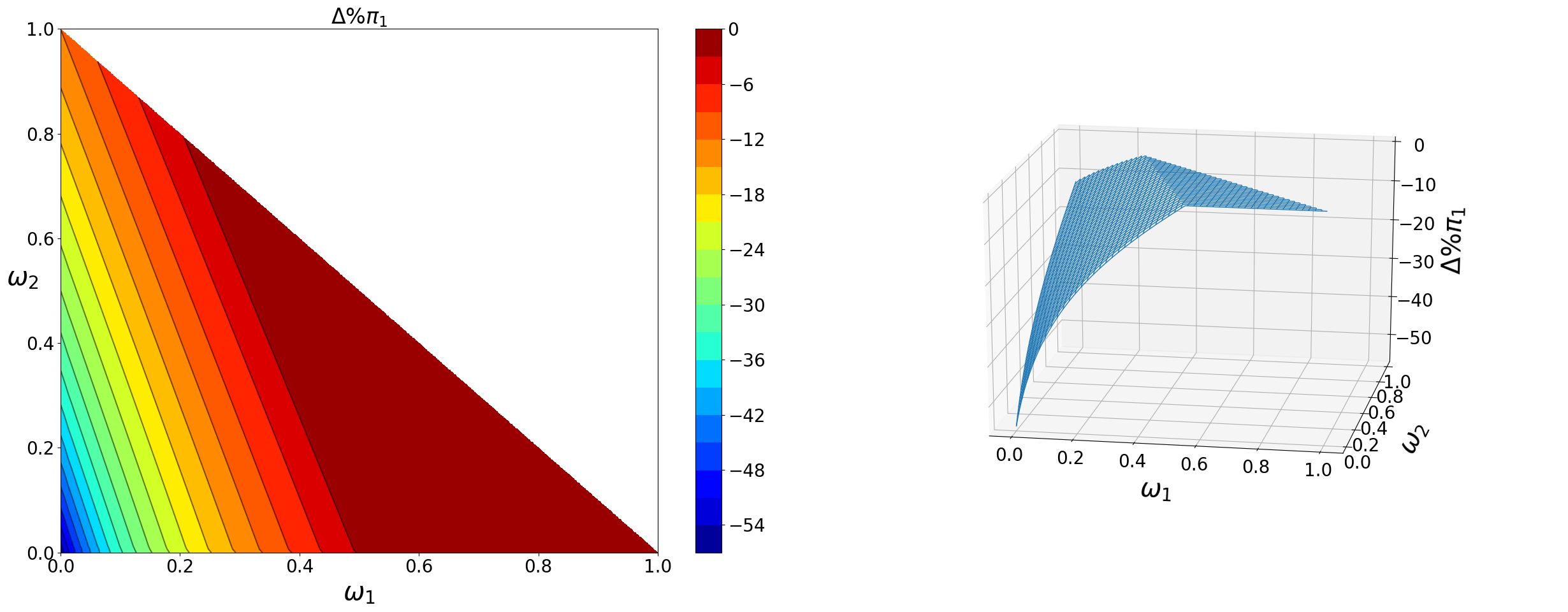}
		\label{fig:three_agents:pi1_dev}
	}
	\subfigure[]{
		\centering
		\includegraphics[trim=175pt 0pt 0pt 0pt, width=1.0\linewidth]{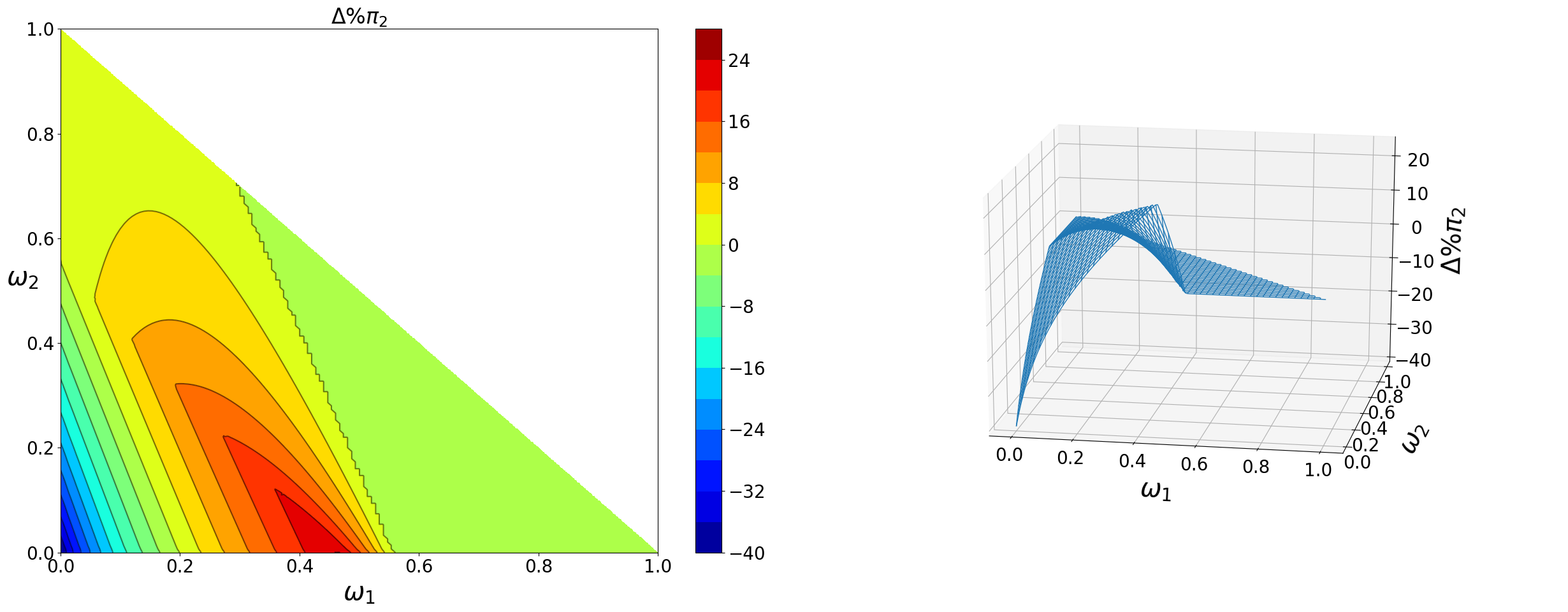}
		\label{fig:three_agents:pi2_dev}
	}
	\subfigure[]{
		\centering
		\includegraphics[trim=175pt 0pt 0pt 0pt, width=1.0\linewidth]{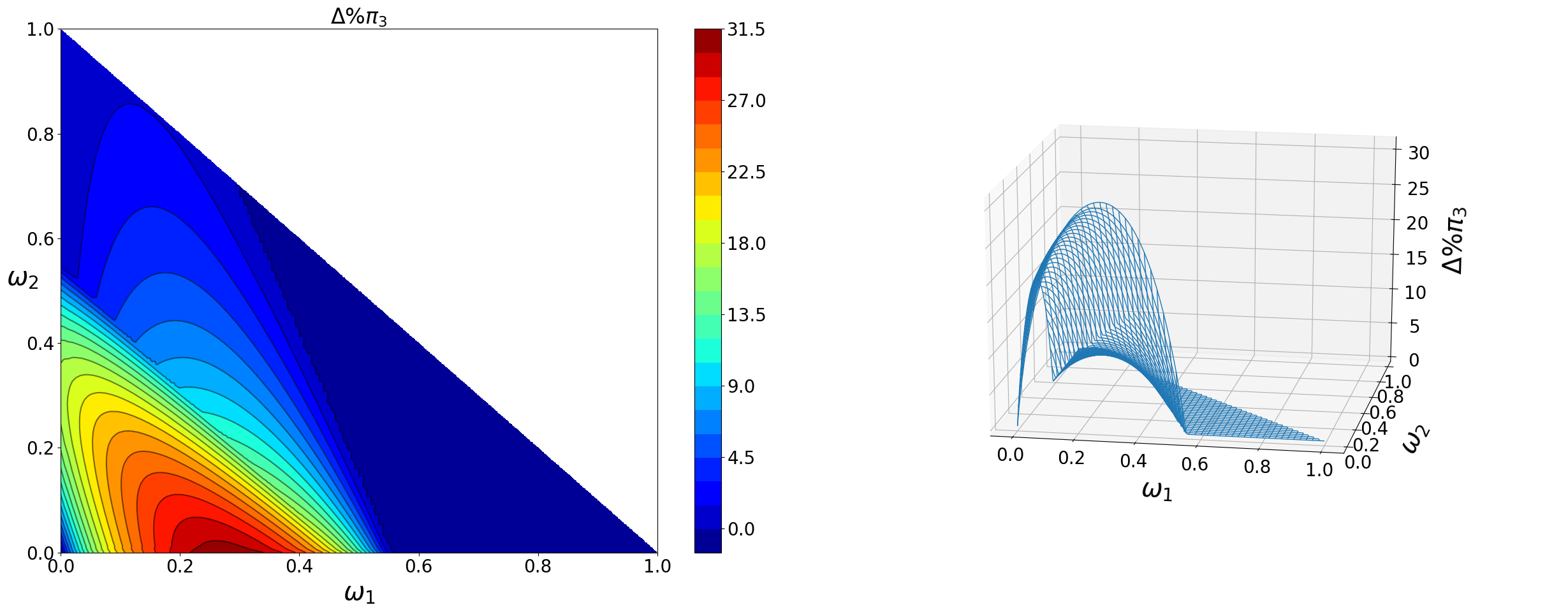}
		\label{fig:three_agents:pi3_dev}
	}
	\caption{Portfolio weights in percentage deviations from complete markets.}
	\label{fig:three_agents:pi_dev}
\end{figure}

These changes in the investment opportunity set are driven by the constraint on portfolios, which are represented in \cref{fig:three_agents:pi}. As you can see, agent $1$ is constrained over a large area of the state space. The portfolio weights of agents $2$ and $3$ are kinked at the interface between regions where agent $1$ is unconstrained and constrained. In the constrained region, the portfolio weights of unconstrained agents are steeper and portfolio weights higher than in the case of complete markets. This can be seen in \cref{fig:three_agents:pi_dev}, which plots percentage deviations from the unconstrained equilibrium. Here we see that unconstrained agents hold substantially more of their wealth in risky assets than they would have in the unconstrained equilibrium. This pushes them closer to their constraint. As agent $1$ becomes more and more constrained, the investment opportunities of agent $2$ improve, causing them to leverage up. Eventually they run into their constraint, creating a sort of cascade. However this increase in portfolio weights does not translate directly into an increase in leverage.

\begin{figure}[h]
	\subfigure[]{
		\centering
		\includegraphics[trim=175pt 0pt 0pt 0pt, width=1.0\linewidth]{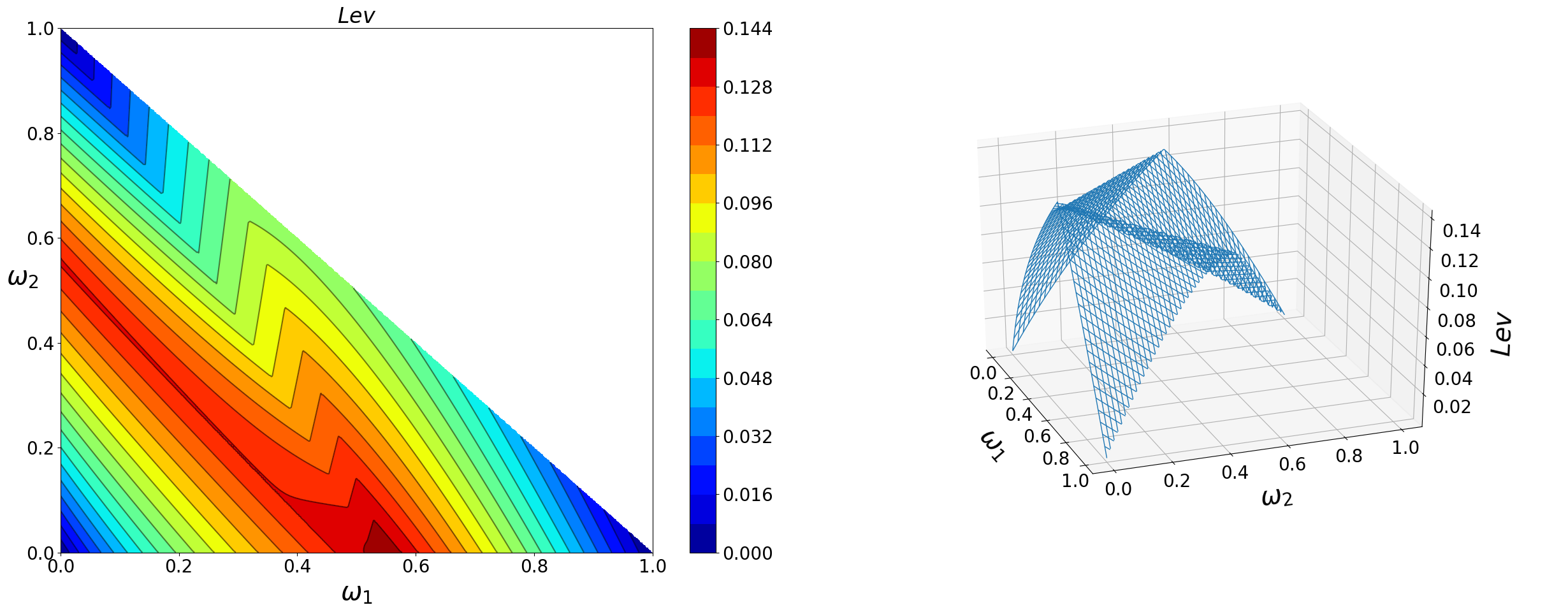}
		\label{fig:three_agents:lev}
	}
	\subfigure[]{
		\centering
		\includegraphics[trim=175pt 0pt 0pt 0pt, width=1.0\linewidth]{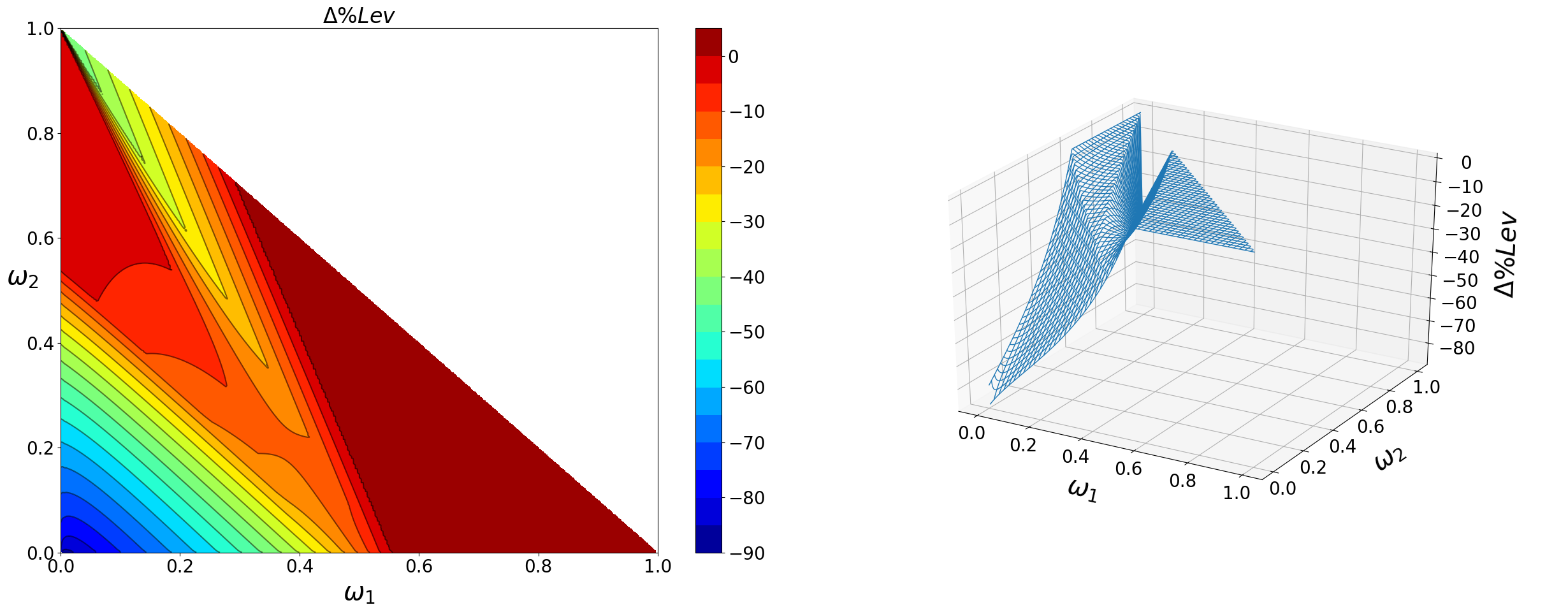}
		\label{fig:three_agents:lev_dev}
	}
	\caption{Leverage in levels (\cref{fig:three_agents:lev}) and in deviations from complete markets (\cref{fig:three_agents:lev_dev}).}
	\label{fig:three_agents:lev_all}
\end{figure}

In the constrained region, leverage is weakly lower than in the unconstrained equilibrium and exhibits non-monotonic and non-linear dynamics. \cref{fig:three_agents:lev_all} shows leverage in both levels and deviations from the unconstrained equilibrium. There are two peaks in leverage, one along the boundary where agent $1$ becomes constrained and another along the boundary where agent $2$ becomes constrained. In the intermediate region, leverage actually recovers back to its unconstrained level, as you can see in \cref{fig:three_agents:lev_dev}, indicated by the dark red region for low values of $\omega_1$ and high values of $\omega_2$. This implies that, even though agent $1$ is constrained, agent two holds a sufficient amount of leverage to completely offset the reduction. At the same time, they do not hold more leverage than necessary to push the economy back to the same amount of aggregate leverage that would prevail without constraint.

\begin{figure}[h]
	\subfigure[]{
		\centering
		\includegraphics[trim=175pt 0pt 0pt 0pt, width=1.0\linewidth]{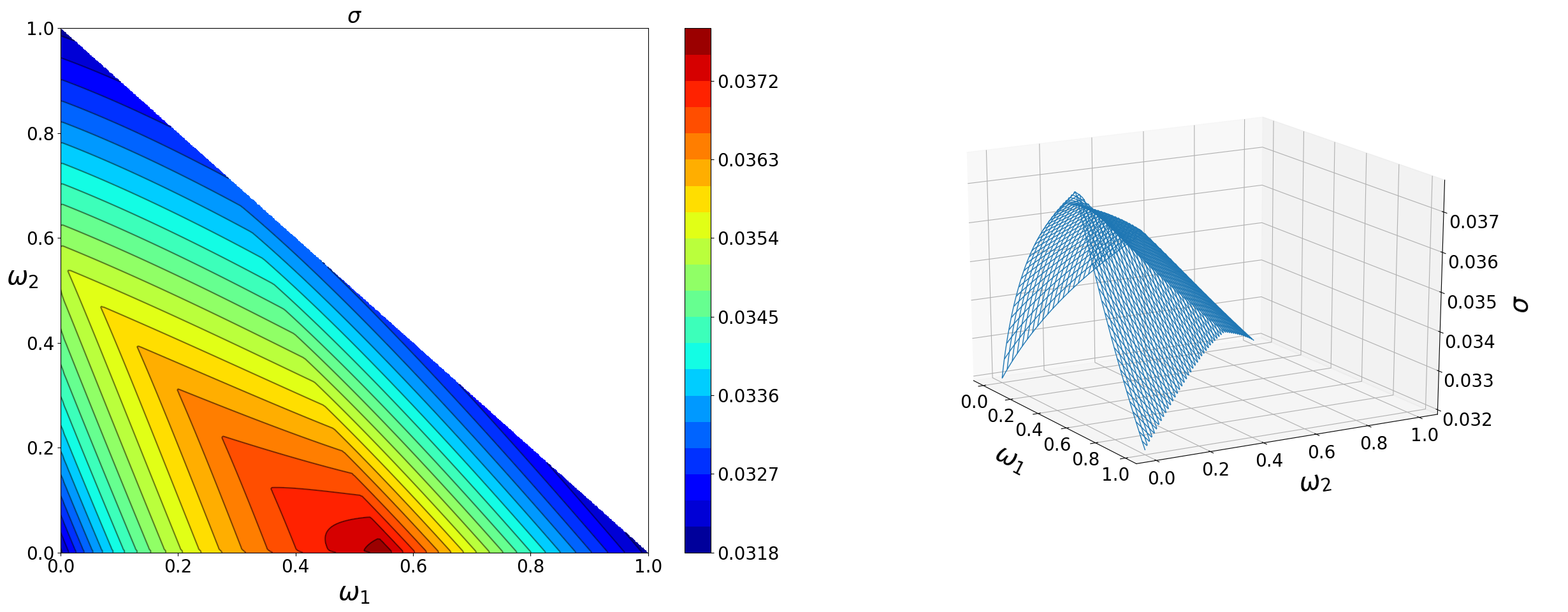}
		\label{fig:three_agents:vol}
	}
	\subfigure[]{
		\centering
		\includegraphics[trim=175pt 0pt 0pt 0pt, width=1.0\linewidth]{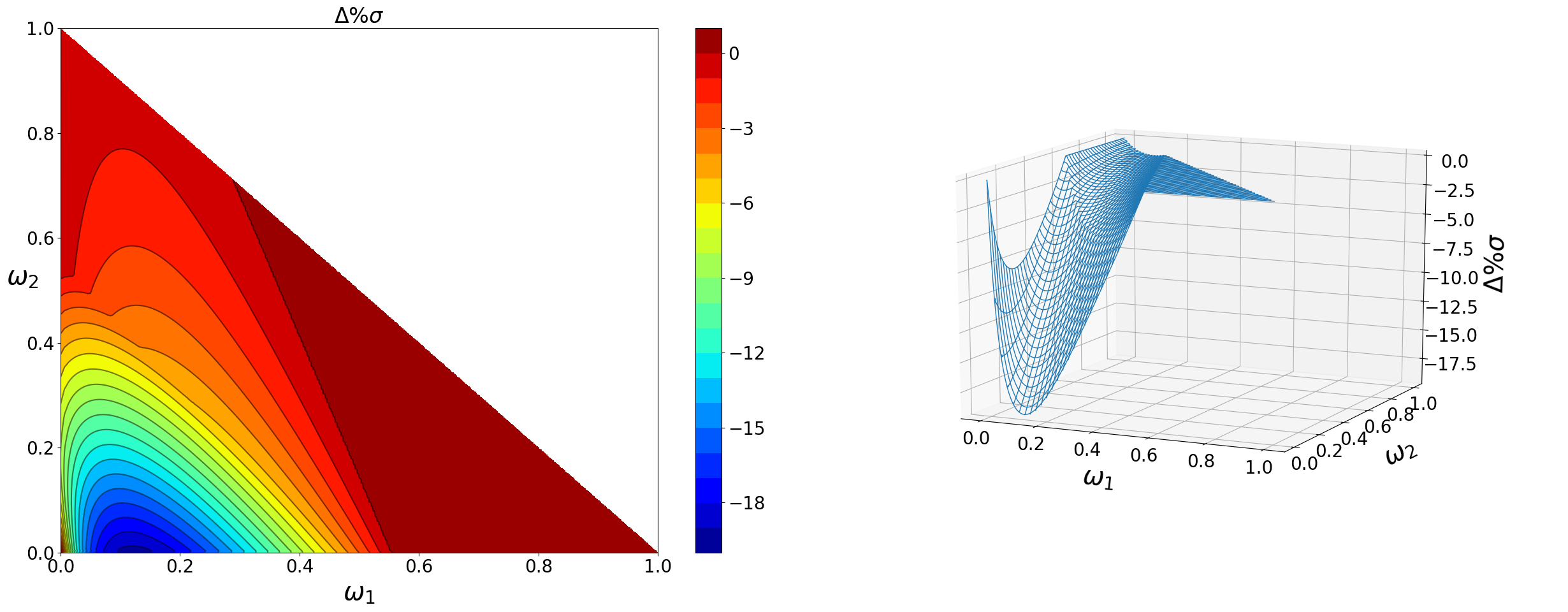}
		\label{fig:three_agents:vol_dev}
	}
	\caption{Volatility in levels (\cref{fig:three_agents:vol}) and in deviations from complete markets (\cref{fig:three_agents:vol_dev}).}
	\label{fig:three_agents:vol_all}
\end{figure}

Finally, the volatility surface shows non-monotonic and non-linear dynamics. As we can see in \cref{fig:three_agents:vol_all}, there is excess volatility above the fundamental volatility $\sigma_D$. However, the constraint reduces this because of a reduction in risk sharing. We can think of the margin constraint as pushing the economy towards the autarkical case, as individuals are unable to trade freely. In the limit when there is no trade whatsoever, the volatility of the asset price is simply the volatility of the underlying dividend. However, the constraint does not quite push the economy to this point, as individuals still exchange in order to remain against their constraint.

All of these observations point to several empirical tests, however the most apparent is that of leverage. We can see that the cyclicality of leverage is varying with other macroeconomic variables. Using this observation we can think about a new way to consider leverage cyclicality.

\section{The Cyclicality of Leverage}\label{sec:data}
\begin{figure}[!h]
	\hspace{-5pt}
	\subfigure[]{
		\includegraphics[width=0.5\textwidth]{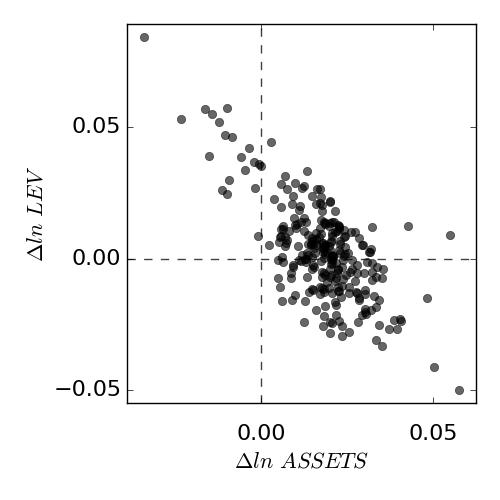}
		\label{fig:data:lev_to_assets_sector_89}
	}\hspace{-10pt}
	\subfigure[]{
		\includegraphics[width=0.5\textwidth]{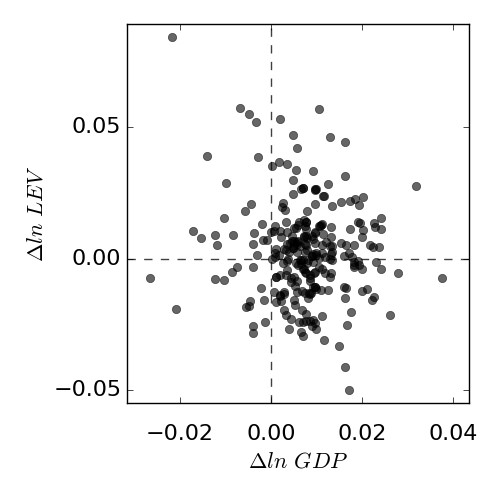}
		\centering
		\label{fig:data:lev_to_gdp_sector_89}
	}
	\caption{Growth rate in leverage plotted against the growth rate in assets (\cref{fig:data:lev_to_assets_sector_89}) and against the growth rate in GDP (\cref{fig:data:lev_to_gdp_sector_89}) for all sectors. Source: FRB Flow of Funds Data.}
	\label{fig:data:scatter_sector_89}
\end{figure}

Beliefs driven leverage cycles are pro-cyclical according to theory. This implication is somewhat contradicted in several empirical studies, including \cite{adrian2010liquidity}. In that paper the authors note that the leverage cycle is only pro-cyclical for a particular sector of the economy, asset broker/dealers. However, those authors plot leverage as a function of total assets, which produces a mechanical correlation. Consider the definition of financial leverage:
\begin{align*}
Leverage = \frac{Liabilities}{Net Wealth} = \frac{Liabilities}{Assets - Liabilities}
\end{align*}
Increases in balance sheet assets produce a negative correlation between leverage and assets\footnote{However, this makes the fact that \cite{adrian2010liquidity} find pro-cyclical leverage cycles for broker/dealers all the more substantial of a finding}(\cite{ang2011hedge}). \cref{fig:data:lev_to_assets_sector_89} plots the rate of growth in leverage against the rate of growth in assets for all sectors over 1952Q1 to 2017Q1 as measured from the US Flow of Funds. As you can see, there is a clear negative relationship.

Consider instead changes in GDP as a proxy for the business cycle. \cref{fig:data:lev_to_gdp_sector_89} plots the rate of growth in leverage for all sectors against the rate of growth in GDP over the same period, again from U.S. Flow of Funds data. The previously clear negative relationship has disappeared, implying the leverage cycle is ambiguous in this sense. However, this ambiguity may simply be that there exists some other explanatory variable which drives the cyclicality of leverage, in particular preference heterogeneity.

\begin{figure}[!h]
	\hspace{-5pt}
	\subfigure[]{
		\includegraphics[width=0.5\textwidth]{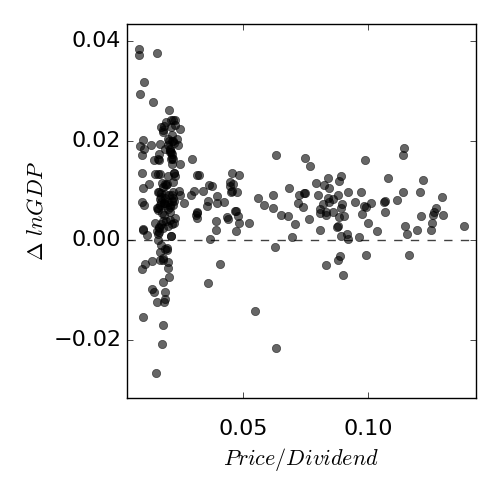}
		\centering
		\label{fig:data:pd_to_gdp_sector_89}
	} \hspace{-10pt}
	\subfigure[]{
		\includegraphics[width=0.5\textwidth]{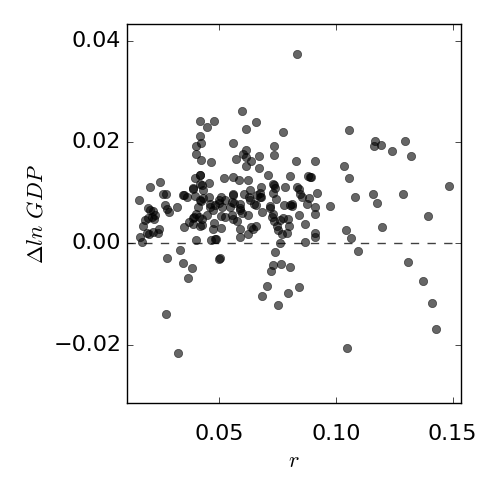}
		\centering
		\label{fig:data:r_to_gdp_sector_89}
	}
	\caption{Growth rate in GDP plotted against the price/dividend ratio (\cref{fig:data:pd_to_gdp_sector_89}), proxied by the price of the S\&P500 divided by GDP, and against the risk free rate (\cref{fig:data:r_to_gdp_sector_89}), proxied by the yield on constant maturity 10-year treasuries. Source: FRB Flow of Funds Data and FRED.}
	\label{fig:data:scatter_sector_89}
\end{figure}

One proxy for preference heterogeneity is the price-dividend ratio. As we saw in \cref{sec:sol}, asset prices will be high relative to dividends and vice-versa when the marginal agent in the economy is less risk averse. \cref{fig:data:pd_to_gdp_sector_89} plots the growth rate in GDP against the price of the S\&P 500 divided by GDP (a measure of the price/dividend ratio of the total economy). Indeed we see that there is substantial dispersion in this measure. The price/dividend ratio bunches towards the origin as asset prices have been rising over time, but there is little evidence for a clear positive or negative relationship with GDP growth. For this reason, we can consider the correlations between these variables, captured by the following regression:
\begin{align*}
\Delta ln \text{ } Lev = \alpha + \beta_1 \Delta ln \text{ } GDP + \beta_2 \Delta ln \text{ } GDP * \frac{S}{D} + \beta_3 \frac{S}{D} 
\end{align*}
The cyclicality of the leverage cycle is then captured by the slope with respect to the growth rate in GDP, that is
\begin{align*}
\partial_{\Delta ln \text{ } GDP} \Delta ln \text{ } Lev = \beta_1 + \beta_2 \frac{S}{D}
\end{align*}
The leverage cycle is pro- or counter-cyclical as this value is positive or negative, respectively \cref{table:reg_pd} reports the results for several specifications, studying different subsamples of the economy.

\begin{table}
	\begin{center}
		\begin{tabular}{ld{4.6}d{4.6}d{4.6}d{4.6}}
			\hline
			& \multicolumn{1}{c}{\begin{tabular}{@{}c@{}} Nonfinancial \\ Corporations \end{tabular}}
			& \multicolumn{1}{c}{\begin{tabular}{@{}c@{}}Nonfinancial \\ Private \\ Business\end{tabular}}
			& \multicolumn{1}{c}{\begin{tabular}{@{}c@{}}HH's and \\ Nonprofits \end{tabular}}
			& \multicolumn{1}{c}{\begin{tabular}{@{}c@{}} All  \\ Sectors \end{tabular}}\\
			& \multicolumn{1}{c}{\begin{tabular}{@{}c@{}} (1) \end{tabular}}
			& \multicolumn{1}{c}{\begin{tabular}{@{}c@{}}(2)\end{tabular}}
			& \multicolumn{1}{c}{\begin{tabular}{@{}c@{}}(3) \end{tabular}}
			& \multicolumn{1}{c}{\begin{tabular}{@{}c@{}} (4)\end{tabular}}\\
			\hline
			\hline
			Intercept & -0.0042  & -0.0078 & -0.0103 & -0.0046  \\
			& (0.0618) & (0.0610)  & (0.0599)   & (0.0590)    \\
			$\Delta ln$ $GDP$ & 0.1034 & 0.2117$**$  & 0.2827$***$  & 0.2397$**$\\
			& (0.0994) & (0.0980)  & (0.0963)   & (0.0949)    \\
			$S/D$  & -0.0045  & -0.0119   & -0.0209    & -0.2107$**$   \\
			& (0.0874) & (0.0862)  & (0.0846)   & (0.0834)    \\
			$\Delta ln$ $GDP * S/D$ & -0.1727  & -0.2726$**$ & -0.3610$***$ & -0.5157$***$\\
			& (0.1070) & (0.1055)  & (0.1036)   & (0.1021)    \\
			\hline
		\end{tabular}
	\end{center}
	\vspace{-10pt}
	\small{Standard errors in parentheses.\\
		$^*$ : $p \leq 0.1$,  $^{**}$ : $p \leq 0.05$,  $^{***}$ : $p \leq 0.01$}
	\caption{Regression results for dependent variable $\Delta ln \text{ } Lev$ for different sectors of the economy. A positive and significant coefficient on $\Delta ln \text{ } GDP$ implies procyclicality, while a negative and significant coefficient on the interaction with $S/D$ implies counter-cyclicality when the price dividend ratio is high. Note: Variables are normalized using z-score.}
	\label{table:reg_pd}
\end{table}

The results imply that the cyclicality of leverage is not the same for all values of the price-dividend ratio. Column 4 gives results for all sectors included in the US Flow of Funds. Leverage growth is positively correlated with GDP growth when the price dividend ratio is low. As asset prices rise the effect changes sign and the correlation becomes negative. Changes in the price-dividend ratio imply changes in the preferences of the marginal agent pricing risky assets. When the price-dividend ratio is low the marginal agent is risk averse, while when the price-dividend ratio is high the marginal agent is more risk neutral. Thus the leverage cycle is pro-cyclical when risk-averse agents dominate and counter-cyclical when risk-neutral agents dominate.

This result is fairly robust to other measures of marginal preferences. One problem could be the heteroscedasticity exhibited by GDP growth over the price/dividend ratio in \cref{fig:data:pd_to_gdp_sector_89}. Consider the risk free rate as a proxy for the marginal agent, which is plotted in \cref{fig:data:r_to_gdp_sector_89} against GDP growth. In this case the dispersion of GDP growth is more uniform over values of the interest rate. Define a similar set of regressions as before, i.e.:
\begin{align*}
\Delta ln \text{ } Lev = \alpha + \beta_1 \Delta ln \text{ } GDP + \beta_2 \Delta ln \text{ } GDP * r + \beta_3 r 
\end{align*}
Again the cyclicality is captured by the slope with respect to the growth rate in GDP:
\begin{align*}
\partial_{\Delta ln \text{ } GDP} \Delta ln \text{ } Lev = \beta_1 + \beta_2 r
\end{align*}
In this case we should expect the sign to flip. The interest rate is high when the marginal agent is risk-averse and low when the marginal agent is risk-neutral. \cref{table:reg_r} reports the results. Leverage growth co-moves positively with GDP growth and the interest rate is high and negatively when the interest rate is low. This result again implies that the cyclicality of the leverage cycle depends in the same way as before on the preferences of the marginal agent.

\begin{table}
	\begin{center}
		\begin{tabular}{ld{4.6}d{4.6}d{4.6}d{4.6}}
			\hline
			& \multicolumn{1}{c}{\begin{tabular}{@{}c@{}} Nonfinancial \\ Corporations \end{tabular}}
			& \multicolumn{1}{c}{\begin{tabular}{@{}c@{}}Nonfinancial \\ Private \\ Business\end{tabular}}
			& \multicolumn{1}{c}{\begin{tabular}{@{}c@{}}HH's and \\ Nonprofits \end{tabular}}
			& \multicolumn{1}{c}{\begin{tabular}{@{}c@{}} All  \\ Sectors \end{tabular}}\\
			& \multicolumn{1}{c}{\begin{tabular}{@{}c@{}} (1) \end{tabular}}
			& \multicolumn{1}{c}{\begin{tabular}{@{}c@{}}(2)\end{tabular}}
			& \multicolumn{1}{c}{\begin{tabular}{@{}c@{}}(3) \end{tabular}}
			& \multicolumn{1}{c}{\begin{tabular}{@{}c@{}} (4)\end{tabular}}\\
			\hline
			\hline
			Intercept  & 0.0170     & 0.0838    & 0.0948     & 0.0227      \\
			& (0.0674)   & (0.0638)  & (0.0674)   & (0.0672)    \\
			$\Delta ln$ $GDP$& -0.5144$***$ & -0.4412$**$ & -0.5330$***$ & -0.9689$***$  \\
			& (0.1907)   & (0.1808)  & (0.1908)   & (0.1904)    \\
			$r$   & 0.0164     & -0.0847   & 0.0018     & 0.0549      \\
			& (0.0789)   & (0.0748)  & (0.0790)   & (0.0788)    \\
			$\Delta ln$ $GDP * r$ & 0.4121$**$   & 0.4095$**$  & 0.5131$***$  & 0.7603$***$   \\
			& (0.1700)   & (0.1611)  & (0.1701)   & (0.1697)    \\
			\hline
		\end{tabular}
	\end{center}
	\vspace{-10pt}
	\small{Standard errors in parentheses.\\
		$^*$ : $p \leq 0.1$,  $^{**}$ : $p \leq 0.05$,  $^{***}$ : $p \leq 0.01$}
	\caption{Regression results for dependent variable $\Delta ln \text{ } Lev$ for different sectors of the economy. A negative and significant coefficient on $\Delta ln \text{ }GDP$ implies counter-cyclicality, while a positive and significant coefficient on the interaction with $r$ implies pro-cyclicality when the interest rate is high. As opposed to \cref{table:reg_pd}, $r$ is high when the risk averse agent dominates, exactly when the price-dividend ratio is low. Note: Variables are normalized using z-score.}
	\label{table:reg_r}
\end{table}

The regression results highlight how the cyclicality of the leverage cycle relates to financial variables and, in turn, preferences. Agents are likely to be constrained when asset prices are low, producing a pro-cyclical leverage cycle. Agents will be far from their constraint when asset prices are high, producing a counter-cyclical the leverage cycle. Asset price movements are explained by changes in the marginal agent in the economy, as seen in \cref{sec:sol}.

\section{Conclusion} \label{sec:conc}
In this paper I've shown how one can solve a model of preference heterogeneity when agents face convex portfolio constraints. The results show how preference heterogeneity and constraint can lead not only to cascade effects, but also high asset prices, high returns, and low interest rates. I also show how leverage cycles can be both pro- or counter-cyclical depending on the underlying assumptions of preference heterogeneity and constraints. In addition I've documented a new stylized fact predicted by the model, namely that leverage is both pro- and counter-cyclical depending on the level of aggregate consumption. Future work on this topic could introduce a stochastic endowment and more general preferences. 

\bibliographystyle{apalike}
\bibliography{/home/tmabbot/MEGA/Writing/mybib}

\begin{thebibliography}{}

\bibitem[Abbot, 2017]{abbot2016heterogeneous}
Abbot, T. (2017).
\newblock Heterogeneous preferences and general equilibrium in financial
  markets.
\newblock {\em Working Paper}.

\bibitem[Adrian and Shin, 2010a]{adrian2010changing}
Adrian, T. and Shin, H.~S. (2010a).
\newblock The changing nature of financial intermediation and the financial
  crisis of 2007--2009.
\newblock {\em Annu. Rev. Econ.}, 2(1):603--618.

\bibitem[Adrian and Shin, 2010b]{adrian2010liquidity}
Adrian, T. and Shin, H.~S. (2010b).
\newblock Liquidity and leverage.
\newblock {\em Journal of financial intermediation}, 19(3):418--437.

\bibitem[Aiyagari, 1994]{aiyagari1994uninsured}
Aiyagari, S.~R. (1994).
\newblock Uninsured idiosyncratic risk and aggregate saving.
\newblock {\em The Quarterly Journal of Economics}, pages 659--684.

\bibitem[Ang et~al., 2011]{ang2011hedge}
Ang, A., Gorovyy, S., and Van~Inwegen, G.~B. (2011).
\newblock Hedge fund leverage.
\newblock {\em Journal of Financial Economics}, 102(1):102--126.

\bibitem[Basak and Cuoco, 1998]{basak1998equilibrium}
Basak, S. and Cuoco, D. (1998).
\newblock An equilibrium model with restricted stock market participation.
\newblock {\em Review of Financial Studies}, 11(2):309--341.

\bibitem[Bernanke et~al., 1999]{bernanke1999financial}
Bernanke, B.~S., Gertler, M., and Gilchrist, S. (1999).
\newblock The financial accelerator in a quantitative business cycle framework.
\newblock {\em Handbook of macroeconomics}, 1:1341--1393.

\bibitem[Bhamra and Uppal, 2014]{bhamra2014asset}
Bhamra, H.~S. and Uppal, R. (2014).
\newblock Asset prices with heterogeneity in preferences and beliefs.
\newblock {\em Review of Financial Studies}, 27(2):519--580.

\bibitem[Brunnermeier and Nagel, 2008]{brunnermeier2008wealth}
Brunnermeier, M.~K. and Nagel, S. (2008).
\newblock Do wealth fluctuations generate time-varying risk aversion?
  micro-evidence on individuals' asset allocation.
\newblock {\em The American Economic Review}, 98(3):713--736.

\bibitem[Brunnermeier and Pedersen, 2009]{brunnermeier2009market}
Brunnermeier, M.~K. and Pedersen, L.~H. (2009).
\newblock Market liquidity and funding liquidity.
\newblock {\em Review of Financial studies}, 22(6):2201--2238.

\bibitem[Brunnermeier and Sannikov, 2014]{brunnermeier2014macroeconomic}
Brunnermeier, M.~K. and Sannikov, Y. (2014).
\newblock A macroeconomic model with a financial sector.
\newblock {\em The American Economic Review}, 104(2):379--421.

\bibitem[Campbell and Cochrane, 1999]{campbell1999force}
Campbell, J.~Y. and Cochrane, J.~H. (1999).
\newblock By force of habit: A consumption-based explanation of aggregate stock
  market behavior.
\newblock {\em The Journal of Political Economy}, 107(2):205--251.

\bibitem[Chabakauri, 2013]{chabakauri2013dynamic}
Chabakauri, G. (2013).
\newblock Dynamic equilibrium with two stocks, heterogeneous investors, and
  portfolio constraints.
\newblock {\em Review of Financial Studies}, 26(12):3104--3141.

\bibitem[Chabakauri, 2015]{chabakauri2015asset}
Chabakauri, G. (2015).
\newblock Asset pricing with heterogeneous preferences, beliefs, and portfolio
  constraints.
\newblock {\em Journal of Monetary Economics}, 75:21--34.

\bibitem[Chiappori et~al., 2012]{chiappori2012aggregate}
Chiappori, P.-A., Gandhi, A., Salani{\'e}, B., and Salani{\'e}, F. (2012).
\newblock From aggregate betting data to individual risk preferences.

\bibitem[Chiappori and Paiella, 2011]{chiappori2011relative}
Chiappori, P.-A. and Paiella, M. (2011).
\newblock Relative risk aversion is constant: Evidence from panel data.
\newblock {\em Journal of the European Economic Association}, 9(6):1021--1052.

\bibitem[Christensen et~al., 2012]{christensen2012equilibrium}
Christensen, P.~O., Larsen, K., and Munk, C. (2012).
\newblock Equilibrium in securities markets with heterogeneous investors and
  unspanned income risk.
\newblock {\em Journal of Economic Theory}, 147(3):1035--1063.

\bibitem[Coen-Pirani, 2004]{coen2004effects}
Coen-Pirani, D. (2004).
\newblock Effects of differences in risk aversion on the distribution of
  wealth.
\newblock {\em Macroeconomic Dynamics}, 8(05):617--632.

\bibitem[Confortola et~al., 2017]{confortola2017backward}
Confortola, F., Cosso, A., and Fuhrman, M. (2017).
\newblock Backward sdes and infinite horizon stochastic optimal control.
\newblock {\em arXiv preprint arXiv:1710.06723}.

\bibitem[Cozzi, 2011]{cozzi2011risk}
Cozzi, M. (2011).
\newblock Risk aversion heterogeneity, risky jobs and wealth inequality.
\newblock Technical report, Queen's Economics Department Working Paper.

\bibitem[Cuoco, 1997]{cuoco1997optimal}
Cuoco, D. (1997).
\newblock Optimal consumption and equilibrium prices with portfolio constraints
  and stochastic income.
\newblock {\em Journal of Economic Theory}, 72(1):33--73.

\bibitem[Cuoco and He, 1994]{cuoco1994dynamic}
Cuoco, D. and He, H. (1994).
\newblock Dynamic equilibrium in infinite-dimensional economies with incomplete
  information.
\newblock Technical report, Working paper, Wharton School, University of
  Pennsylvania.

\bibitem[Cuoco et~al., 2001]{cuoco2001dynamic}
Cuoco, D., He, H., et~al. (2001).
\newblock Dynamic aggregation and computation of equilibria in
  finite-dimensional economies with incomplete financial markets.
\newblock {\em Annals of Economics and Finance}, 2(2):265--296.

\bibitem[Cvitani{\'c} et~al., 2011]{cvitanic2011financial}
Cvitani{\'c}, J., Jouini, E., Malamud, S., and Napp, C. (2011).
\newblock Financial markets equilibrium with heterogeneous agents.
\newblock {\em Review of Finance}, page rfr018.

\bibitem[Cvitani{\'c} and Karatzas, 1992]{cvitanic1992convex}
Cvitani{\'c}, J. and Karatzas, I. (1992).
\newblock Convex duality in constrained portfolio optimization.
\newblock {\em The Annals of Applied Probability}, pages 767--818.

\bibitem[Dumas, 1989]{dumas1989two}
Dumas, B. (1989).
\newblock Two-person dynamic equilibrium in the capital market.
\newblock {\em Review of Financial Studies}, 2(2):157--188.

\bibitem[Epstein et~al., 2014]{epstein2014much}
Epstein, L.~G., Farhi, E., and Strzalecki, T. (2014).
\newblock How much would you pay to resolve long-run risk?
\newblock {\em The American Economic Review}, 104(9):2680--2697.

\bibitem[Epstein and Zin, 1989]{epstein1989substitution}
Epstein, L.~G. and Zin, S.~E. (1989).
\newblock Substitution, risk aversion, and the temporal behavior of consumption
  and asset returns: A theoretical framework.
\newblock {\em Econometrica: Journal of the Econometric Society}, pages
  937--969.

\bibitem[G{\^a}rleanu and Panageas, 2015]{garleanu2015young}
G{\^a}rleanu, N. and Panageas, S. (2015).
\newblock Young, old, conservative and bold: The implications of heterogeneity
  and finite lives for asset pricing.
\newblock {\em Journal of Political Economy}, 123(3):670--685.

\bibitem[Garleanu and Pedersen, 2011]{garleanu2011margin}
Garleanu, N. and Pedersen, L.~H. (2011).
\newblock Margin-based asset pricing and deviations from the law of one price.
\newblock {\em Review of Financial Studies}, 24(6):1980--2022.

\bibitem[Geanakoplos, 1996]{geanakoplos1996promises}
Geanakoplos, J. (1996).
\newblock Promises promises.

\bibitem[Geanakoplos, 2010]{geanakoplos2010leverage}
Geanakoplos, J. (2010).
\newblock The leverage cycle.
\newblock In {\em NBER Macroeconomics Annual 2009, Volume 24}, pages 1--65.
  University of Chicago Press.

\bibitem[Guvenen, 2006]{guvenen2006reconciling}
Guvenen, F. (2006).
\newblock Reconciling conflicting evidence on the elasticity of intertemporal
  substitution: A macroeconomic perspective.
\newblock {\em Journal of Monetary Economics}, 53(7):1451--1472.

\bibitem[Guvenen, 2009]{guvenen2009parsimonious}
Guvenen, F. (2009).
\newblock A parsimonious macroeconomic model for asset pricing.
\newblock {\em Econometrica}, 77(6):1711--1750.

\bibitem[Halling et~al., 2016]{halling2016leverage}
Halling, M., Yu, J., and Zechner, J. (2016).
\newblock Leverage dynamics over the business cycle.
\newblock {\em Journal of Financial Economics}, 122(1):21--41.

\bibitem[Hardouvelis and Peristiani, 1992]{hardouvelis1992margin}
Hardouvelis, G.~A. and Peristiani, S. (1992).
\newblock Margin requirements, speculative trading, and stock price
  fluctuations: The case of japan.
\newblock {\em The Quarterly Journal of Economics}, 107(4):1333--1370.

\bibitem[Hardouvelis and Theodossiou, 2002]{hardouvelis2002asymmetric}
Hardouvelis, G.~A. and Theodossiou, P. (2002).
\newblock The asymmetric relation between initial margin requirements and stock
  market volatility across bull and bear markets.
\newblock {\em Review of Financial Studies}, 15(5):1525--1559.

\bibitem[He and Pages, 1993]{he1993labor}
He, H. and Pages, H.~F. (1993).
\newblock Labor income, borrowing constraints, and equilibrium asset prices.
\newblock {\em Economic Theory}, 3(4):663--696.

\bibitem[He and Krishnamurthy, 2013]{he2013intermediary}
He, Z. and Krishnamurthy, A. (2013).
\newblock Intermediary asset pricing.
\newblock {\em The American Economic Review}, 103(2):732--770.

\bibitem[Hugonnier, 2012]{hugonnier2012rational}
Hugonnier, J. (2012).
\newblock Rational asset pricing bubbles and portfolio constraints.
\newblock {\em Journal of Economic Theory}, 147(6):2260--2302.

\bibitem[Ishii, 1989]{ishii1989uniqueness}
Ishii, H. (1989).
\newblock On uniqueness and existence of viscosity solutions of fully nonlinear
  second-order elliptic pde's.
\newblock {\em Communications on pure and applied mathematics}, 42(1):15--45.

\bibitem[Karatzas et~al., 1987]{karatzas1987optimal}
Karatzas, I., Lehoczky, J.~P., and Shreve, S.~E. (1987).
\newblock Optimal portfolio and consumption decisions for a “small
  investor” on a finite horizon.
\newblock {\em SIAM journal on control and optimization}, 25(6):1557--1586.

\bibitem[Karatzas et~al., 2003]{karatzas2003optimal}
Karatzas, I., {\v{Z}}itkovi{\'c}, G., et~al. (2003).
\newblock Optimal consumption from investment and random endowment in
  incomplete semimartingale markets.
\newblock {\em The Annals of Probability}, 31(4):1821--1858.

\bibitem[Kimball, 1990]{kimball1990precautionary}
Kimball, M.~S. (1990).
\newblock Precautionary saving in the small and in the large.
\newblock {\em Econometrica: Journal of the Econometric Society}, pages 53--73.

\bibitem[Kiyotaki and Moore, 1997]{kiyotaki1997credit}
Kiyotaki, N. and Moore, J. (1997).
\newblock Credit cycles.
\newblock {\em Journal of political economy}, 105(2):211--248.

\bibitem[Kogan et~al., 2007]{kogan2007equity}
Kogan, L., Makarov, I., and Uppal, R. (2007).
\newblock The equity risk premium and the riskfree rate in an economy with
  borrowing constraints.
\newblock {\em Mathematics and Financial Economics}, 1(1):1--19.

\bibitem[Korajczyk and Levy, 2003]{korajczyk2003capital}
Korajczyk, R.~A. and Levy, A. (2003).
\newblock Capital structure choice: macroeconomic conditions and financial
  constraints.
\newblock {\em Journal of financial economics}, 68(1):75--109.

\bibitem[Krusell and Smith, 1998]{krusell1998income}
Krusell, P. and Smith, Jr, A.~A. (1998).
\newblock Income and wealth heterogeneity in the macroeconomy.
\newblock {\em Journal of Political Economy}, 106(5):867--896.

\bibitem[Longstaff and Wang, 2012]{longstaff2012asset}
Longstaff, F.~A. and Wang, J. (2012).
\newblock Asset pricing and the credit market.
\newblock {\em Review of Financial Studies}, 25(11):3169--3215.

\bibitem[Lucas, 1978]{lucas1978asset}
Lucas, R.~E. (1978).
\newblock Asset prices in an exchange economy.
\newblock {\em Econometrica: Journal of the Econometric Society}, pages
  1429--1445.

\bibitem[Mehra and Prescott, 1985]{mehra1985equity}
Mehra, R. and Prescott, E.~C. (1985).
\newblock The equity premium: A puzzle.
\newblock {\em Journal of monetary Economics}, 15(2):145--161.

\bibitem[Merton, 1971]{merton1971optimum}
Merton, R.~C. (1971).
\newblock Optimum consumption and portfolio rules in a continuous-time model.
\newblock {\em Journal of economic theory}, 3(4):373--413.

\bibitem[Prieto, 2010]{prieto2010dynamic}
Prieto, R. (2010).
\newblock Dynamic equilibrium with heterogeneous agents and risk constraints.

\bibitem[Radner, 1972]{radner1972existence}
Radner, R. (1972).
\newblock Existence of equilibrium of plans, prices, and price expectations in
  a sequence of markets.
\newblock {\em Econometrica: Journal of the Econometric Society}, pages
  289--303.

\bibitem[Rytchkov, 2014]{rytchkov2014asset}
Rytchkov, O. (2014).
\newblock Asset pricing with dynamic margin constraints.
\newblock {\em The Journal of Finance}, 69(1):405--452.

\bibitem[Santos and Veronesi, 2010]{santos2010habit}
Santos, T. and Veronesi, P. (2010).
\newblock Habit formation, the cross section of stock returns and the cash-flow
  risk puzzle.
\newblock {\em Journal of Financial Economics}, 98(2):385--413.

\bibitem[Stiglitz and Weiss, 1981]{stiglitz1981credit}
Stiglitz, J.~E. and Weiss, A. (1981).
\newblock Credit rationing in markets with imperfect information.
\newblock {\em The American economic review}, 71(3):393--410.

\bibitem[Weil, 1989]{weil1989equity}
Weil, P. (1989).
\newblock The equity premium puzzle and the risk-free rate puzzle.
\newblock {\em Journal of Monetary Economics}, 24(3):401--421.

\end{thebibliography}

\clearpage

\appendix
\crefalias{section}{appsec}

\section{Proofs}

\begin{proof}[Proof of \cref{prop:r_theta}]
	Take the market clearing condition in consumption and divide through by agent $i$'s consumption
	\begin{align*}
	{\sum_j} c_{jt} = D_t \Leftrightarrow \ind{c} = \frac{\ind{c}}{\sum_j c_{jt}}D_t = \left ( \frac{\left( e^{\rho t}\Lambda_i \ind{H} \right )^{\frac{-1}{\gamma_i}}}{\sum_j \left( e^{\rho t}\Lambda_j H_{jt} \right )^{\frac{-1}{\gamma_j}}} \right)D_t = \omega_{it}D_{it}
	\end{align*}
	where $\omega_{it}$ represents an individual's consumption weight and is given by
	\begin{align*}
	\omega_{it} = \frac{\left( e^{\rho t}\Lambda_iH_{it} \right )^{\frac{-1}{\gamma_i}}}{\sum_{j=1}^N \left( e^{\rho t}\Lambda_jH_{jt} \right )^{\frac{-1}{\gamma_j}}} 
	\end{align*}
	Assume individual consumption follows an It\^o process such that
	\begin{align}\label{cons}
	\frac{dc_{it}}{c_{it}} = \mu_{cit}dt + \sigma_{cit}dW(t)
	\end{align}
	Apply It\^o's lemma to \cref{eq:foc} and solve for $\mu_{cit}$ and $\sigma_{cit}$
	\begin{align*}
	\mu_{cit} = \frac{r_t - \rho + \delta_i(\ind{\nu})}{\gamma_i} + \frac{1 + \gamma_i}{\gamma_i^2}\frac{1}{2}\left( \theta_t + \frac{\nu_{it}}{\sigma_t} \right)^2
	\hspace{10pt} \text{ , } \hspace{10pt}
	\sigma_{c^i}(t) = \frac{1}{\gamma_i} \left( \theta + \frac{\nu_{it}}{\sigma_t} \right)
	\end{align*}
	Apply It\^o's lemma to the market clearing condition for consumption and match coefficients to find
	\begin{align*}
	\mu_D = {\sum_{i=1}^N} \omega_{it}\mu_{cit}
	\hspace{10pt} \text{ , } \hspace{10pt}
	\sigma_D = {\sum_{i=1}^N} \omega_{it}\sigma_{cit}
	\end{align*}
	Now substitute the values for consumption drift and diffusion and solve for the interest rate and the market price of risk:
	\begin{align*}
	\theta_t &= \frac{1}{\sum_i \frac{\ind{\omega}}{\gamma_i}}\left(\sigma_D - \frac{1}{\sigma_t } \sum_i \frac{\ind{\omega}\ind{\nu}}{\gamma_i} \right) \\
	r_t &= \frac{1}{\sum_i \frac{\ind{\omega}}{\gamma_i}} \left( \mu_D + \rho\sum_i \frac{\ind{\omega}}{\gamma_i} -  \sum_i \frac{\ind{\omega}}{\gamma_i}\delta_i(\ind{\nu}) - \frac{1}{2}\sum_i \frac{1 + \gamma_i}{\gamma_i} \left(\theta_t + \frac{\ind{\nu}}{\sigma_t} \right)^2 \ind{\omega} \right) 
	\end{align*}

\end{proof}

\begin{proof}[Proof of \cref{prop:weights}]
	Apply It\^o's lemma to $\omega_{it} = \frac{c_{it}}{D_t}$ and match coefficients to find the dynamics of consumption weights in \cref{eq:prop:mu_w} and \cref{eq:prop:sig_w}.
\end{proof}

\begin{proof}[Proof of \cref{prop:PDE,prop:portfolios}]
	Assume there exists a Markovian equilibrium in $\bm{\omega}_t = [\omega_{1t}, \dots, \omega_{(N-1)t}]$. In the individual's fictitious financial market, they solve the following optimization problem
	\begin{equation*} \label{eq:maximization}
	\begin{aligned}
	J_i(t, x, \bm{\omega}) =& \underset{\{ c_{it},\pi_{it} \}_{u=t}^\infty}{\text{max}}
	& & \mathbb{E} \int_{t}^{\infty} e^{-\rho (u - t)}\frac{c_{iu}^{1-\gamma_i}}{1-\gamma_i}du \\
	& \text{s.t.}
	& &  dX_{it} = \left [ X_{it} \left ( r_t + + \delta_i(\nu_{it}) + \pi_{it} \sigma_t \left( \theta + \frac{\nu_{it}}{\sigma_t} \right) \right ) - c_{it} \right ] dt + X_{it} \pi_{it} \sigma_t dW_t\\
	& & & d\bm{\omega}_t = \bm{\mu_\omega}dt + \bm{\sigma_\omega} dW_t\\
	& & & \delta_i(\nu_{it}) + \nu_{it}\ind{\pi} = 0\\
	& & & \ind{X} = x \text{ , } \bm{\omega}_t = \bm{\omega}
	\end{aligned}
	\end{equation*}
	Then an individual's Hamilton-Jacobi-Bellman (HJB) equation writes\footnote{In the following, the $t$ subscript denotes dependence on the state. For all values other than $J_{it}$, this implies only $\bm{\omega}$.}
	\begin{equation}
	\begin{aligned}\label{eq:proof:prop3_hjb}
	0 = \max_{c, \pi} \left\{ e^{-\rho t}\frac{c^{1 - \gamma_i }}{1 - \gamma_i} + \frac{\partial J_{it}}{\partial t} + \left[ X \left( r_t + \delta_i(\nu_{it}) + \pi \sigma_t \left( \theta_t + \frac{\nu_{it}}{\sigma_t} \right) \right) - c \right] \frac{\partial J_{it}}{\partial X} \right. \\
	\left. + \sum_{j=1}^{N-1}\mu_{\omega j t} \omega_{jt} \frac{\partial J_{it}}{\partial \omega_j} + \pi\sum_{j=1}^{N-1}\sigma_{\omega jt} \sigma_t  \omega_j X \frac{\partial^2 J_{it}}{\partial X \partial \omega_j} + \sum_{j=1}^{N-1}\sum_{k < j} \sigma_{\omega jt}\sigma_{\omega kt}\omega_j \omega_k \frac{\partial^2 J_{it}}{\partial \omega_j \partial \omega_k} \right.\\
	\left. + \frac{1}{2} \left[ X^2 \pi^2 \sigma_t^2 \frac{\partial^2 J_{it}}{\partial X^2} + \sum_{j=1}^{N-1} \sigma_{\omega jt}^2 \omega_j^2 \frac{\partial^2 J_{it}}{\partial \omega_j^2} \right]\right\}
	\end{aligned}
	\end{equation}
	subject to the transversality condition $\mathbb{E}_t J_{it} \rightarrow 0$ for all $i$. First order conditions imply
	\begin{align}
	c &= \left(e^{\rho t} \frac{\partial J_{it}}{\partial X}\right)^{\frac{-1}{\gamma_i}} \label{eq:proof_prop2_opt_c}\\
	\pi &= -\left( X\sigma_t \frac{\partial^2\ind{J}}{\partial X^2}\right)^{-1} \left[ \left( \theta_t + \frac{\ind{\nu}}{\sigma_t} \right) \frac{\partial\ind{J}}{\partial X} + \sum_{j=1}^{N-1} \sigma_{\omega jt} \omega_{j} \frac{\partial^2\ind{J}}{\partial X \partial \omega_j} \right] \label{eq:proof_prop2_opt_pi}
	\end{align}
	Assume that the value function is separable as
	\begin{align}\label{eq:proof_prop2_seperable}
	J_i(t, x, \bm{\omega}) = e^{-\rho t} \frac{x^{1 - \gamma_i} V_i(\bm{\omega})^{\gamma_i}}{1 - \gamma_i}
	\end{align}
	Substituting \cref{eq:proof_prop2_seperable} into \cref{eq:proof_prop2_opt_c,eq:proof_prop2_opt_pi} gives
	\begin{align}
	c&= \frac{x}{V_i(\bm{\omega})} \label{eq:proof_prop3_opt_sep_c}\\
	\pi &= \frac{1}{\gamma_i \sigma_t} \left( \theta_t + \frac{\ind{\nu}}{\sigma_t} +  \frac{\gamma_i}{V_i(\bm{\omega})} \sum_{j=1}^{N-1} \sigma_{\omega jt} \omega_j \frac{\partial V_i(\bm{\omega})}{\partial \omega_j} +  \right) \label{eq:proof_prop3_opt_sep_pi}
	\end{align}
	which shows that $V_i(\bm{\omega})$ is the wealth-consumption ratio as a function of the vector of consumption weights. Define $\nabla$ as the gradient operator and use \cref{eq:prop:bmwdynamics}, then \cref{eq:proof_prop3_opt_sep_pi} rewrites as
	\begin{align*}
	\pi &= \frac{1}{\gamma_i \sigma_t} \left( \theta_t + \frac{\ind{\nu}}{\sigma_t} + \gamma_i  \frac{\bm{\sigma_\omega}^T\nabla V_i(\bm{\omega})}{V_i(\bm{\omega})} \right)
	\end{align*}
	as in \cref{prop:portfolios}. Next, substitute \cref{eq:proof_prop2_seperable,eq:proof_prop3_opt_sep_c,eq:proof_prop3_opt_sep_pi} into \cref{eq:proof:prop3_hjb} and simplify to find
	\begin{equation}
	\begin{aligned}\label{eq:proof:pde}
	0 =& 1 + \frac{1}{2} \bm{\sigma_\omega}^T HV_i(\bm{\omega})\bm{\sigma_\omega} + \left[ \frac{1 - \gamma_i}{\gamma_i} \left( \theta_t + \frac{\ind{\nu}}{\sigma_t} \right) \bm{\sigma_\omega}^T + \bm{\mu_\omega}^T \right] \nabla V_i(\bm{\omega}) \\
	&+ \frac{1}{\gamma_i} \left[ (1 - \gamma_i)(r_t + \delta_i(\ind{\nu})) - \rho + \frac{1 - \gamma_i}{2\gamma_i}\left( \theta_t + \frac{\ind{\nu}}{\sigma_t} \right)^2 \right] V_i(\bm{\omega})
	\end{aligned}
	\end{equation}
	Where $\bm{\mu_\omega}$ and $\bm{\sigma_\omega}$ are as in \cref{prop:weights}, and where $H$ represents the Hessian operator. The boundary conditions are given by recognizing that the limits in $\omega_i \rightarrow \{0, 1\}$ is an economy where agent $i$ has zero weight, while the remaining agents determine prices.
\end{proof}

\begin{proof}[Proof of \cref{prop:asset_prices}]
	Define the price-dividend ratio as a function of consumption weights: $\pd(\bm{\omega}) = \pd_t = \frac{S_t}{D_t}$. Taking the market clearing condition for wealth
	\begin{align*}
	S_t &=  \sum_{i} \ind{X} \Leftrightarrow  \frac{S_t}{D_t}  =  \sum_{i} \frac{\ind{X}}{D_t} =  \sum_{i} \frac{\ind{X}}{\ind{c}}\frac{\ind{c}}{D_t} = \sum_{i} V_i(\bm{\omega})\omega_{i} = \pd(\bm{\omega})
	\end{align*}		
	To find volatility, apply It\^o's lemma to $D_t\sum_i V_i(\bm{\omega}_t)\omega_{it} = S_t$ and match coefficients to find the expression in \cref{eq:prop:eq_sigmat}.
\end{proof}

\begin{proof}[Proof of \cref{prop:adjustment_margin}]
	For a homogeneous margin constraint, $\ind{\nu} \leq 0$ and $m \geq 0$, thus $\ind{\nu}m \leq 0$ (\cite{cvitanic1992convex,chabakauri2015asset}). Additionally, $\ind{\pi} \leq m$. Substituting the solution for $\ind{\pi}$ from \cref{eq:prop2_opt_sep_pi} into the latter inequality and recognizing that, by the Kuhn-Tucker conditions at least one of the inequalities holds with equality gives the result.
\end{proof}

\begin{proof}[Proof of \cref{prop:verification}]
	This proof shows that the present setting satisfies the assumptions necessary to apply Proposition 5.1 and Theorem 5.1 from \cite{confortola2017backward}, namely that the problem admits a dynamic programming representation and that the solution (in the viscosity sense) to the HJB in \cref{eq:proof:prop3_hjb} is indeed the value function. The proof proceeds in three steps: first the state is defined as a Markov diffusion, second the assumptions from \cite{confortola2017backward} are verified, and third the assumptions from \cite{ishii1989uniqueness} are verified, giving uniqueness of the viscosity solution.
	\paragraph{Markovian}
	Define $Y_t = [X_{it}, \omega_{1t}, \dots, \omega_{(N-1)t}]^T \in \mathbb{R^+} \times \Delta^{N-1} = \mathcal{X}$ as the state vector of an individual and $\alpha_t = [c_{it}, \pi_{it}]^T \in \mathbb{R}^+ \times \Pi_i = \mathcal{A}$ as the control vector. By \cref{prop:r_theta,prop:weights,prop:PDE,prop:portfolios,prop:asset_prices,prop:adjustment_margin} the state vector is Markovian and has controlled dynamics
	\begin{align*}
	dY_t = b(Y_t, \alpha_t)dt + \sigma(Y_t, \alpha_t)dW_t
	\end{align*}
	\paragraph{Bellman Principle and Existence of Viscosity Solution}
	Define the felicity function $f(y, a) = u_i(c)$ and notice that this is not a function of the state and only a function of one of the controls. The following assumptions must be verified:
	\begin{assumption}\label{assumption1}
		The dynamics of wealth, the dynamics of consumption weights, and the utility function must be continuous. 
	\end{assumption}	
	\begin{assumption}\label{assumption2}
		The dynamics of wealth must be jointly Lipschitz continuous in the state variable. 
	\end{assumption}
	\begin{assumption}\label{assumption3}
		There must exist some constants $M>0$ and $r \geq 0$ such that
		\begin{align*}
		\| f(y, a) \| \leq M(1 + \| y \|^r)
		\end{align*}
		for all $y \in \mathbb{R}^N$ and all $a \in \mathcal{A}$.
	\end{assumption}
	\begin{assumption}\label{assumption4}
		$\rho > \overline{\rho}$, where $\overline{\rho} = 0$ if $r$ from \cref{assumption3} is zero, otherwise if $r > 0$, $\overline{\rho}>0$ is such that
		\begin{align*}
		\mathbb{E}\left[ \sup_{s\in[0, t]} \| Y_s \| \right] \leq Ce^{\overline{\rho}t}(1 + |x|^r)
		\end{align*}
		for some constant $C \geq 0$, with $\overline{\rho}$ and $C$ independent of $t$, $\alpha$, and $x$.
	\end{assumption}
	\begin{assumption}\label{assumption5}
		$f(x, a)$ is continuous in $x$ uniformly with respect to $a$.
	\end{assumption}
	
	\noindent\Cref{assumption5} is trivially satisfied as the utility function is not a function of the state, but only of the control.
	
	\noindent\Cref{assumption3} is easily shown to be satisfied since the utility function is of power form:
	\begin{align*}
	\left| \frac{c^{1 - \gamma_i}}{1 - \gamma_i} \right| \leq M(1 + \| y \|^r)
	\end{align*}
	taking $r=0$ we have
	\begin{align*}
	\left| \frac{c^{1 - \gamma_i}}{2(1 - \gamma_i)} \right| \leq M
	\end{align*}
	which is satisfied in finite $t$ by taking $M \geq max_t\{D_t\}^{1 - \gamma}/(2(1 - \gamma_i))$ in any finite $t$, thanks to market clearing. In the limit when $t \rightarrow \infty$, we have $D_t$ going to $\infty$, a.s. In this case, the admissible set goes to a singleton.
	
	\noindent \Cref{assumption4} is satisfied for all non-zero rates of time preference, given we took $r=0$ above.
	
	\noindent \Cref{assumption1} is satisfied for the utility function. For the dynamics of wealth, the functions are continuous up to $\nu_i(\cdot)$ and $\nabla V_i(\cdot)$. If we assume that $V_i(\cdot)$ is once continuously differentiable, then we satisfy \cref{assumption1}.
	
	\noindent \Cref{assumption2} Is satisfied if $V_i(\cdot)$ has bounded first derivative.
	
	\paragraph{Uniqueness of Viscosity Solution} (To be completed) Outline of proof: according to \cite{ishii1989uniqueness}, if the PDE given in \cref{eq:prop:pde} is uniform elliptic, if sub- and super-solutions have exponential growth, and under the above assumptions on continuity of the dynamics of the state, then the viscosity solution is unique.

\end{proof}

\section{Numerical Method}\label{appendix:numerical}
In this appendix I discuss the numerical solution of the problem of three preference types under margin constraints. This problem has several difficult features which make it challenging from a numerical perspective. First, the system of PDE's in \cref{eq:prop:ode} represents a highly non-linear  system, as the coefficients depend in a non-trivial way on the solution, as well as on the Jacobian. In addition, the coefficients are not smooth at the point where the constraint binds. Finally, the state space is the $2$ dimensional simplex, making finite difference tedious. 

\subsection{The Discretized State Space}\label{appendix:numerical:discretize}
\begin{figure}[!h]
	\hspace{-5pt}
	\subfigure[]{
		\centering
		\includegraphics[trim=0.0 0pt 0.0 0.0, width=0.5\linewidth]{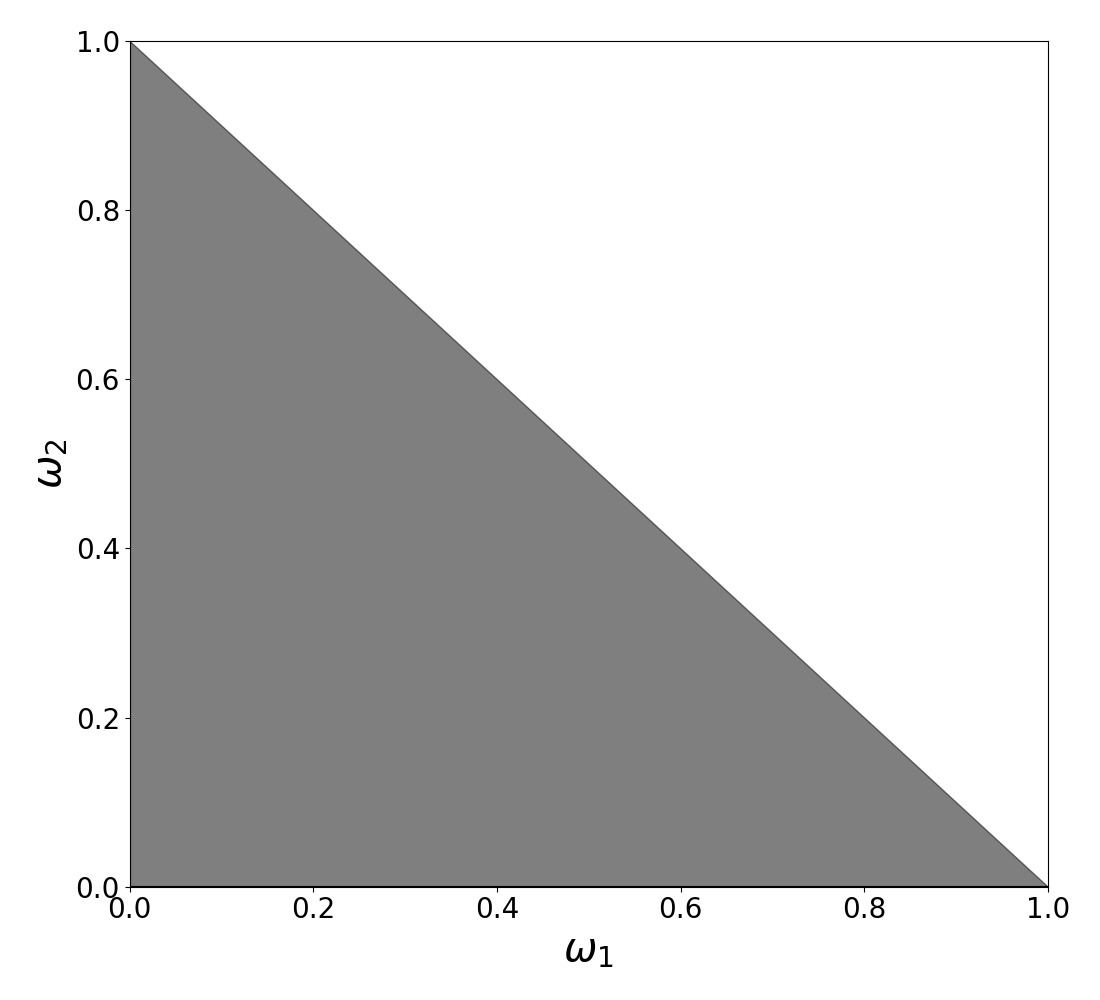}
		\label{fig:statespace}
	}\hspace{-10pt}
	\subfigure[]{
		\centering
		\includegraphics[trim=0.0 0pt 0.0 0.0, width=0.5\linewidth]{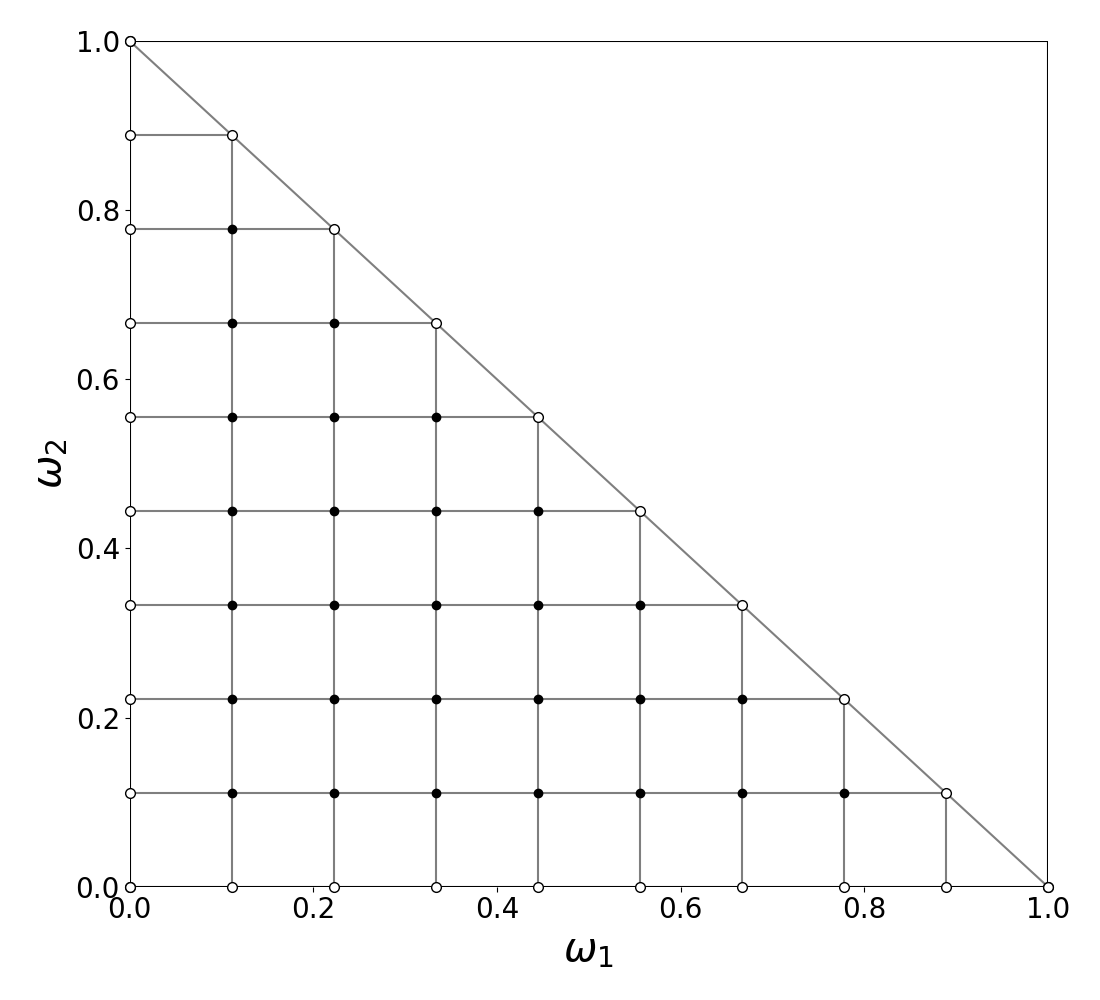}
		\label{fig:statespacedisc}
	}
	\caption{The two dimensional standard simplex (\cref{fig:statespace}) and its discretized version (\cref{fig:statespacedisc}) with $K=10$ points along each axis. Black points are interior points and white points are boundary points}
	\label{fig:statespace_combined}
\end{figure}

Consider first the state space when $N=3$. In this case $\bm{\omega} \in \{\bm{x} \in [0, 1]^2 \hspace{10pt} | \hspace{10pt} \| \bm{x} \| \leq 1 \}$. \cref{fig:statespace} plots this space as the shaded region, simply the lower triangle of the $[0, 1]$ square in the first quandrant.
To solve the PDE with finite difference we must discretize this space. I take the convention of specifying a discretization by the number $K$ of points along each axis. \cref{fig:statespacedisc} gives an example of such a discretization when $K=10$. If we define $L$ as the number of points in the state space we have $L = K*(K+1)/2$ points. Notice that along each edge of the simplex we have boundary conditions, so the number of interior points is $(K-1)(K-1)/2$.

\subsection{Boundary Vertices}\label{appendix:numerical:vertices}
The boundary conditions in the state space are very non-trivial. At each extreme point in the simplex, a single agent dominates. That is
\begin{align*}
(\omega_1, \omega_2) &= (1, 0) \Rightarrow \text{Agent $1$ dominates.}\\
(\omega_1, \omega_2) &= (0, 1) \Rightarrow \text{Agent $2$ dominates.}\\		(\omega_1, \omega_2) &= (0, 0) \Rightarrow \text{Agent $3$ dominates.}	
\end{align*}
For each $\bm{\omega_{vj}}$ in this set of vertices (subscript denotes vertex with dominant agent $j$), the dominant agent's price dominates. We have the following aggregate variables
\begin{align*}
\theta(\bm{\omega_{vj}}) &= \sigma_D \gamma_j\\
r(\bm{\omega_{vj}}) &= \rho + \mu_D\gamma_j - \gamma_j(1+\gamma_j)\frac{\sigma_D^2}{2}\\
\sigma(\bm{\omega_{vj}}) &= \sigma_D
\end{align*}
In addition, individuals may be constrained on the vertices. It can be shown that their adjustments satisfy
\begin{align*}
\nu_i(\bm{\omega_{vj}}) = \min\left\lbrace 0, (m_i\gamma_i - \gamma_j)\sigma_D^2 \right\rbrace
\end{align*}
Finally, the wealth/consumption ratios on these vertices are given by
\begin{equation*}
\begin{aligned}
V_i(\bm{\omega_{vj}}) = \frac{\gamma_i}{\rho - (1 - \gamma_i)\left( \frac{\left( \sigma_D + \nu_i(\bm{\omega_{vj}})/\sigma_D \right)^2}{2\gamma_i} + r(\bm{\omega_{vj}}) -m_i\nu_i(\bm{\omega_{vj}}) \right)} 
\end{aligned}	
\end{equation*}

\subsection{Boundary Edges}\label{appendix:numerical:edges}
Along a boundary edge, we are in a case where one agent has zero consumption weight and the other two agents have weight that varies. There are three cases:
\begin{align*}
(\omega_1, \omega_2) &\in \lbrace (x, y) \hspace{5pt} | \hspace{5pt} x \in [0, 1], y=0 \rbrace\\
(\omega_1, \omega_2) &\in \lbrace (x, y) \hspace{5pt} | \hspace{5pt} x=0, y \in [0, 1] \rbrace\\
(\omega_1, \omega_2) &\in \lbrace (x, y) \in [0, 1]^2 \hspace{5pt} | \hspace{5pt} x +y=1\rbrace
\end{align*}
If we take the state variable to be the consumption weight of the individual with the lowest index $j$ and who has non-zero weight along the edge, the PDEs become ODEs:
\begin{equation}
\begin{aligned}\label{eq:app:numerical:ode}
&\hspace{10pt}\frac{1}{\gamma_i} \left[ (1 - \gamma_i)(r(\omega_j) - m_i\nu_i(\omega_j)) - \rho + \frac{1 - \gamma_i}{2\gamma_i} \left(\theta(\omega_j) + \frac{\nu_i(\omega_j)}{\sigma(\omega_j)} \right)^2 \right] V_i(\omega_j) + \\
&\left[ \frac{1 - \gamma_i}{\gamma_i} \left(\theta(\omega_j) + \frac{\nu_i(\omega_j)}{\sigma(\omega_j)} \right) \sigma_{\omega j}\omega_j + \mu_{\omega j}\omega_j\right] V_i'(\omega_j) + \frac{1}{2} \sigma_{\omega j}^2 V_i''(\omega_j) + 1 = 0
\end{aligned}
\end{equation}
The vertex values in \cref{appendix:numerical:vertices} give boundary conditions for solving the two agent problem. Already this ODE problem is highly non-linear. To solve it I use an implicit scheme and Picard iteration. For the implicit method, we can add a time derivative term and consider the long run level of the wealth/consumption ratio to be $\Delta_t$, the length of the discrete time step. Discretize the state space along the edge into $P$ points $\omega_j^p = j/P$ $\forall$ $j \in \lbrace1, \dots, P\rbrace$. Denote $V_i(t, \omega_j^p)$ as $V_i^{t, p}$. In addition, define the coefficients in \cref{eq:app:numerical:ode} such that
\begin{equation*}
\begin{aligned}\label{eq:app:numerical:ode}
\partial_t V_i^{t, p} + a_i(\omega_j^p, \bm{V}^{t, p})V_i^{t, p} + b(\omega_j^p, \bm{V}^{t, p})\partial_\omega V_i^{t, p} + c(\omega_j^p, \bm{V}^{t, p})\partial^2_{\omega \omega}V_i^{t, p}  + 1 = 0
\end{aligned}
\end{equation*}
To carry out Picard iteration, initialize the solution at the terminal value, then evaluate the coefficients using the current solution guess and treat the derivatives as unknowns. Using the following second-order-accurate central difference schemes:
\begin{align*}
\partial_t V_i^{t, p} &\approx \frac{V_i^{t+1, p} - V_i^{t, p}}{\Delta_t}\\
\partial_\omega V_i^{t, p} &\approx \frac{V_i^{t, p+1} - V_i^{t, p-1}}{2h}\\
\partial^2_{\omega \omega}V_i^{t, p} &\approx \frac{V_i^{t, p+1} - 2V_i^{t,p} + V_i^{t, p-1}}{h^2}
\end{align*}
where $h = 1/P$, and using a similar superscripting scheme for the coefficients, the discretized scheme can be rearranged as
\begin{align*}
\left[\frac{c^{t+1, p}}{h^2} - \frac{b^{t+1, p}}{2h} \right]V_i^{t, p-1} + \left[ a^{t+1, p} - \frac{2c^{t+1, p}}{h^2} - \frac{1}{\Delta_t} \right]V_i^{t, p-1} + \left[ \frac{c^{t+1, p}}{h^2} + \frac{b^{t+1, p}}{2h} \right]V_i^{t, p-1}\\
= -\left[ 1 + \frac{V_i^{t+1, p}}{\Delta_t} \right]
\end{align*}
This system of linear equations can be written as
\begin{align*}
Ax = b
\end{align*}
where $A$ is tridiagonal. At each time step this system is solved and the algorithm stops when two consecutive steps are sufficiently close. An example of a solution to the two agent problem is given in \cref{appendix:twoagents}

\subsection{Full State Space}
Given the numerical solution in \cref{appendix:numerical:vertices,appendix:numerical:edges}, we have boundary values and can now turn to the full PDE. For two agents the PDE in \cref{prop:PDE} becomes
\begin{equation}
\begin{aligned}\label{eq:app:numerical:pde}
&\hspace{10pt}\frac{1}{\gamma_i} \left[ (1 - \gamma_i)(r(\bm{\omega}) + \delta_i(\nu_i(\bm{\omega}))) - \rho + \frac{1 - \gamma_i}{2\gamma_i} \left(\theta(\bm{\omega}) + \frac{\nu_i(\bm{\omega})}{\sigma(\bm{\omega})} \right)^2 \right] V_i(\bm{\omega}) + \\
&\left[ \frac{1 - \gamma_i}{\gamma_i} \left(\theta(\bm{\omega}) + \frac{\nu_i(\bm{\omega})}{\sigma(\bm{\omega})} \right) \sigma_{\omega 1}(\bm{\omega})\omega_1 + \mu_{\omega 1}(\bm{\omega})\omega_1\right] \partial_{\omega_1}V_i(\bm{\omega}) +  \\
&\left[ \frac{1 - \gamma_i}{\gamma_i} \left(\theta(\bm{\omega}) + \frac{\nu_i(\bm{\omega})}{\sigma(\bm{\omega})} \right) \sigma_{\omega 2}(\bm{\omega})\omega_2 + \mu_{\omega 2}(\bm{\omega})\omega_2\right] \partial_{\omega_2}V_i(\bm{\omega}) + \\
&\frac{1}{2}\left[ \sigma_{\omega 1}^2\omega_1^2 \partial^2_{\omega_1 \omega_1} V_i(\bm{\omega}) + \sigma_{\omega 2}^2\omega_2^2 \partial^2_{\omega_2 \omega_2} V_i(\bm{\omega}) + 2\sigma_{\omega 1}\sigma_{\omega 2}\omega_1\omega_2 \partial^2_{\omega_1 \omega_2} V_i(\bm{\omega}) \right] + 1 = 0
\end{aligned}
\end{equation}
As stated in \cref{appendix:numerical:discretize}, we discretize each edge into $K$ points, giving a total of $K(K+1)/2$ points in the state space. If we denote $\bm{\omega}^{jk} = [\omega_1^j, \omega_2^k]^T$ the $(j, k)$'th point in the state space and if we use a similar superscripting scheme as in \cref{appendix:numerical:edges}, we can use the following set of second-order-accurate central difference schemes:
\begin{align*}
\partial_t V_i^{t,j,k} &\approx \frac{V_i^{t+1, j,k} - V_i^{t, j,k}}{\Delta_t}\\
\partial_{\omega_1} V_i^{t,j,k} &\approx \frac{V_i^{t,j+1,k} - V_i^{t,j-1,k}}{2h}\\
\partial_{\omega_2} V_i^{t,j,k} &\approx \frac{V_i^{t,j,k+1} - V_i^{t,j,k-1}}{2h}\\
\partial^2_{\omega_1 \omega_1}V_i^{t,j,k} &\approx \frac{V_i^{t,j+1,k} - 2V_i^{t,j,k} + V_i^{t, j-1,k}}{h^2}\\
\partial^2_{\omega_2 \omega_2}V_i^{t,j,k} &\approx \frac{V_i^{t,j,k+1} - 2V_i^{t,j,k} + V_i^{t, j,k-1}}{h^2}\\
\partial^2_{\omega_1 \omega_2}V_i^{t,j,k} &\approx \frac{V_i^{t,j+1,k+1} - V_i^{t,j-1,k+1} - V_i^{t,j+1,k-1} + V_i^{t, j-1,k-1}}{4h^2}
\end{align*}
If in addition we use the same Picard iteration scheme, we can discretize the equation as
\begin{align*}
&a_{i,1}^{t+1,j,k} V_i^{t,j,k} + a_{i,2}^{t+1,j,k} V_i^{t,j+1,k} + a_{i,3}^{t+1,j,k} V_i^{t,j-1,k} + a_{i,4}^{t+1,j,k} V_i^{t,j,k+1} + a_{i,5}^{t+1,j,k} V_i^{t,j,k-1} + \\
&a_{i,6}^{t+1,j,k} V_i^{t,j+1,k+1} + a_{i,7}^{t+1,j,k} V_i^{t,j+1,k-1} + a_{i,8}^{t+1,j,k} V_i^{t,j-1,k+1} + a_{i,9}^{t+1,j,k} V_i^{t,j-1,k-1} = b_i^{t+1,j,k}
\end{align*}
where the coefficients are given by
\begin{equation*}
\begin{aligned}\label{eq:app:numerical:pde}
a_{i,1}^{t,j,k} &= \frac{1}{\gamma_i} \left[ (1 - \gamma_i)(r^{t+1,j,k} - m_i\nu_i^{t+1,j,k}) - \rho + \frac{1 - \gamma_i}{2\gamma_i} \left(\theta^{t+1,j,k} + \frac{\nu_i^{t+1,j,k}}{\sigma^{t+1,j,k}} \right)^2 \right] \\
& - \frac{(\sigma_{\omega 1}^{t+1,j,k}\omega_1^{j,k})^2}{h^2} - \frac{(\sigma_{\omega 2}^{t+1,j,k}\omega_2^{j,k})^2}{h^2} - \frac{1}{\Delta_t}\\
a_{i,2}^{t,j,k} &= \frac{(\sigma_{\omega 1}^{t+1,j,k}\omega_1^{j,k})^2}{2h^2} + \frac{1}{2h}\left[ \frac{1 - \gamma_i}{\gamma_i} \left(\theta^{t+1,j,k} + \frac{\nu_i^{t+1,j,k}}{\sigma^{t+1,j,k}} \right) \sigma_{\omega 1}^{t+1,j,k}\omega_1^{j,k} + \mu_{\omega 1}^{t+1,j,k}\omega_1^{j,k} \right]\\ 
a_{i,3}^{t,j,k} &= \frac{(\sigma_{\omega 1}^{t+1,j,k}\omega_1^{j,k})^2}{2h^2} - \frac{1}{2h}\left[ \frac{1 - \gamma_i}{\gamma_i} \left(\theta^{t+1,j,k} + \frac{\nu_i^{t+1,j,k}}{\sigma^{t+1,j,k}} \right) \sigma_{\omega 1}^{t+1,j,k}\omega_1^{j,k} + \mu_{\omega 1}^{t+1,j,k}\omega_1^{j,k} \right] \\ 
a_{i,4}^{t,j,k} &= \frac{(\sigma_{\omega 2}^{t+1,j,k}\omega_2^{j,k})^2}{2h^2} + \frac{1}{2h}\left[ \frac{1 - \gamma_i}{\gamma_i} \left(\theta^{t+1,j,k} + \frac{\nu_i^{t+1,j,k}}{\sigma^{t+1,j,k}} \right) \sigma_{\omega 2}^{t+1,j,k}\omega_2^{j,k} + \mu_{\omega 2}^{t+1,j,k}\omega_2^{j,k} \right] \\ 
a_{i,5}^{t,j,k} &= \frac{(\sigma_{\omega 2}^{t+1,j,k}\omega_2^{j,k})^2}{2h^2} - \frac{1}{2h}\left[ \frac{1 - \gamma_i}{\gamma_i} \left(\theta^{t+1,j,k} + \frac{\nu_i^{t+1,j,k}}{\sigma^{t+1,j,k}} \right) \sigma_{\omega 2}^{t+1,j,k}\omega_2^{j,k} + \mu_{\omega 2}^{t+1,j,k}\omega_2^{j,k} \right] \\ 
a_{i,6}^{t,j,k} &= \frac{\sigma_{\omega 1}^{t+1,j,k}\sigma_{\omega 2}^{t+1,j,k}\omega_1^{j,k}\omega_2^{j,k}}{4h^2}\\
a_{i,7}^{t,j,k} &= -\frac{\sigma_{\omega 1}^{t+1,j,k}\sigma_{\omega 2}^{t+1,j,k}\omega_1^{j,k}\omega_2^{j,k}}{4h^2}\\
a_{i,8}^{t,j,k} &= -\frac{\sigma_{\omega 1}^{t+1,j,k}\sigma_{\omega 2}^{t+1,j,k}\omega_1^{j,k}\omega_2^{j,k}}{4h^2}\\
a_{i,9}^{t,j,k} &= \frac{\sigma_{\omega 1}^{t+1,j,k}\sigma_{\omega 2}^{t+1,j,k}\omega_1^{j,k}\omega_2^{j,k}}{4h^2}\\
b_i^{t,j,k} &= -1 - \frac{V_i^{t+1,j,k}}{\Delta_t}
\end{aligned}
\end{equation*}

\subsection{The Diagonal Boundary}
One issue is that across the diagonal boundary the finite difference approximation will reach beyond the state space. 
\begin{figure}[!h]
	\centering
	\includegraphics[trim=0.0 0pt 0.0 0.0, width=0.9\linewidth]{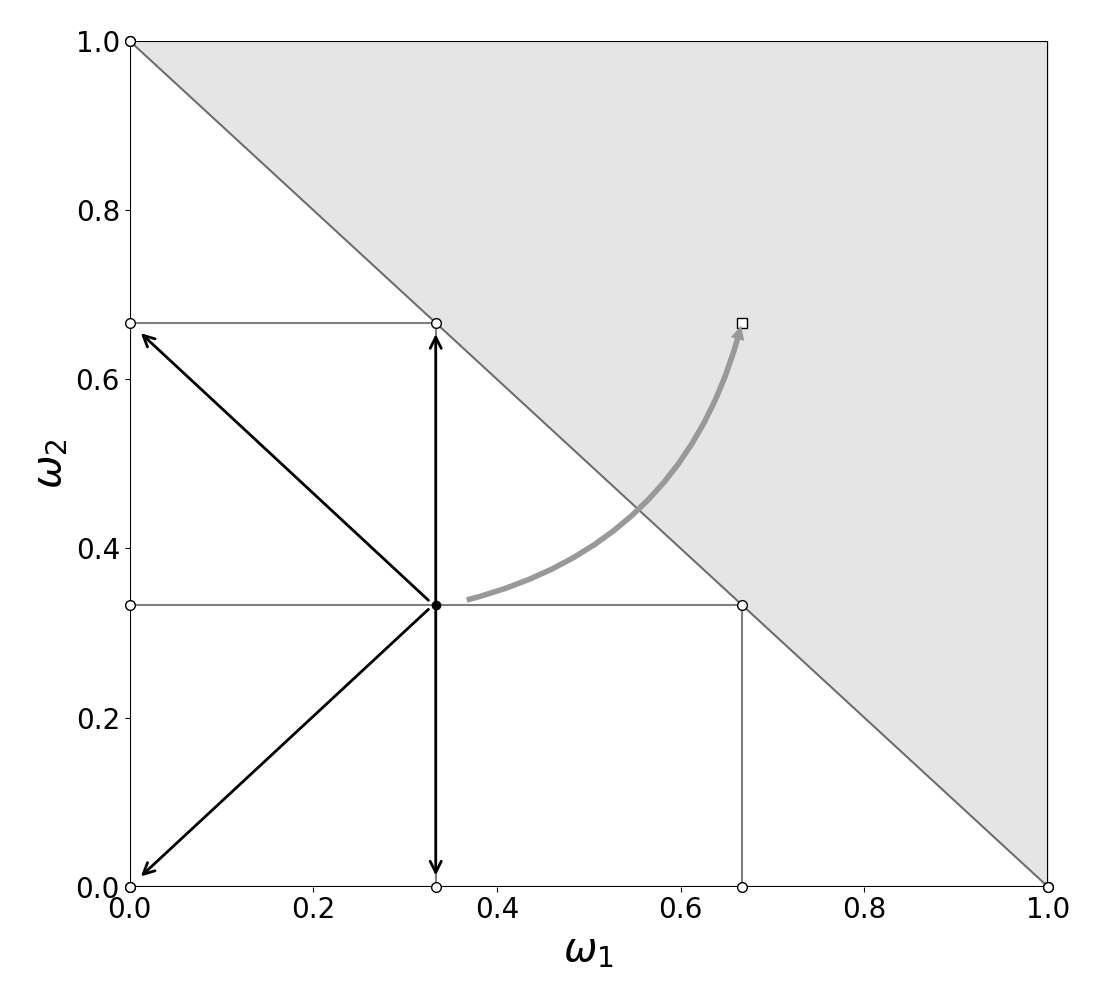}
	\caption{When $K=4$ we have the minimum number of points to have an interior point. When we attempt to approximate the cross partial at this point the finite difference approximation attempts to reach across the diagonal (curved arrow). To deal with this we can introduce a particular difference approximation (\cref{eq:diag_bound_difference}), which instead makes use of the points which exist in the state space (straight arrows).}
	\label{fig:diagonal_boundary}
\end{figure}
To deal with this issue, we can define an additional set of boundary points along the diagonal boundary and use a special finite difference scheme to approximate this cross partial. Since $\partial^2_{xy}f(x,y) = \partial_x(\partial_yf(x,y))$, we can use a central difference in one direction and a backward difference in the other. This will allow us to avoid crossing the boundary. This approximation is given by
\begin{align}
\partial^2_{xy} f(x, y) \approx \frac{1}{2h^2}\left[ f_{x,y+h} - f_{x-h,y+h} - f_{x,y-h} + f_{x-h,y-h} \right] \label{eq:diag_bound_difference}
\end{align}
Using this discretization for the cross-partial at the diagonal boundary, we have the following discretization at each index $(j, K-j)$:
If in addition we use the same Picard iteration scheme, we can discretize the equation as
\begin{align*}
&\tilde{a}_{i,1}^{t+1,j,K-j} V_i^{t,j,K-j} + \tilde{a}_{i,2}^{t+1,j,K-j} V_i^{t, j+1, K-j} +  \tilde{a}_{i,3}^{t+1,j,K-j} V_i^{t,j-1,K-j} + \tilde{a}_{i,4}^{t+1,j,K-j} V_i^{t,j,K-j+1} \\
&+ \tilde{a}_{i,5}^{t+1,j,K-j} V_i^{t,j,K-j-1} + \tilde{a}_{i,6}^{t+1,j,K-j}  V_i^{t,j-1,K-j-1} + \tilde{a}_{i,7}^{t+1,j,K-j}V_i^{t,j-1,K-j+1} = b_i^{t+1,j,k}
\end{align*}
where the coefficients are given by
\begin{equation*}
\begin{aligned}\label{eq:app:numerical:pde}
\tilde{a}_{i,1}^{t,j,K-j} &= \frac{1}{\gamma_i} \left[ (1 - \gamma_i)(r^{t+1,j,K-j} - m_i\nu_i^{t+1,j,K-j}) - \rho + \frac{1 - \gamma_i}{2\gamma_i} \left(\theta^{t+1,j,K-j} + \frac{\nu_i^{t+1,j,K-j}}{\sigma^{t+1,j,K-j}} \right)^2 \right] \\
& - \frac{(\sigma_{\omega 1}^{t+1,j,K-j}\omega_1^{j,K-j})^2}{h^2} - \frac{(\sigma_{\omega 2}^{t+1,j,K-j}\omega_2^{j,K-j})^2}{h^2} - \frac{1}{\Delta_t} \\
&= a_{i,1}^{t,j,K-j}\\
\tilde{a}_{i,2}^{t,j,K-j} &= \frac{1}{2h}\left[ \frac{1 - \gamma_i}{\gamma_i} \left(\theta^{t+1,j,K-j} + \frac{\nu_i^{t+1,j,K-j}}{\sigma^{t+1,j,K-j}} \right) \sigma_{\omega 1}^{t+1,j,K-j}\omega_1^{j,K-j} + \mu_{\omega 1}^{t+1,j,K-j}\omega_1^{j,K-j} \right] \\
& + \frac{(\sigma_{\omega 1}^{t+1,j,K-j}\omega_1^{j,K-j})^2}{h^2} \\
& = a_{i,2}^{t,j,K-j}\\
\tilde{a}_{i,3}^{t,j,K-j} &= \frac{(\sigma_{\omega 1}^{t+1,j,K-j}\omega_1^{j,K-j})^2}{2h^2}\\ 
& - \frac{1}{2h}\left[ \frac{1 - \gamma_i}{\gamma_i} \left(\theta^{t+1,j,K-j} + \frac{\nu_i^{t+1,j,K-j}}{\sigma^{t+1,j,K-j}} \right) \sigma_{\omega 1}^{t+1,j,K-j}\omega_1^{j,K-j} + \mu_{\omega 1}^{t+1,j,K-j}\omega_1^{j,K-j} \right] \\
&= a_{i,3}^{t,j,K-j} \\
\tilde{a}_{i,4}^{t,j,K-j}	&= \frac{1}{2h}\left[ \frac{1 - \gamma_i}{\gamma_i} \left(\theta^{t+1,j,K-j} + \frac{\nu_i^{t+1,j,K-j}}{\sigma^{t+1,j,K-j}} \right) \sigma_{\omega 2}^{t+1,j,K-j}\omega_2^{j,K-j} + \mu_{\omega 2}^{t+1,j,K-j}\omega_2^{j,K-j} \right] \\
& + \frac{(\sigma_{\omega 2}^{t+1,j,K-j}\omega_2^{j,K-j})^2}{h^2} + \frac{\sigma_{\omega 1}^{t+1,j,K-j}\sigma_{\omega 2}^{t+1,j,K-j}\omega_1^{j,K-j}\omega_2^{j,K-j}}{2h^2}\\
&= a_{i,4}^{t,j,K-j} + 2a_{i,6}^{t,j,K-j} \\
\tilde{a}_{i,5}^{t,j,K-j}	&= \frac{(\sigma_{\omega 2}^{t+1,j,K-j}\omega_2^{j,K-j})^2}{h^2} - \frac{\sigma_{\omega 1}^{t+1,j,K-j}\sigma_{\omega 2}^{t+1,j,K-j}\omega_1^{j,K-j}\omega_2^{j,K-j}}{2h^2}\\
&- \frac{1}{2h}\left[ \frac{1 - \gamma_i}{\gamma_i} \left(\theta^{t+1,j,K-j} + \frac{\nu_i^{t+1,j,K-j}}{\sigma^{t+1,j,K-j}} \right) \sigma_{\omega 2}^{t+1,j,K-j}\omega_2^{j,K-j} + \mu_{\omega 2}^{t+1,j,K-j}\omega_2^{j,K-j} \right] \\
&= a_{i,5}^{t,j,K-j} - 2a_{i,6}^{t,j,K-j}\\
\tilde{a}_{i,6}^{t,j,K-j} &= \frac{\sigma_{\omega 1}^{t+1,j,K-j}\sigma_{\omega 2}^{t+1,j,K-j}\omega_1^{j,K-j}\omega_2^{j,K-j}}{2h^2} = 2a_{i,6}^{t,j,K-j}\\
\tilde{a}_{i,7}^{t,j,K-j} &= -\tilde{a}_{i,6}^{t,j,K-j} = 2a_{i,7}^{t,j,K-j}\\
b_i^{t,j,K-j} &= -1 - \frac{V_i^{t+1,j,k}}{\Delta_t} \\
\end{aligned}
\end{equation*}
where the constant is much larger because points where $j + k = K+1$ are boundary points.

\subsection{Linear Equations and Sparsity Pattern}
\begin{figure}[!h]
	\centering
	\includegraphics[trim=0.0 0pt 0.0 0.0, width=0.9\linewidth]{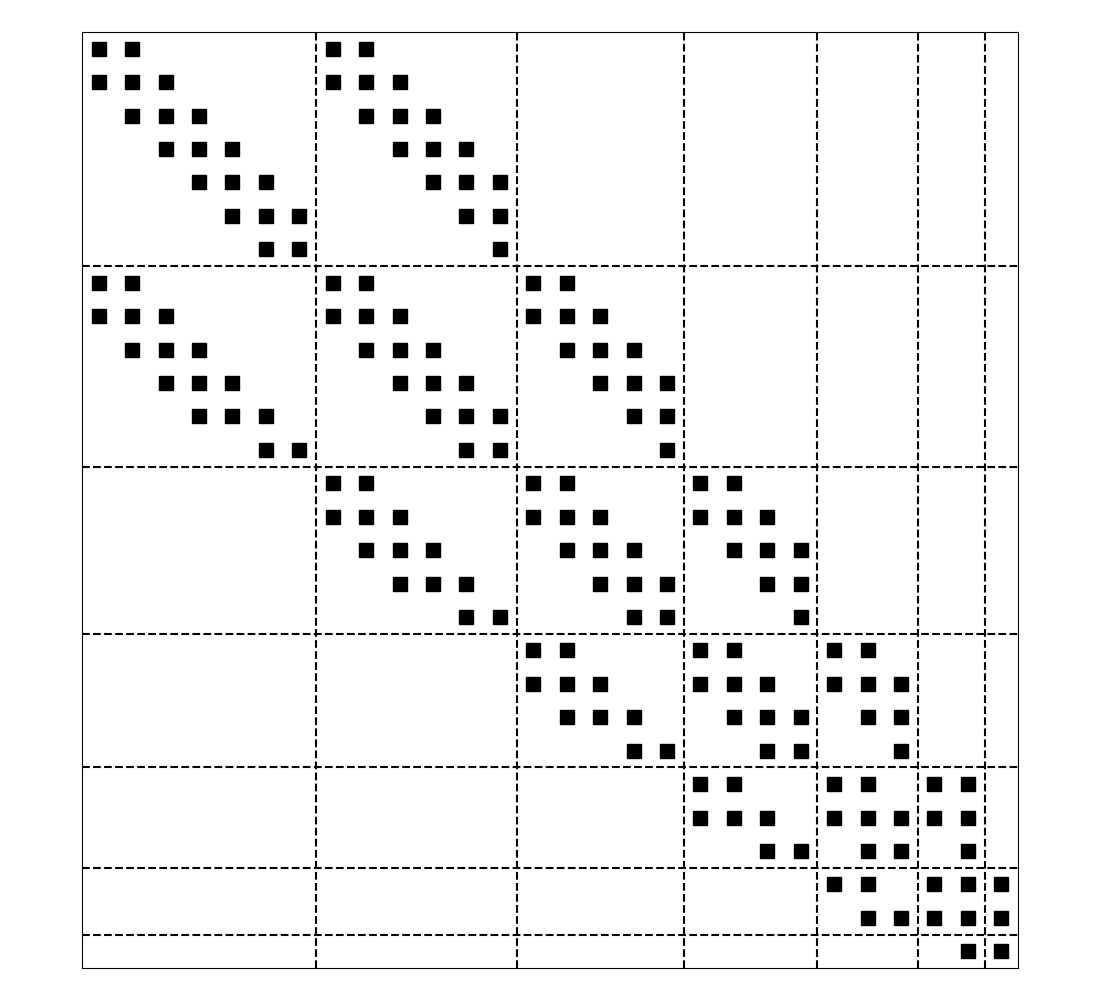}
	\caption{Sparsity pattern for the $A$ matrix in generated by the finite difference discretization of the PDE for three agents when the number of points on each axis is $K=10$.}
	\label{fig:sparsity}
\end{figure}

As in the two agent case we again have a system of linear equations given by
\begin{align*}
Ax = b
\end{align*}
However, now $A$ is block tri-diagonal and exhibits a special sparsity pattern. Because each subsecquent row or column in the state space is one point shorter, the blocks of $A$ are shrinking in size. As one rolls across columns, one also encounters boundary conditions. To give you an idea of the shape of $A$, here is the upper left corner of the matrix(if $j$ corresponds to rows and $k$ to columns of $A$, implying a C-style ordering of the points in the state space)
\begin{align*}
\hspace{-50pt}
\begin{bmatrix}[ccccc:cccc:cc]
a_{i,1}^{t+1,1,1} & a_{i,2}^{t+1,1,1} & 0 & \dots & \dots & a_{i,4}^{t+1,1,1} & a_{i,6}^{t+1,1,1}  & 0 & \dots & \dots \\
a_{i,3}^{t+1,1,1} & a_{i,1}^{t+1,1,1} & a_{i,2}^{t+1,1,1} & 0 & \dots  & a_{i,8}^{t+1,1,1} & a_{i,4}^{t+1,1,1} & a_{i,6}^{t+1,1,1}  & 0 & \dots \\
0 & \ddots & \ddots& \ddots & \ddots & 0 & \ddots & \ddots & \ddots & \ddots \\
0 & \dots & a_{i,3}^{t+1,K-2,1} & a_{i,1}^{t+1,K-2,1} & a_{i,2}^{t+1,K-2,1}  & 0 & \dots &a_{i,8}^{t+1,K-2,1} & a_{i,4}^{t+1,K-2,1} &  \\
0 & \dots & 0 & a_{i,3}^{t+1,K-1,1} & a_{i,1}^{t+1,K-1,1} & 0 & \dots & \dots &a_{i,8}^{t+1,K-1,1} & \\
\hdashline
a_{i,5}^{t+1,1,2} & a_{i,7}^{t+1,1,2} & 0 & \dots & \dots & a_{i,1}^{t+1,1,2} & a_{i,2}^{t+1,1,2}  & 0 & \dots & a_{i,4}^{t+1,1,2} \\
a_{i,9}^{t+1,2,2} & a_{i,5}^{t+1,2,2} & a_{i,7}^{t+1,2,2} & 0 & \dots  & a_{i,3}^{t+1,2,2} & a_{i,1}^{t+1,2,2} & a_{i,2}^{t+1,2,2}  & 0 & \dots \\
0 & \ddots & \ddots& \ddots & \ddots & 0 & \ddots & \ddots & \ddots & \ddots \\
\vdots  & a_{i,9}^{t+1,K-3,2} & a_{i,5}^{t+1,K-3,2} & a_{i,7}^{t+1,K-3,2} & 0 & \dots & a_{i,3}^{t+1,K-3,2} & a_{i,1}^{t+1,K-3,2} & a_{i,2}^{t+1,K-3,2} &  \\
\vdots & 0 & a_{i,9}^{t+1,K-2,2} & a_{i,5}^{t+1,K-2,2} & a_{i,7}^{t+1,K-2,2} & \dots & \dots & a_{i,3}^{t+1,K-3,2} & a_{i,1}^{t+1,K-3,2} &  \\	
\hdashline
0 & \ddots & \ddots& \ddots & \ddots & a_{i,5}^{t+1,1,3} & a_{i,7}^{t+1,1,3} & \ddots & \ddots & \ddots \\
\end{bmatrix}
\end{align*}
The dotted lines represent moves from one column of the state space to another. You can see that the diagonal blocks have a semetric sparsity pattern, while the off diagonal blocks do not. Moving down a column of blocks, each block is one row smaller. This pattern can be seen more clearly in \cref{fig:sparsity}.

\subsection{Computing Adjustments}
The final detail to be taken care of is how to compute the coefficients. The adjustments $\lbrace \nu_{it} \rbrace_i$ satisfy a system of non-linear equations. In addition, conditional on knowledge of $\bm{V}$ and $\bm{J_V}$, one must simultaneously solve for volatility. Thus, on each iteration, given a solution $\bm{V}^{t,j,k}$, compute the price dividend ratio $\mathcal{S}^{t,j,k}$, then gather the equations for adjustments in \cref{prop:adjustment_margin} and for volatility in \cref{prop:asset_prices}. Using a numerical solver, look for a root of this system of $N+1$ equations. Currently I am using the Scipy optimization module's root function and the "hybrid" algorithm, an implimentation of the Minpack "hybrd" function, which uses a modified Powell method.

\end{document}